\documentclass{aastex62}
\usepackage[utf8]{inputenc}
\usepackage{graphicx}
\usepackage{rotating}
\usepackage{subfigure}

\begin{document}
\title[Superorbital modulations in three supergiant HMXBs]{Investigating the superorbital modulations in 4U 1909+07, IGR J16418--4532 and IGR J16479--4514 with Swift XRT, BAT and NuSTAR observations}

\author{Nazma Islam}
\affil{Center for Space Science and Technology, University of Maryland, Baltimore County, 1000 Hilltop Circle, Baltimore, MD 21250, USA}
\affil{X-ray Astrophysics Laboratory, NASA Goddard Space Flight Center, Greenbelt, MD 20771, USA}

\author{Robin~H.~D. Corbet}
\affil{CRESST and CSST, University of Maryland, Baltimore County, 1000 Hilltop Circle, Baltimore, MD 21250, USA}
\affil{X-ray Astrophysics Laboratory, NASA Goddard Space Flight Center, Greenbelt, MD 20771, USA}
\affil{Maryland Institute College of Art, 1300 W Mt Royal Ave, Baltimore, MD 21217, USA}

\author{Joel~B. Coley}
\affil{CRESST/Department of Physics and Astronomy, Howard University, Washington DC 20059, USA}
\affil{Code 661 Astroparticle Physics Laboratory, NASA Goddard Space Flight Center, Greenbelt, MD 20771, USA}

\author{Katja Pottschmidt}
\affil{CRESST and CSST, University of Maryland, Baltimore County, 1000 Hilltop Circle, Baltimore, MD 21250, USA}
\affil{Code 661 Astroparticle Physics Laboratory, NASA Goddard Space Flight Center, Greenbelt, MD 20771, USA}

\author{Felix Fuerst}
\affil{Quasar Science Resources S.L for European Space Agency (ESA), European Space Astronomy Centre (ESAC), Camino Bajo del Castillo s/n, 28692 Villanueva de la Ca$\tilde{n}$ada, Madrid, Spain }

\correspondingauthor{Nazma Islam}
\email{nislam@umbc.edu}

\begin{abstract}
A puzzling variety of superorbital modulations have been discovered in several supergiant High-Mass X-ray binaries (sgHMXBs). To investigate the mechanisms driving these superorbital modulations, we have analyzed long-term Neil Gehrels Swift Observatory (Swift) Burst Alert Telescope (BAT) observations of three sgHMXBs: 4U 1909+07, IGR J16418--4532 and IGR J16479--4514 and constructed their dynamic power spectra and superorbital intensity profiles. These Swift BAT observations are complemented by pointed Swift X-ray Telescope (XRT) and Nuclear Spectroscopic Telescope Array (NuSTAR) observations performed near the predicted maximum and minimum phase of a single superorbital cycle for each of these sources. The BAT dynamic power spectra show changes in the strength of the superorbital modulation on timescales of years, with either the peak at the fundamental frequency and/or the second harmonic present at different times for all three sources. The pointed Swift XRT and NuSTAR observations show no significant differences between the pulse profiles and spectral parameters at the superorbital maximum and minimum phase. This is likely due to the fact the superorbital modulation had weakened significantly during the times when the NuSTAR observations were carried out for all three sources. The results from the Swift XRT, BAT and NuSTAR analysis indicate the possible presence of multiple co-rotating interaction regions (CIRs) in the stellar winds of the supergiant stars, although a structured stellar wind from the supergiant star due to tidal oscillations cannot be ruled out.

\end{abstract}

\section{Introduction}
In addition to pulsations and orbital periods, a third type of periodicity is seen in many neutron star X-ray binaries. While the first two periodicities are attributed to the spin period of the neutron star and the binary orbit of the system, the causes of the third type of periodicity which is longer than the orbital period and therefore called superorbital modulation, are still not fully understood. For systems where the accretion occurs by a Roche lobe overflow via an accretion disk onto the neutron star like Her X--1 \citep{leahy2002,staubert2009,brumback2021}, SMC X--1 \citep{wojdowski1998,clarkson2003a,brumback2020}, LMC X--4 \citep{lang1981,brumback2021}, the superorbital modulation is attributed to the precession of the accretion disks, powered by the irradiation of the accretion disks by the NS \citep{ogilvie2001} or due to a change in the torque from an accretion disk \citep{Hu2019}. For High-Mass X-ray binaries (HMXBs) consisting of a Be star as a companion (Be-HMXBs), superorbital variability may be linked to the formation and dissipation of a circumstellar `decretion’ disk around these systems \citep{rajoelimanana2011}. Superorbital variability has also been discovered in Ultra-luminous X-ray sources (ULXs) {\it e.g.} in M82 X--2 ($\sim$ 60 d; \citealt{brightman2019}), NGC 5907 ULX--1 ($\sim$ 78 d; \citealt{walton2016}), where a strong correlation is found between their orbital and superorbital periods \citep{townsend2020}.
\par
 Supergiant High Mass X-ray binaries (sgHMXBs) are X-ray binaries with mostly a neutron star as the compact object and an early type O or B supergiant star as an optical companion. Supergiant Fast X-ray Transients (SFXTs) are a subclass of sgHMXBs which are characterized by a rapid flaring behavior during which the X-ray intensity increases a factor of $\sim 10^{3-4}$ or more. The accretion in sgHMXBs is mostly driven by strong stellar winds which are stochastic in nature and provide a less regularly structured environment compared to an accretion disk. However superorbital modulations have been discovered in 6 sources at about three to four times their orbital period: 2S 0114+650 \citep{farrell2006}, IGR J16493--4348, IGR J16418--4532, IGR J16479--4514, 4U 1909+07 \citep{corbet2013, drave2013a} and 4U 1538--522 \citep{corbet2021}. Since these systems do not have a `decretion’ disk like Be-HMXBs and may or may not have a transient accretion disk, the mechanisms driving these superorbital modulations are still unknown.
 \par
Previous observations of 2S 0114+650 and IGR J16493--4348 as a function of their superorbital cycles showed a hardening in the spectral shape, which is indicative of a change in the mass transfer rate driving these superorbital modulations \citep{farrell2008, coley2019}. A possible mechanism driving these changes in the mass transfer rate could be tidal oscillations induced in non-synchronously rotating stars \citep{koenigsberger2006}. \cite{bozzo2017} proposed the presence of `Corotating Interaction Regions’ (CIRs) in the winds of these OB supergiant stars, which are characterized by spiral shaped higher density and velocity structures in the winds of the OB supergiant stars, extending up to tens of stellar radii. The beat frequency between a CIR and the NS orbit would cause this variation in X-ray intensities on superorbital timescales in sgHMXBs. Other models invoked to explain these intensity variations on superorbital timescales are neutron star precession models \citep{postnov2013}, and the presence of a third star in a  hierarchical system \citep{chou2001}, though they are unlikely to account for the superorbital modulations seen in HMXBs \citep{coley2019}. 
\par
Superorbital modulations have been detected in 4U 1909+07, IGR J16418--4532 and IGR J16479--4514 using long-term Swift BAT lightcurves in 15--50 keV energy-band \citep{corbet2013}. The superorbital period estimated for 4U 1909+07, IGR J16418--4532 and IGR J16479--4514 is $\sim$15.18 d, $\sim$14.73 d and $\sim$11.88 d respectively. In this paper, we analyzed the long-term lightcurves of Swift BAT and pointed NuSTAR and Swift XRT observations of the above three sgHMXBs carried out at their predicted superorbital maximum and minimum phases from the ephemerides given in \citet{corbet2013}. The results from this analysis are used to investigate the possible mechanism driving these superorbital modulations. 

\subsection{Sample of supergiant HMXBs showing superorbital modulations}
4U 1909+07 is an accreting pulsar with a spin period of $\sim$ 605 s and an orbital period of $\sim$ 4.4 d \citep{levine2004}. The optical companion is an early B-type (B0-B3) star, probably in its supergiant phase and the distance to the system is estimated to be 4.85$\pm$0.5 kpc \citep{martinez2015}. The system does not show the presence of an X-ray eclipse however, the X-ray flux is strongly modulated on its 4.4 d orbital period. A previous Chandra observation suggested the presence of a Compton shoulder on the Fe K$\alpha$ line, which indicates the system is embedded in a Compton-thick environment \citep{torrejon2010}. The spectrum of 4U 1909+07 from Rossi X-ray Timing Explorer (RXTE) Proportional Counter Array (PCA), Suzaku and INTErnational Gamma-Ray Astrophysics Laboratory (INTEGRAL) Soft Gamma Ray Imager (ISGRI) observations revealed the presence of a large neutral hydrogen column density $N_{\rm{H}} \sim 10^{23}$ cm$^{-2}$ and a power-law spectrum with a photon index of $\sim$ 1 with a high energy cut-off of $\sim$ 20 keV, typical of a HMXB pulsar \citep{furst2011,furst2012}. 
\par
IGR J16418--4532 is a candidate intermediate SFXT characterized by short X-ray flares (on the timescales of an hour), reaching a dynamic X-ray flux range of $\sim 10^{2}$ \citep{sguera2006,ducci2010}. Bright X-ray flares have been observed from this system reaching an X-ray luminosity of 10$^{37}$ ergs \,s$^{-1}$ \citep{romano2012,romano2012a,krimm2013a,romano2015}. From Swift XRT observations of this source, \cite{romano2012} found evidence of an inhomogenous clumpy wind in the system, which could give rise to short X-ray flares seen in the lightcurve. It is an eclipsing X-ray binary of orbital period $\sim$ 3.73 d \citep{corbet2006} and spin period of $\sim$ 1210 s \citep{sidoli2012}. The minimum distance to this binary system is estimated to be 13 kpc for a companion of the spectral type BN0.5Ia \citep{coleiro2013} although, from an eclipse timing analysis, \citet{coley2015} found the spectral type of the optical companion star to be O8.5I or earlier. The X-ray spectrum of IGR J16418--4532 is highly absorbed with $N_{\rm{H}} \sim 10^{23}$ cm$^{-2}$ and a power-law spectrum with a photon index of $\sim$ 1 \citep{drave2013, sidoli2012}. Using a 40\,ks XMM-Newton observation, \cite{sidoli2012} proposed that the observed X-ray variability and the quasi-periodic flaring activity are indicative of a neutron star accreting in a Transitional Roche Lobe Overflow regime i.e between pure wind accretion and Roche Lobe overflow.  
\par
IGR J16479--4514 is an eclipsing SFXT with an orbital period of 3.32 d \citep{jain2009}. The nature of the compact object is unknown due to an absence of pulsations detected in the system. However, its spectrum is similar to that of accreting HMXB pulsars, suggesting a neutron star as the putative compact object. The optical companion is likely to be an O7 star and this suggests a distance of 4.5 kpc to the binary system \citep{chaty2008,coley2015}. In addition to short low luminosity X-ray flares lasting a few thousand seconds, there are orbital phase-locked X-ray flares present in the lightcurve, with an occasional bright X-ray flare lasting few hours and reaching an X-ray flux of $10^{-9}$ ergs \,cm$^{-2}$ \,s$^{-1}$ \citep{sguera2008,romano2008, sidoli2013, sguera2020}. These flares indicate the presence of large-scale structures in the stellar wind of the supergiant star \citep{bozzo2009,sguera2020}.

\section{Data and Analysis}

\subsection{Long-term observations with Swift BAT}

\begin{table}
\caption{Sample of sgHMXBs with their orbital periods, superorbital periods, orbital and superorbital ephemerides}
\centering
\begin{tabular}{c c c c c c}
\hline
Name & $P_{\rm{orb}}$ & $T_{\rm{mid}}$ & $P_{\rm{super}}^{\dagger}$ & $T_{0}^{\dagger}$ & Spectral Type \\ 
\hline
4U 1909+07 & 4.4$^{(a)}$ & - & 15.196$\pm$0.004 & 59502.1$\pm$0.1 & B0-B3$^{(c)}$ \\
IGR J16418--4532 & 3.73880$^{(b)}$ & 55087.714$^{(b)}$ & 14.702$\pm$0.004 & 59506.9$\pm$0.1 & O8.5I$^{(b)}$ \\
IGR J16479--4514 & 3.31961$^{(b)}$ & 55081.571$^{(b)}$ & 11.894$\pm$0.006 & 59507.7$\pm$0.1 & O7$^{(b)}$ \\
\hline
\end{tabular} \\
 Notes: (a)--\cite{levine2004}; (b)--\cite{coley2015}; (c)--\cite{martinez2015}; $\dagger$--This work; $P_{\rm{orb}}$ is the orbital period in days;  $P_{\rm{super}}$ is the superorbital period in days; $T_{\rm{mid}}$ is the mid-eclipse time; $T_{0}$ is the time of the maximum flux of the average superorbital intensity profile.
\end{table}

\begin{figure*}
    \centering
    \includegraphics[scale=0.5, angle=-90]{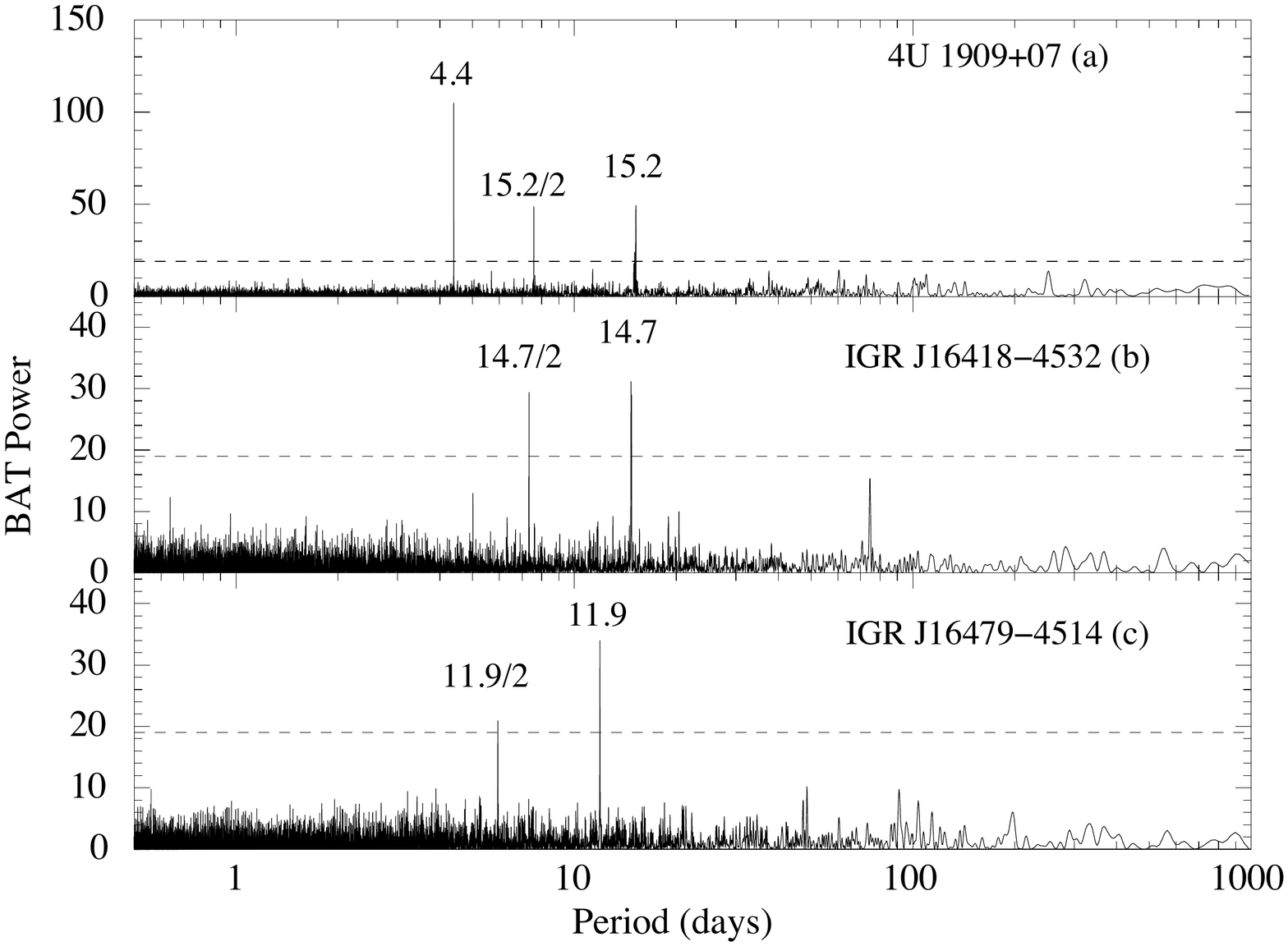}
    \caption{Power spectra from 16.7 years of Swift BAT lightcurves of (a) 4U 1909+07, (b) IGR J16418--4532, (c) IGR J16479--4514 in the energy band of 15--50 keV. The data-points around the X-ray eclipse were removed from the lightcurves of IGR J16418--4532 and IGR J16479--4514, for the computation of the power spectra. The dashed lines indicate ``white noise" 99.99\% significance levels. The power spectrum of 4U 1909+07 (a) exhibits prominent peaks at the orbital period of $\sim$4.4 d, superorbital period of $\sim$15.2 d and the second harmonic of the superorbital period. The power spectrum of IGR J16418--4532 (b) exhibits prominent peaks at the superorbital period of $\sim$14.7 d and the second harmonic. The power spectrum of IGR J16479--4514 (c) exhibits prominent peaks at the superorbital period of $\sim$11.9 d and the second harmonic}.
    \label{bat_power}
\end{figure*}

The Swift BAT is a hard X-ray telescope that uses a coded mask and operates in the 14--195\, keV energy band \citep{barthelmy2005}. We used lightcurves from the BAT Transient Monitor\footnote{\url{https:// swift.gsfc.nasa.gov/results/transients/}} \citep{krimm2013} in the 15--50\, keV energy band from MJD 53,416 to MJD 59,515. The lightcurves were further screened to exclude bad quality points and only use the data where the data quality flag ``DATA$\_$FLAG'' was set to 0. A small number of data points with very low fluxes and unrealistically small uncertainties were also removed from the lightcurves \citep{corbet2013}. The Swift BAT lightcurves of 4U 1909+07, IGR J16418--4532 and IGR J16479--4514 were used to construct the power spectra from MJD 53,415 to MJD 59,515 (16.7 years, compared to $\sim$ 8 years of Swift BAT lightcurve used in \citealt{corbet2013}). The power spectrum was calculated using the ``semi-weighting'' technique where the error bar on each data point and the excess variability in the lightcurve was taken into account \citep{corbet2007,corbet2013}. The significance of the peak at the superorbital period was estimated using the false-alarm probability (FAP; \citealt{scargle1982}). The uncertainty in the superorbital period measurement was obtained using the expression given in \cite{horne1986}. The ``white noise" 99.99\% significance level was calculated by fitting the continuum power levels in a narrow frequency range around the peak of interest. To investigate the contribution of the red noise to the power spectrum, we fitted the logarithm of the power as a function of the logarithm of the frequency with a quadratic function \citep{vaughan2005,corbet2008}. We found that the continuum is flat and therefore the contribution of the red noise to the power spectrum of these sources is negligible in the superorbital period range of $\sim$10 days.
\par
The power spectrum of 4U 1909+07 plotted in panel (a) of Figure \ref{bat_power} exhibit prominent peaks at the orbital period of $\sim$4.4 d, superorbital period of $\sim$15.2 d and the second harmonic of the superorbital period at $\sim$7.6 d. The peaks at the superorbital period and the second harmonic are of equal strengths with FAPs of $\sim 10^{-7}$. For IGR J16418--4532 and IGR J16479--4514 which have X-ray eclipses present in their lightcurves, the power spectra were constructed after removing the data-points lying inside the X-ray eclipse to mitigate the effects of the orbital modulation. We use the ephemerides given in Table 1 to remove the X-ray eclipses for IGR J16418--4532 and IGR J16479--4514 between the orbital phase of 0.85 to 1.15, where phase 1.0 was chosen to be the center of the X-ray eclipse. The power spectrum of IGR J16418--4532 plotted in panel (b) of Figure \ref{bat_power} exhibit prominent peaks at the superorbital period of $\sim$14.7 d and the second harmonic of $\sim$7.4 d with a FAP of $\sim 10^{-6}$. The power spectrum of IGR J16479--4514 plotted in panel (c) of Figure \ref{bat_power} exhibits a prominent peak at the superorbital period of $\sim$11.9 d with a FAP of $\sim 10^{-7}$ and a smaller peak at the second harmonic of $\sim$5.9 d. Removing X-ray eclipses from the lightcurve can cause possible aliasing effects in the power spectra of IGR J16418--4532 and IGR J16479--4514. To investigate this, the orbital intensity profiles were fitted with a sinusoidal modulation with an orbital period of 3.73 d and 3.32 d for IGR J16418--4532 and IGR J16479--4514 respectively. These fitted sine curves were used to simulate lightcurves using the times of the Swift BAT data and the power spectra were constructed from them. No aliasing at their superorbital modulation periods were found in the power spectra of IGR J16418--4532 and IGR J16479--4514. The refined superorbital periods of 4U 1909+07, IGR J16418--4532 and IGR J16479--4514 are given in Table 1.
\par
\begin{figure*}
    \centering
    \includegraphics[scale=0.4]{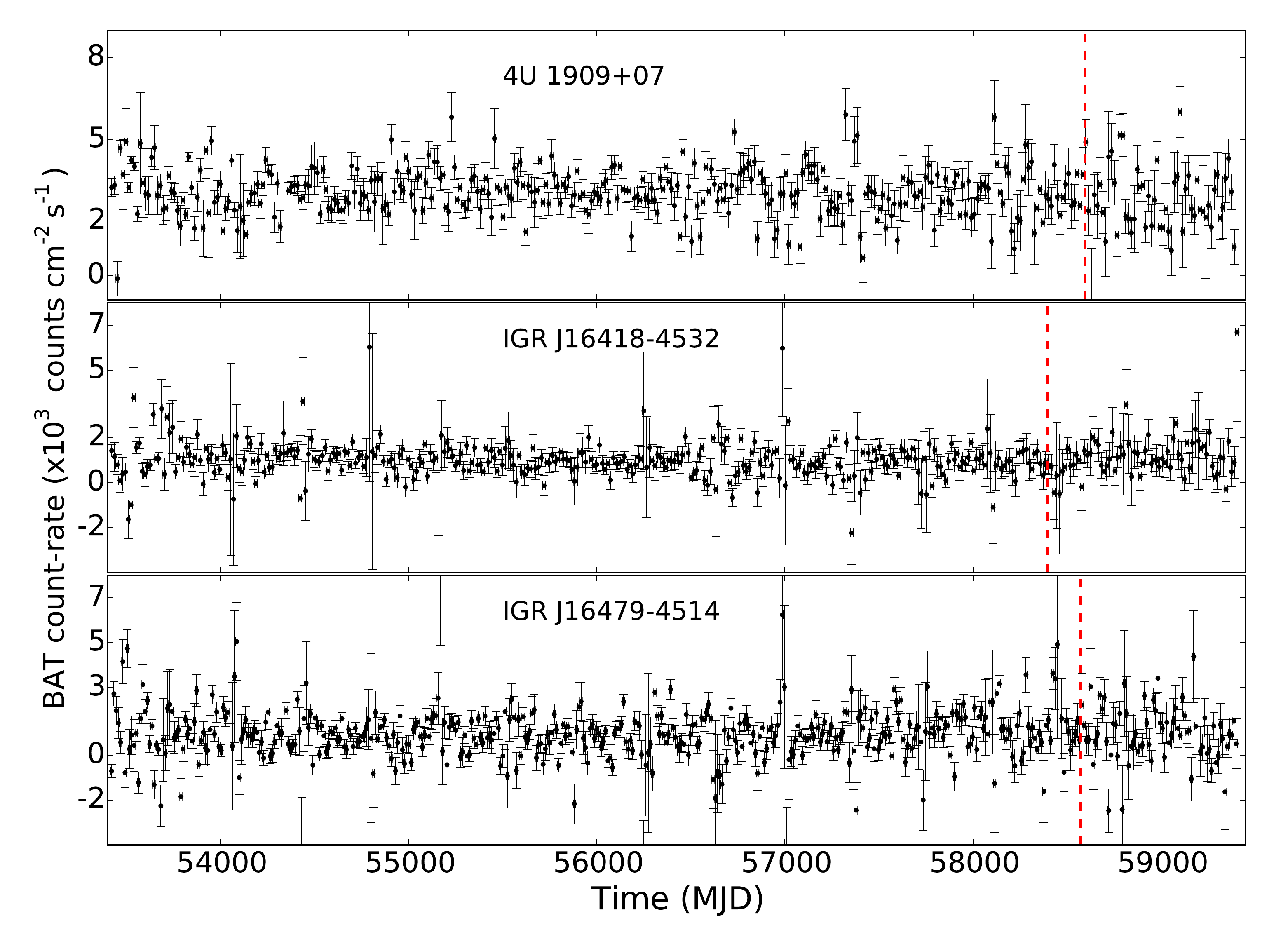}
    \caption{Swift BAT 15--50 keV lightcurves of 4U 1909+07, IGR J16418--4534 and IGR J16479--4514 binned with their superorbital periods of 15.196 d, 14.702 d and 11.894 d respectively. The red dashed lines mark the epoch of the NuSTAR observations analyzed in this paper: MJD 58,595 for 4U 1909+07, MJD 58,393 for IGR J16418--4532 and MJD 58,574 for IGR J16479--4514.}
    \label{bat_lc}
\end{figure*}

In Figure \ref{bat_lc}, we show the Swift BAT lightcurves of 4U 1909+07, IGR J16418--4532 and IGR J16479--4514 binned with their 15.196 d, 14.702 d and 11.894 d superorbital periods respectively. We find no significant long-term intensity variations between different superorbital cycles for any of the sources.  
\par
\begin{figure}
\hspace{-5mm}
\includegraphics[scale=0.7, angle=0]{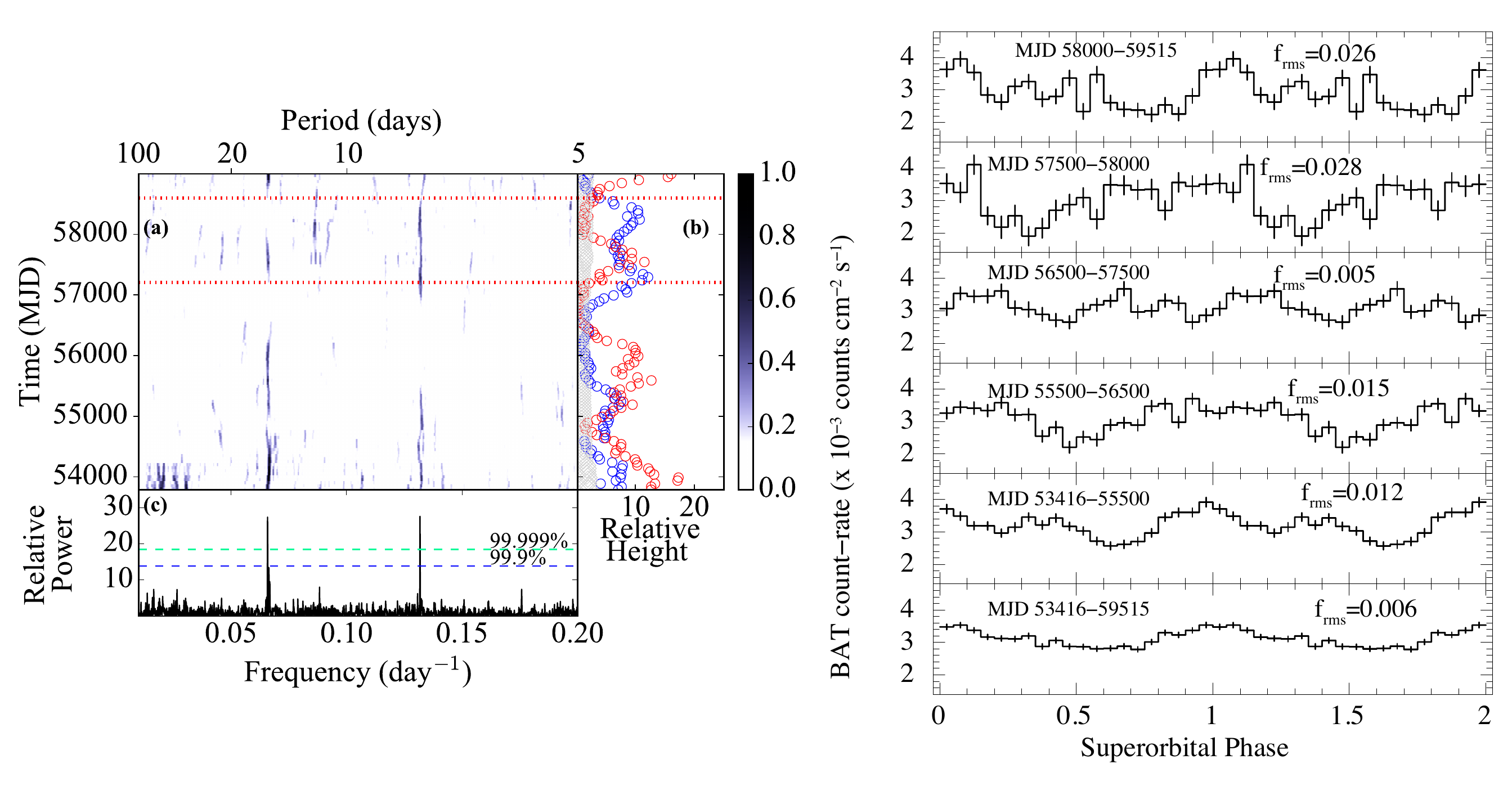}
%\hspace{-300mm}
\caption{{\it Left panel}: Panel (a) shows the dynamic power spectrum of 4U 1909+07 constructed using 16.7 years of Swift BAT lightcurve in a 15--50 keV energy band. The power spectrum was calculated using 750 day intervals, which were successively shifted in time by 50 days relative to each others. The red dashed line corresponds to the NuSTAR observations taken in 2019 analyzed in this paper. The second red dashed line corresponds to the NuSTAR observation taken in 2015. The red points in panel (b) show the relative height of the fundamental peak of the superorbital period to the mean power (in grey points) and the blue points are the relative height of the second harmonic to the mean power. Panel (c) shows the power spectrum of the Swift BAT lightcurve normalized to the average power, with 99.9\% and 99.99\% significance levels indicated by the blue and green dashed lines respectively. {\it Right panel}: Different segments of Swift BAT lightcurve where the peak at the fundamental and the second harmonic were stronger and/or weaker, folded on the superorbital period of 15.196 d. Phase zero was chosen as the time of the maximum flux MJD 59,502.1$\pm$0.1. The fractional RMS ($f_{\rm{rms}}$) was calculated using the formulation of \cite{vaughan2003} for the superorbital intensity profiles. The y-axes are plotted with the same range}
\label{4U1909_power}
\end{figure}
\begin{figure}
\includegraphics[scale=0.7, angle=0]{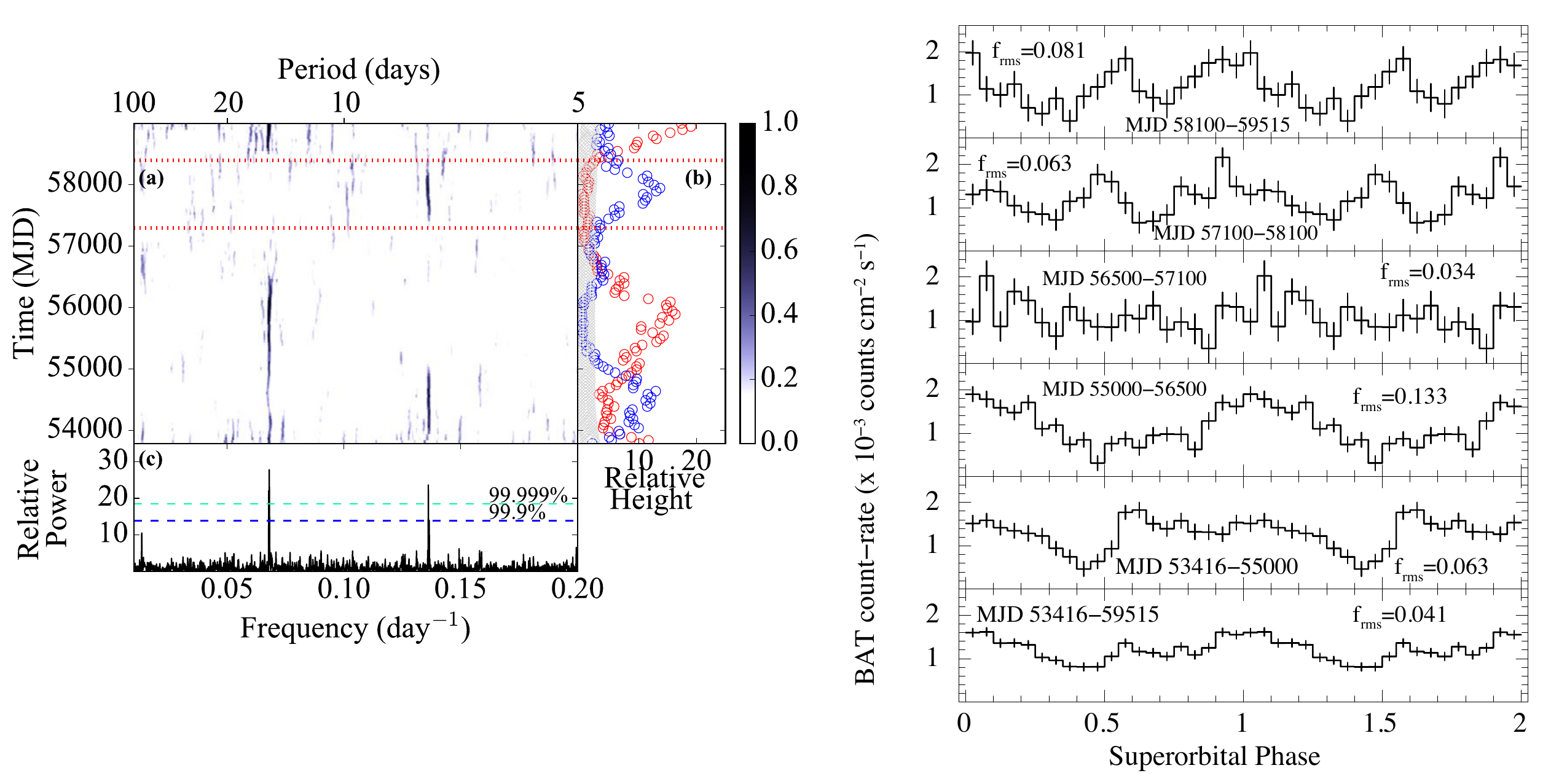}
\caption{{\it Left panel}: Panel (a) shows the dynamic power spectrum of IGR J16418--4532 constructed using 16.7 years of Swift BAT lightcurve in a 15--50 keV energy band. The power spectrum was calculated using 1000 day intervals, which were successively shifted in time by 50 days relative to each other. The red dashed line corresponds to the NuSTAR observations taken in 2018 analyzed in this paper. The second red dashed line corresponds to the NuSTAR observation taken in 2015. The red points in panel (b) show the relative height of the fundamental peak of the superorbital period to the mean power (in grey points) and the blue points are the relative height of the second harmonic to the mean power. Panel (c) shows the power spectrum of the Swift BAT lightcurve normalized to the average power, with 99.9\% and 99.99\% significance levels indicated by the blue and green dashed lines respectively. {\it Right panel}: Different segments of Swift BAT lightcurve where the peak at the fundamental and second harmonic were stronger and/or weaker, folded on the superorbital period of 14.702 d. Phase zero was chosen as the time of the maximum flux MJD 59,506.9$\pm$0.1. The superorbital intensity profiles were constructed excluding the X-ray eclipses present in the Swift BAT lightcurve. The fractional RMS ($f_{\rm{rms}}$) was calculated using the formulation of \cite{vaughan2003} for the superorbital intensity profiles. The y-axes are plotted with the same range.}
\label{igrj16418_power}
\end{figure}
\begin{figure}
\includegraphics[scale=0.7, angle=0]{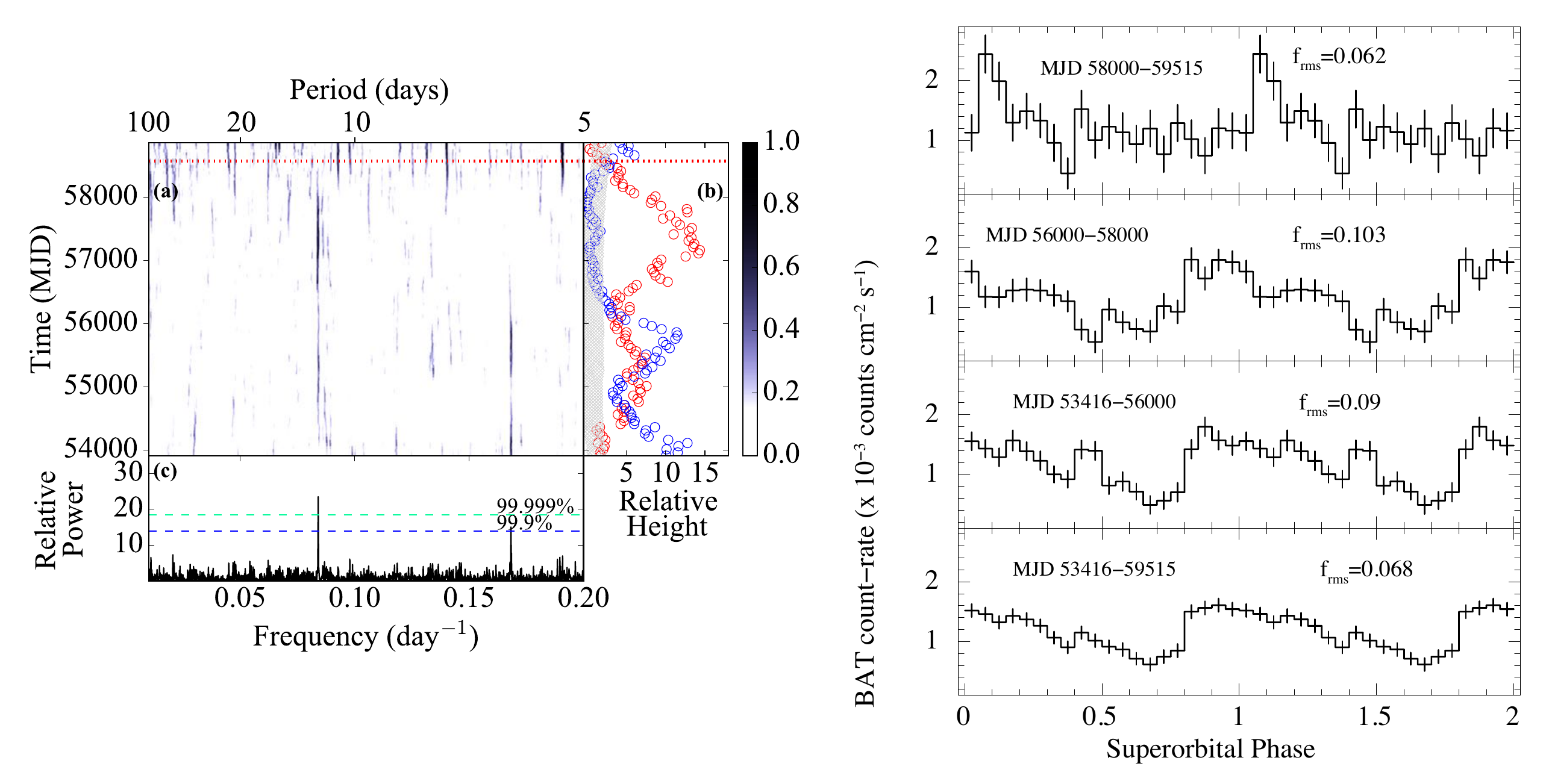}
\caption{{\it Left panel}: Panel (a) shows the dynamic power spectrum of IGR J16479--4514 constructed using 16.7 years of Swift BAT lightcurve in a 15--50 keV energy band. The power spectrum was calculated using 1000 day intervals, shifted in time by 50 days relative to each other. The red dashed line corresponds to the NuSTAR observations taken in 2019 analyzed in this paper. The red points in panel (b) show the relative height of the fundamental peak of the superorbital period to the mean power (in grey points) and the blue points are the relative height of the second harmonic to the mean power. Panel (c) shows the power spectrum of the Swift BAT lightcurve normalized to the average power, with 99.9\% and 99.99\% significance levels indicated by the blue and green dashed lines respectively. {\it Right panel}: Different segments of Swift BAT lightcurve where the peak at the fundamental and second harmonic were stronger and/or weaker, folded on the superorbital period of 11.894 d. Phase zero was chosen as the time of the maximum flux MJD 59,507.7$\pm$0.1. The superorbital intensity profiles were constructed excluding the X-ray eclipses present in the Swift BAT lightcurve. The fractional RMS ($f_{\rm{rms}}$) was calculated using the formulation of \cite{vaughan2003} for the superorbital intensity profiles. The y-axes are plotted with the same range}
\label{igrj16479_power}
\end{figure}

To monitor the changes in the strengths of the superorbital modulations, we constructed the dynamic power spectra using the Swift BAT lightcurves for the three sources. The power spectra were calculated using 750 day time intervals, which were successively shifted in time by 50 days relative to each other for 4U 1909+07, whereas for IGR J16418--4532 and IGR J16479--4514, 1000 day time intervals were used which where successively shifted in time by 50 days relative to each other. The left panels (a) of Figures \ref{4U1909_power}, \ref{igrj16418_power} and \ref{igrj16479_power} show the dynamic power spectra constructed for the three sources using 16.7 years of Swift BAT lightcurves. The red points in the left panels (b) of Figures \ref{4U1909_power}, \ref{igrj16418_power} and \ref{igrj16479_power} show the relative height of the peak at the superorbital period to the mean power (in grey points) and the blue points show the relative height of the peak at the second harmonic to the mean power. The left panels (c) of Figures \ref{4U1909_power}, \ref{igrj16418_power} and \ref{igrj16479_power} show the power spectra of the Swift BAT lightcurves, exhibiting the peaks in the power spectrum corresponding to the superorbital period and the second harmonic, along with the significance levels of the noise, similar to Figure \ref{bat_power} but with a smaller range of time around the superorbital period. 

\subsubsection{Superorbital intensity profiles of 4U 1909+07}
The dynamic power spectrum of 4U 1909+07 shown in the left panel (a) of Figure \ref{4U1909_power} exhibit peaks at the fundamental superorbital period of $\sim$15.2 d and the second harmonic of $\sim$7.6 d. The strengths of these peaks are varying on different timescales of years. At the epoch of the NuSTAR observations analyzed in this paper as well as the previous NuSTAR observation in 2015, the superorbital modulation at the fundamental frequency was weaker and consistent with the mean power. The mean power is defined as the average power in the power spectrum excluding any peaks and is similar to the noise. The relative height of the fundamental peak of the superorbital period (red points) and the second harmonic (blue points) to the mean power, shown in the left panel (b) of Figure \ref{4U1909_power} are stronger by a factor of 10 or more than the mean power (grey points) during the times when the superorbital modulation at the fundamental period and/or the second harmonic was present. 
\par
To illustrate the changes in the superorbital intensity profiles associated with a change in the strength of the superorbital modulation, we divided the Swift BAT lightcurve into different segments where the peaks corresponding to the fundamental superorbital period and the second harmonic were stronger and/or weaker in the dynamic power spectrum. As seen in the dynamic power spectrum of 4U 1909+07 (left panel of Figure \ref{4U1909_power}a), both peaks at the fundamental frequency and the second harmonic were present from MJD 53,416 to MJD 55,500, with relative heights of 10. Only the peak at the fundamental frequency was present from MJD 55,500 to MJD 56,500 with a relative height of 10. From MJD 56,500 to MJD 57,500, both the peaks at the fundamental frequency and the second harmonic were not strongly present and was consistent with the mean power. The peak at the second harmonic was present from MJD 57,500 to MJD 58,000, with the peak the fundamental frequency briefly present. Around  MJD 59,500, the peak at the fundamental frequency reappeared and at the second harmonic became weaker, however continued monitoring is required to confirm this. The Swift BAT lightcurves were folded with the superorbital period of 15.192 d and phase zero was chosen as the time of the maximum flux of the average superorbital intensity profile (constructed using data from MJD 53,416 to MJD 59,515) given in Table 1. The right panel of Figure \ref{4U1909_power} show the superorbital intensity profiles constructed from these different segments of the lightcurve. We estimated the fractional root mean square amplitude ($f_{\rm{rms}}$) for the superorbital intensity profiles, using the formulation in \cite{vaughan2003}. The $f_{\rm{rms}}$ for the superorbital intensity profiles are noted inside the panels of the right plot of Figure \ref{4U1909_power}. The $f_{\rm{rms}}$ was lower for the superorbital intensity profile constructed from the MJD 56,500 to MJD 57,500 when the peaks at the fundamental frequency and the second harmonic were not strongly present, compared to other segments where atleast one of the peaks was present in the dynamic power spectrum (left panel of Figure \ref{4U1909_power}a). 

\subsubsection{Superorbital intensity profiles of IGR J16418--4532}
The dynamic power spectrum of IGR J16418--4532 shown in the left panel (a) of Figure \ref{igrj16418_power} exhibit peaks at the fundamental superorbital period of $\sim$14.7 d and the second harmonic at $\sim$7.4 d, with varying strengths on the timescales of years. At the epoch of the NuSTAR observations analyzed in this paper and a previous NuSTAR observation in 2015, the superorbital modulation both at the fundamental frequency and the second harmonic was weaker and consistent with the mean power. The relative height of the fundamental peak of the superorbital period (red points) and the second harmonic (blue points) to the mean power, shown in the left panel (b) of Figure \ref{igrj16418_power} is stronger by a factor of 10 or more than the mean power (grey points) during the times when the superorbital modulation at the fundamental period and/or the second harmonic was present. 
\par
To illustrate the changes in the superorbital intensity profiles associated with a change in the strength of the superorbital modulation, we excluded the X-ray eclipses present in the lightcurve and divided the Swift BAT lightcurve into different segments where the peaks corresponding to the fundamental superorbital period and the second harmonic were stronger and/or weaker in the dynamic power spectrum. As seen in the dynamic power spectrum of IGR J16418--4532 (left panel of Figure \ref{igrj16418_power}a), both peaks at the fundamental frequency and the second harmonic were present from MJD 53,416 to MJD 55,000, with relative heights of 10. Only the peak at the fundamental frequency was present with a relative height of 20 from MJD 55,000 to MJD 56,500. From MJD 56,500 to MJD 57,100, both the peaks at the fundamental frequency and the second harmonic were not strongly present and was consistent with the mean power. Only the peak at the second harmonic was present from MJD 57,100 to MJD 58,100 with a relative height of 10. Around  MJD 59,500, the peak at the fundamental frequency reappeared and at the second harmonic became weaker, however continued monitoring is required to confirm this. The Swift BAT lightcurves were folded with the superorbital period of 14.702 d and phase zero was chosen as the time of the maximum flux of the average superorbital intensity profile (constructed using data from MJD 53,416 to MJD 59,515) given in Table 1. The right panel of Figure \ref{igrj16418_power} shows the superorbital intensity profiles constructed from these different segments of the lightcurve, along with their $f_{\rm{rms}}$. The $f_{\rm{rms}}$ was lower for the superorbital intensity profile constructed from the MJD 56,500 to MJD 57,100 when the peaks at the fundamental frequency and the second harmonic were not strongly present. The $f_{\rm{rms}}$ was highest for the superorbital intensity profile constructed from MJD 55,000 to MJD 56,500 when only the peak at the fundamental frequency was present with a relative height of 20.  

\subsubsection{Superorbital intensity profiles of IGR J16479--4514}
From the dynamic power spectrum of IGR J16479--4514 in the left panel (a) of Figure \ref{igrj16479_power}, we see peaks corresponding to the fundamental of the superorbital period at $\sim$11.9 d and the second harmonic at $\sim$5.9 d. At the epoch of the NuSTAR observations analyzed in this paper, the superorbital modulation at both the fundamental frequency and the second harmonic were weak and consistent with the mean power. The relative height of the fundamental peak of the superorbital period (red points) and the second harmonic (blue points) to the mean power, shown in the left panel (b) of Figure \ref{igrj16479_power} is stronger by a factor of 15 or more than the mean power (grey points) during the times when the superorbital modulation at the fundamental and/or the harmonic was present. 
\par
Similar to the procedure described for IGR J16418--4532, we excluded the X-ray eclipses present in the lightcurve and divided the Swift BAT lightcurves into different segments where the peaks corresponding to the fundamental superorbital period and the second harmonic were stronger and/or weaker in the dynamic power spectrum. As seen in the dynamic power spectrum of IGR J16479--4514 (left plot of Figure \ref{igrj16479_power}a), both peaks at the fundamental frequency and the second harmonic were present from MJD 53,416 to MJD 56,000. Only the peak at the fundamental frequency was present from MJD 56,000 to MJD 58,000 with a relative height of 15. The Swift BAT lightcurves were folded with the superorbital period of 11.894 d and phase zero was chosen as the time of the maximum flux of the average superorbital intensity profile (constructed using data from MJD 53,416 to MJD 59,515) given in Table 1. The right panel of Figure \ref{igrj16479_power} shows the superorbital intensity profiles constructed from these different segments of the lightcurve, along with their $f_{\rm{rms}}$. The $f_{\rm{rms}}$ was lower for the superorbital intensity profile constructed from the MJD 58,000 to MJD 59,515 when the peaks at the fundamental frequency and the second harmonic were not strongly present and was consistent with the mean power.
\par
For IGR J16418--4532 and IGR J16479--4514, there are other peaks seen in the dynamic power spectra which are not related to either the orbital or superorbital modulations (panel (a) of the left plots of Figure \ref{igrj16418_power} and \ref{igrj16479_power}). However, none of the peaks have a high statistical significance as the fundamental frequency of the superorbital modulation or the second harmonic. The dynamic power spectrum was calculated using a 1000 days sliding window, shifted by 50 days and the modulations at these periods last for shorter durations than the more prominent superorbital modulation. As seen in the panel (c) of the left plots of Figures \ref{igrj16418_power} and \ref{igrj16479_power}, the statistical significance of these short peaks are lower than statistical significance levels of 99.9\% and 99.99\% and have a FAP of $\sim 10^{-1}$. Hence these peaks are most likely spurious and might be due to statistical noise.

\subsection{Pointed observations with Swift XRT and NuSTAR}
\begin{sidewaystable}
%\centering
\caption{Summary of X-ray observations} 
\label{obs}
\begin{tabular}{c c c c c c c c c c}
\hline
Source & Telescope & ObsID  & Start Time  & End Time & Orbital & Superorbital & Total Exposure & Avg count rate \\
 & & & & & Phase & Phase & (ks) & (3--50 keV) \\
\hline
4U 1909+07 & NuSTAR & 30402026002 & 2019-04-22T00:11:09  & 2019-04-22T11:36:09  & 0.37--0.48 & 0.68--0.71 & 18.8 & 4.7(FPMA)/4.5(FPMB) \\
    & Swift XRT & 00088747001 & 2019-04-22T01:04:34 & 2019-04-22T02:53:52 & 0.4 & 0.7 & 1.8 & 0.64 \\
\hline
 & NuSTAR & 30402026004 & 2019-04-26T13:41:09 & 2019-04-27T20:56:09  & 0.41--0.52 & 0.98--1.02 & 23.2 & 6.9(FPMA)/6.6(FPMB)\\
\hline
\hline
IGR J16418--4532 & NuSTAR & 30402027002 & 2018-10-02T20:51:09  & 2018-10-03T07:51:09  & 0.84--0.96 & 0.88--0.91 & 19.85 &  0.59(FPMA)/0.55(FPMB) \\
    & Swift XRT & 00088748001 & 2018-10-02T21:32:47 & 2018-10-02T23:25:52 & 0.85 & 0.89 & 1.8 & 0.13 \\
\hline
 & NuSTAR & 30402027004 & 2018-10-10T18:51:09 & 2018-10-11T12:21:09 & 0.96--1.15 & 1.42--1.4 & 31.02 & 1.2(FPMA)/1.1(FPMB)\\
    & Swift XRT & 00088748002 & 2018-10-10T20:45:56 & 2018-10-10T22:39:52 & 0.98 & 1.44 & 1.8 & 0.24\\
\hline
\hline
IGR J16479--4514 & NuSTAR & 30402028002 & 2019-04-01T19:06:09  & 2019-04-02T07:36:09  & 0.49--0.65 & 0.07--0.11  & 21.86 & 1.5(FPMA)/1.4(FPMB) \\
    & Swift XRT & 00088749001 & 2019-04-01T20:28:34 & 2019-04-01T22:21:53 & 0.55 & 0.095 & 1.9 & 0.30\\
\hline
 & NuSTAR & 30402028004 & 2019-04-09T09:11:09 & 2019-04-10T15:07:18 & 0.78--0.98 & 0.71--0.76 & 31.02 & 2.3(FPMA)/2.1(FPMB)\\
    & Swift XRT & 00088749002 & 2019-04-09T10:11:19 & 2019-04-09T12:13:54 & 0.8 & 0.73 & 1.9 & 0.06\\
\hline
\end{tabular}
\end{sidewaystable}

\subsubsection{NuSTAR}
NuSTAR is a hard X-ray telescope operating in the 3--79 keV energy band \citep{harrison2013}. It carries two co-aligned grazing incidence Wolter I imaging telescopes that focus onto two independent Focal Plane Modules FPMA and FPMB. The NuSTAR observations were performed around the predicted superorbital maximum and minimum of a single superorbital cycle for the three sources according to the superorbital phase ephemerides given in \cite{corbet2013}. In addition to the NuSTAR observations, Swift XRT observations with short exposures ($\sim$ 2 ks) were simultaneously carried out, except for the NuSTAR observations of 4U 1909+07 during its predicted superorbital maximum phase. Table 2 gives a summary of the NuSTAR and Swift XRT observations for the three sources which are analyzed in this paper, along with their predicted superorbital and orbital phase. 
\par
The NuSTAR data were reduced and analyzed using NuSTAR Data Analysis Software (NuSTARDAS) v.2.1.1 package provided under {\it HEAsoft v.6.29} and calibration files 2021-10-26. The event files were reprocessed using {\tt nupipeline} using the standard filtering procedure and the default screening criteria. The event times were corrected to the solar system barycenter using {\tt nuproducts} and the FTOOL {\tt barycorr} with the DE200 solar system ephemeris. The source spectra, response matrices, ancillary matrices files and energy-resolved lightcurves were extracted in SCIENCE mode (01) from a circular region of radius 60'' centered on the source using {\tt nuproducts}. The background spectra and lightcurves were extracted from a circular region of 60'' in a source-free region on the same chip. For carrying out analysis with user input good time intervals especially for analysis of the X-ray flares present in the lightcurves of 4U 1909+07 and IGR J16418--4532, we use {\tt usrgti} as an input to {\tt nuproducts}. We corrected the event times for the orbital motion of the neutron star in 4U 1909+07 using the ephemerides given in \cite{levine2004} for a circular orbit. Since the orbital ephemerides of IGR J16418--4532 and IGR J16479--4514 are unknown, we do not correct the event files for the orbital motion of the compact object.  

\subsubsection{Swift XRT}
Swift XRT is a Wolter I imaging telescope sensitive to X-rays ranging from 0.3 to 10 keV \citep{burrows2005}. The data were reduced and analyzed with {\it HEASoft v.6.29} and calibration files 20210915. The event files were reprocessed using {\tt xrtpipeline} using the standard filtering procedure and the default screening criteria\footnote{https://www.swift.ac.uk/analysis/xrt/}. We used the observations in Photon Counting (PC) mode, applying the standard filtering procedures. The Swift XRT observation of 4U 1909+07 carried out at the predicted superorbital minimum phase has a moderate pileup. To mitigate the effects of the pileup, we used the pileup threads given in the XRT analysis guide\footnote{https://www.swift.ac.uk/analysis/xrt/pileup.php}. The source region was taken as an annulus with an outside radius of 60'' and an inner radius of 7''. The background spectrum was extracted from a circular region of radius 60'' from a source-free region.  
\par
The Swift XRT observations of IGR J16418--4532 and IGRJ 16479--4514 were not affected by pileup. The source spectra were extracted from a circular region of radius 60'' centered on the source and background spectra were extracted with the same radius from a source-free region. The exposure maps and ancillary response matrices were extracted using {\tt xrtexpomap} and {\tt xrtmkarf} respectively. 
\par
The NuSTAR and Swift XRT source spectra were grouped to give a minimum of 30 counts in each spectral bin. In addition to the ObsIDs listed in Table 1, we also used an archival NuSTAR observation (ObsID:30101036002, obs date:2015-09-26) and an archival XMM-Newton observation (ObsID:0823990401, obs date:2019-02-21) of IGR J16418--4532, to investigate the long-term behavior of the source. The NuSTAR observation was analyzed using the same procedure mentioned in Section 2.2.1. The XMM-Newton observation was carried out in Full Window Imaging mode and was analyzed using SAS v19.1.0 with the latest calibration files. The lightcurves were extracted using the {\tt XMM-SAS} data analysis threads\footnote{https://www.cosmos.esa.int/web/xmm-newton/sas-threads}, which filtered out the background flares and applied the latest calibration files. A circular source region was selected with a radius of 40'' centered on the source and a similar size background region was selected in a source-free region on the chip. This XMM observation was used to study the spin period of the source, hence only EPIC-PN data were used. 

\subsubsection{Lightcurves and hardness ratios}
\begin{figure*}
    \centering
    \includegraphics[scale=0.4, angle=0]{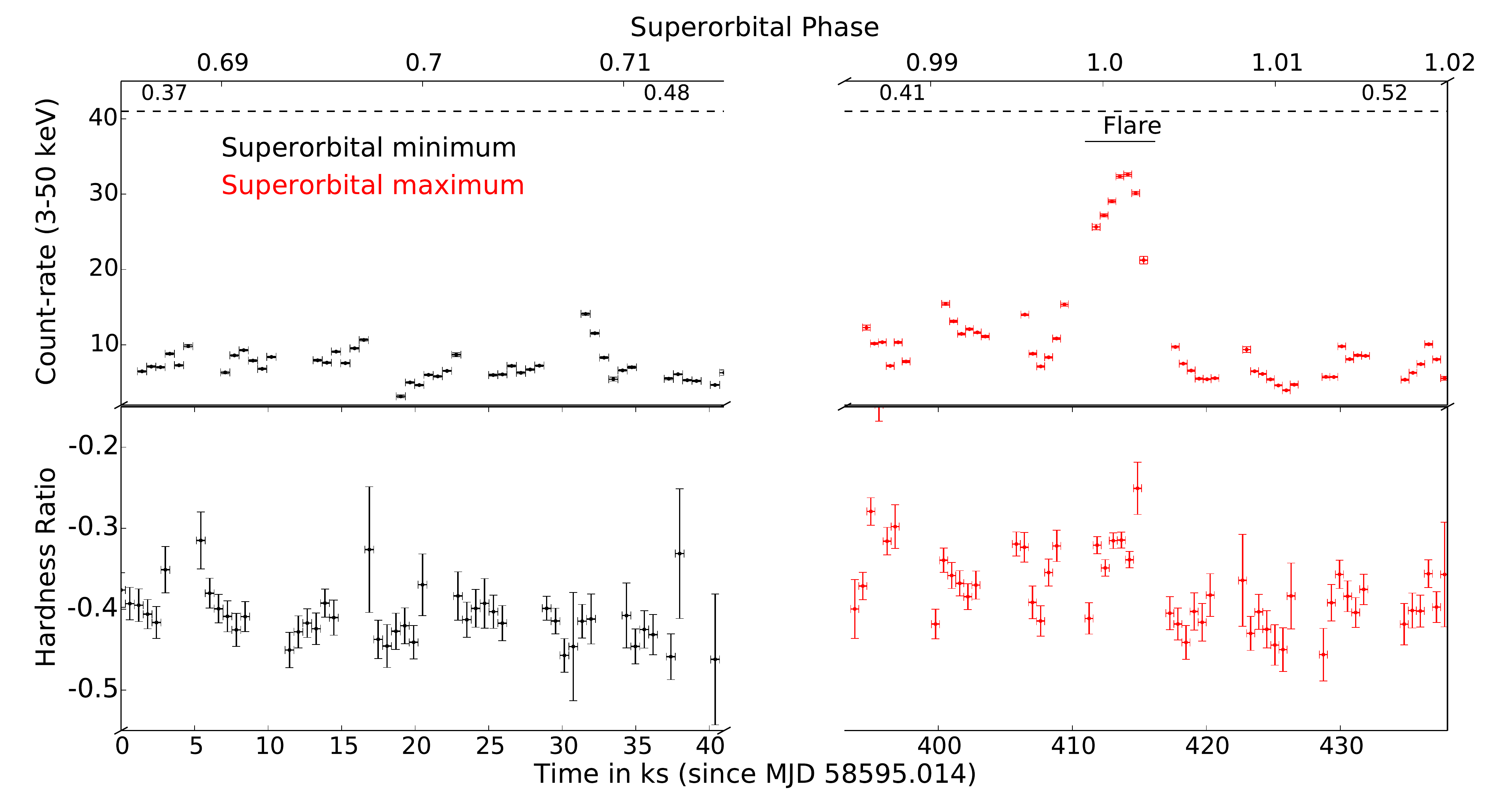}
    \caption{NuSTAR FPMA+FPMB lightcurves of 4U 1909+07 in 3--50 keV energy band, binned with the pulsar spin period of 603 s. The hardness ratios were constructed using lightcurves in 3--10 keV and 10--50 keV. The top x-axis displays the superorbital phase of the observations calculated using the ephemerides given in \cite{corbet2013}. The dashed line on the top x-axis displays the orbital phase of the observations calculated using the ephemerides given in \cite{levine2004}. The red points correspond to the observation carried out during the predicted superorbital maximum phase and the black points correspond to the observation carried out during the predicted superorbital minimum phase. The segment `flare' denotes a short X-ray flare in the observation when the count rate increased by a factor of 5.}
    \label{4U1909_lc}
\end{figure*}

\begin{figure*}
    \centering
    \includegraphics[scale=0.4,angle=0]{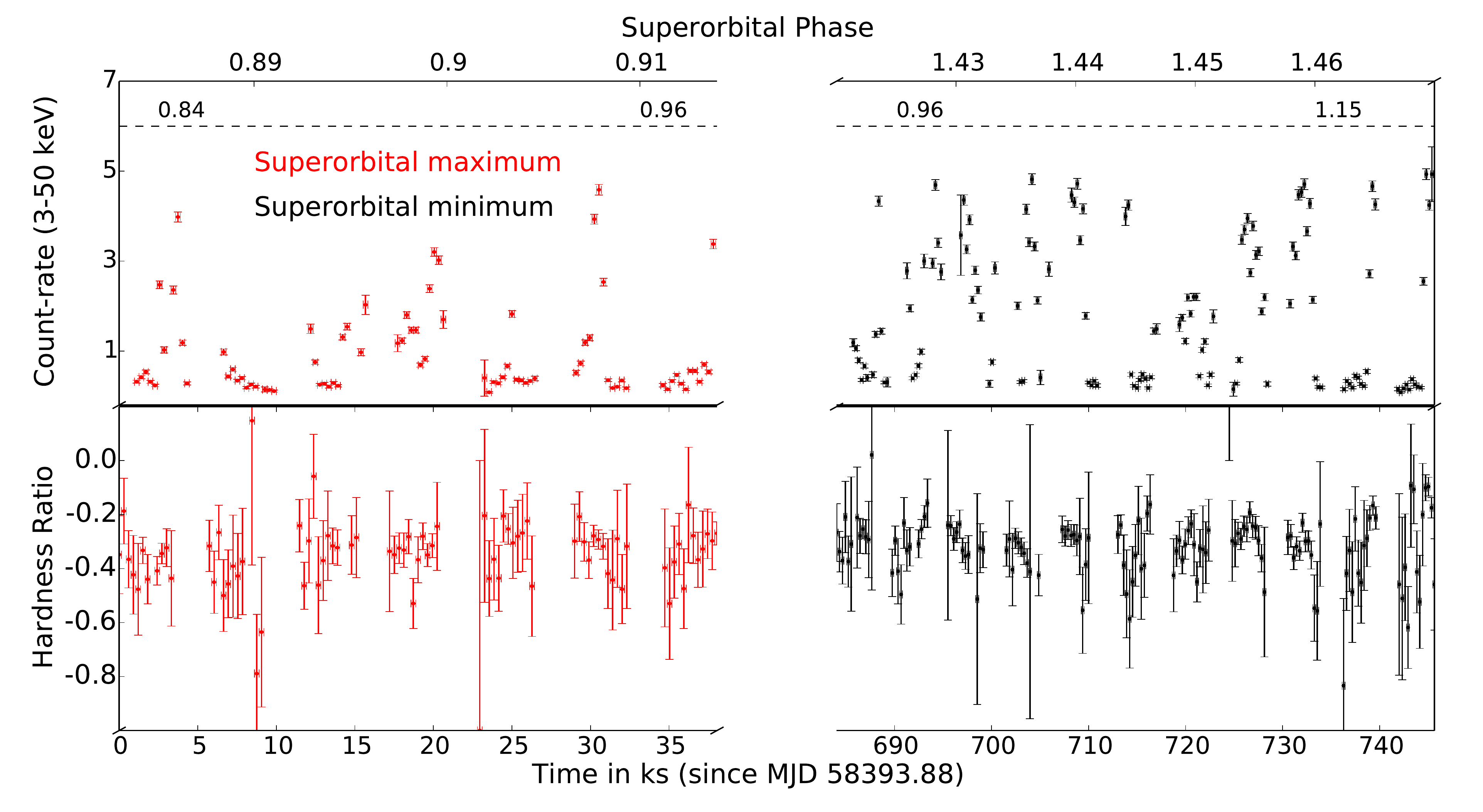}
    \caption{NuSTAR FPMA+FPMB lightcurves of IGR J16418--4532 in 3--50 keV energy band, binned with 302 s, one-fourth of the spin period of the pulsar (1208 s). The hardness ratios were constructed using lightcurves in 3--10 keV and 10--50 keV. The top x-axis displays the predicted superorbital phase of the observations calculated using the ephemerides given in \cite{corbet2013}. The dashed line on the top x-axis displays the orbital phase of the observations calculated using the ephemerides given in \cite{coley2015}. The red points correspond to the observation carried out during the predicted superorbital maximum phase and the black points correspond to the observation carried out during the predicted superorbital minimum phase.}
    \label{igrj16418_lc}
\end{figure*}

\begin{figure*}
    \centering
    \includegraphics[scale=0.4,angle=0]{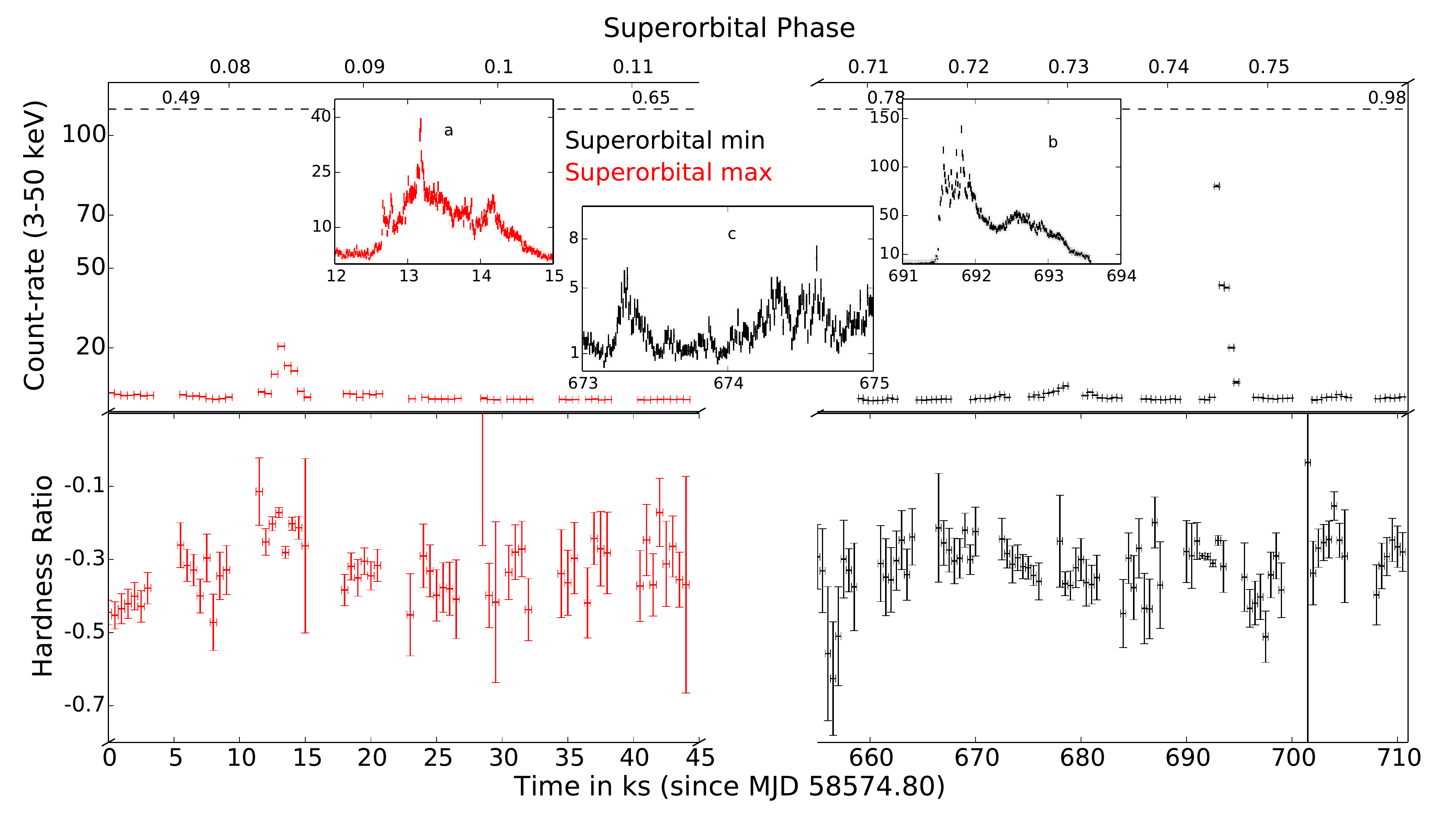}
    \caption{NuSTAR FPMA+FPMB lightcurves of IGR J16479--4514 in 3--50 keV energy band, binned by 500 s. The hardness ratios were constructed using lightcurves in 3--10 keV and 10--50 keV. The top x-axis displays the predicted superorbital phase of the observations calculated using the ephemerides given in \cite{corbet2013}. The dashed line on the top x-axis displays the orbital phase of the observations calculated using the ephemerides given in \cite{coley2015}. The red points correspond to the observation carried out during the predicted superorbital maximum phase and the black points correspond to the observation carried out during the predicted superorbital minimum phase. The inset plot (a) display an enlarged view of the X-ray flare at the superorbital phase 0.08, (b) display an enlarged view of the X-ray flare at the superorbital phase 0.74 and (c) display an enlarged view of the non-flaring part of the lightcurve. The inset plots are binned with 10 s.}
    \label{igrj16478_lc}
\end{figure*}

The energy-resolved lightcurves of the NuSTAR observations were constructed in 3--50 keV, 3--10 keV and 10--50 keV energy bands to study the short-term variability and changes in hardness ratio during an observation. The hardness ratio is defined for the lightcurves in the energy ranges 3--10 keV (with count rate C$_{S}$ ) and 10--50 keV (with count rate C$_{H}$) as:\\
\begin{equation}
    HR = \frac{C_{H}-C_{S}}{C_{H}+C_{S}} \nonumber
\end{equation}

Figure \ref{4U1909_lc} shows the NuSTAR FPMA+FPMB lightcurves in 3--50 keV energy band of the two 4U 1909+07 observations, binned with the pulsar spin period of 603 s. There is a short 3 ks X-ray flare marked as `flare' around superorbital phase $\sim$ 1.0 and orbital phase of $\sim$ 0.5, where the count rate increases by a factor of 5. This short X-ray flare is most likely related to a short-term variability arising due to accretion from a clumpy stellar wind. To disentangle this short-term variability from a clumpy stellar wind from the long-term variability from the superorbital modulations, we remove the short X-ray flare from the pulse profiles and spectral analysis and focus only on the non-flaring part of the observations. With the exception of the short X-ray flare, the count rate remains similar for both observations, which is indicative of the weakened amplitude of the superorbital modulations seen with Swift BAT in the left panel of Figure \ref{4U1909_power}a. We find no significant change in the hardness ratios between the predicted superorbital maximum and minimum observations.
\par 
In Figure \ref{igrj16418_lc}, the NuSTAR FPMA+FPMB 3--50 keV energy band lightcurves of IGR J16418--4532 were binned with 302 s, one-fourth of the spin period of the pulsar (1208 s), to show the presence of several short X-ray flares where the count rate increases by a factor of 5--10. The hardness ratios suggest a spectral softening during these X-ray flares. There are several more short X-ray flares in the predicted superorbital minimum observation compared to that present in the predicted superorbital maximum observation, likely due to the longer duration of the observation. This could also explain the higher X-ray fluxes seen in Table 4 for the observation at the predicted superorbital minimum phase, compared to that of the observation at the predicted superorbital maximum phase.
\par
Figure \ref{igrj16478_lc} shows the NuSTAR FPMA+FPMB 3--50 keV energy band lightcurves of IGR J16479--4514, binned with 500 s. We see two large X-ray flares in the lightcurve: at superorbital phase $\sim$0.08 (orbital phase of $\sim$0.5), where the count rate increases by a factor of 10 (Figure \ref{igrj16478_lc}a) and at superorbital phase $\sim$0.74 (orbital phase of $\sim$0.9) where the count rate increase by a factor of 100 (Figure \ref{igrj16478_lc}b). These large X-ray flares are most likely related to the orbital phase-locked X-ray flares previously detected in this system \citep{sidoli2012, sguera2020}. To disentangle the effects of the orbital phase-locked matter in a clumpy stellar wind from the long-term variability from the superorbital modulations, we remove these large X-ray flares for the spectral analysis and focus only on the non-flaring part of the observations. The non-flaring part of the lightcurve in Figure \ref{igrj16478_lc}c exhibits small fluctuations in the count rates. The average count rate in the non-flaring part of the lightcurve is similar for both observations. The hardness ratios suggest that the spectral shape softens during the X-ray flares and remains constant outside the flares.

\subsubsection{Pulsations and energy-resolved pulse profiles}

\begin{figure}
    \centering
    \includegraphics[scale=0.38]{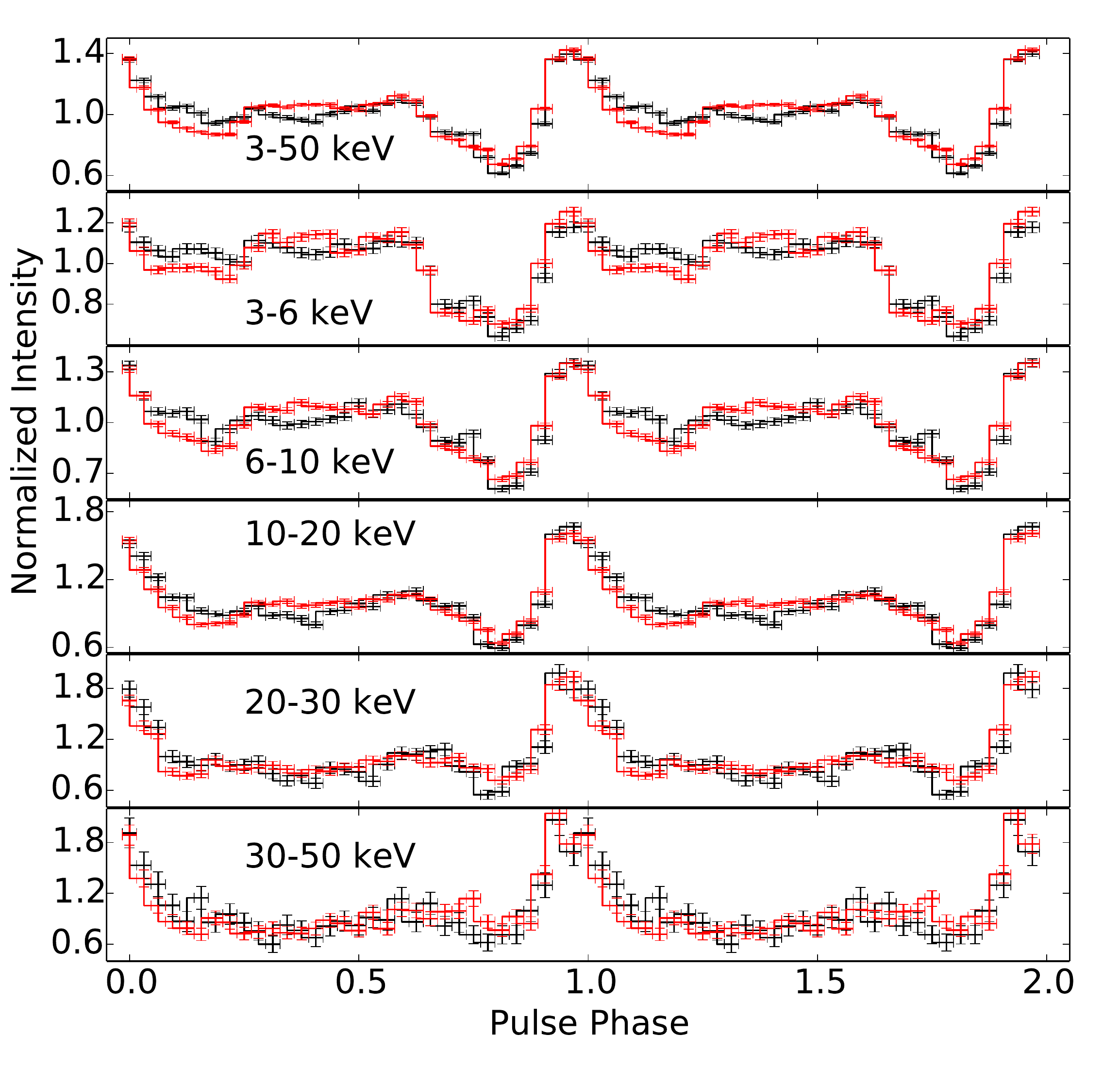}
    \includegraphics[scale=0.35]{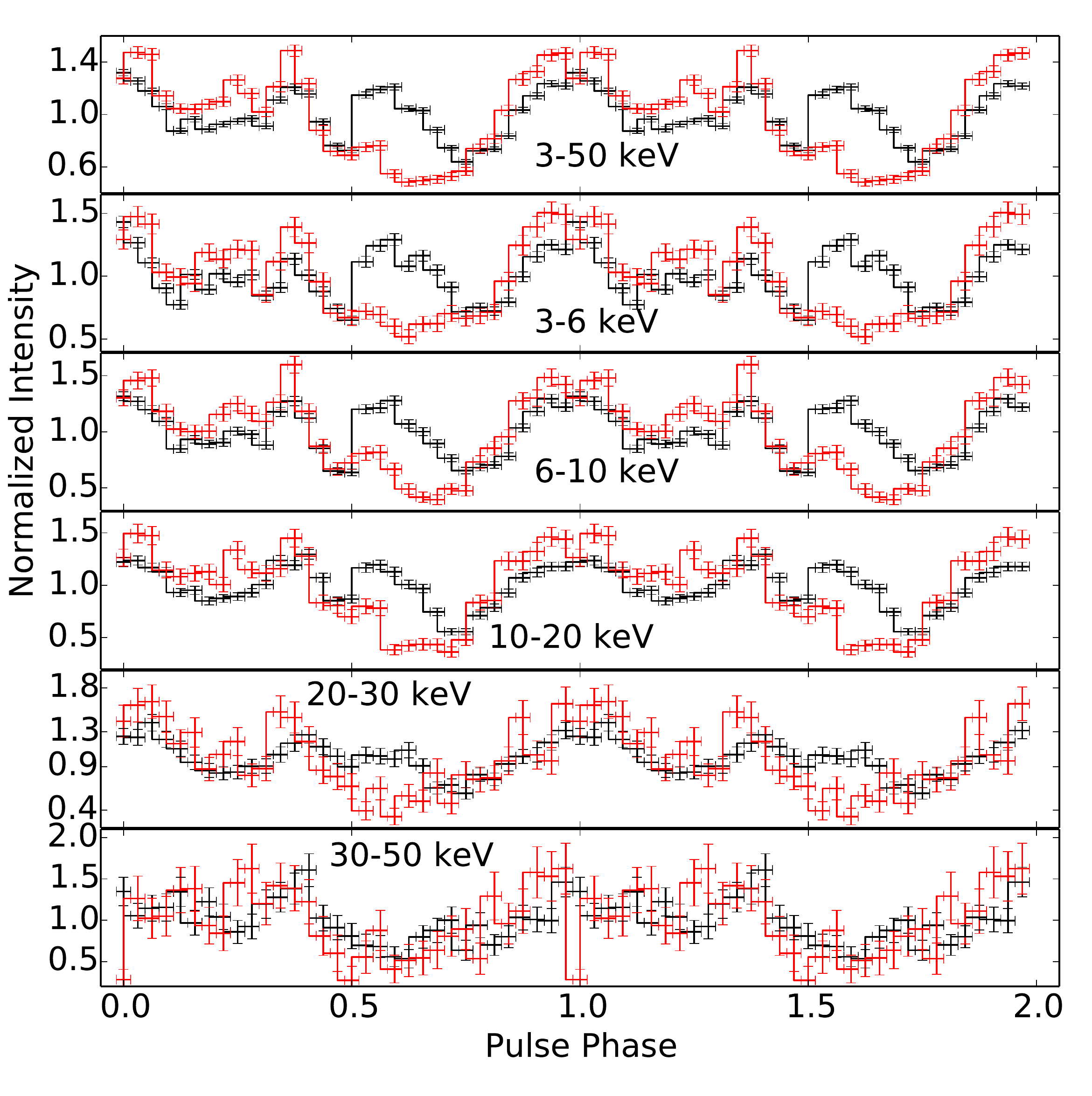}
    \caption{Energy-resolved pulse profiles created with NuSTAR FPMA+FPMB lightcurves of 4U 1909+07 (left panel) and IGR J16418--4532 (right panel). The pulse profiles for the predicted superorbital minimum (black lines) and maximum (red lines) observations are overlaid on the same plot for comparison.}
    \label{pulse}
\end{figure}

\begin{figure}
    \centering
    \includegraphics[scale=0.3, angle=-90]{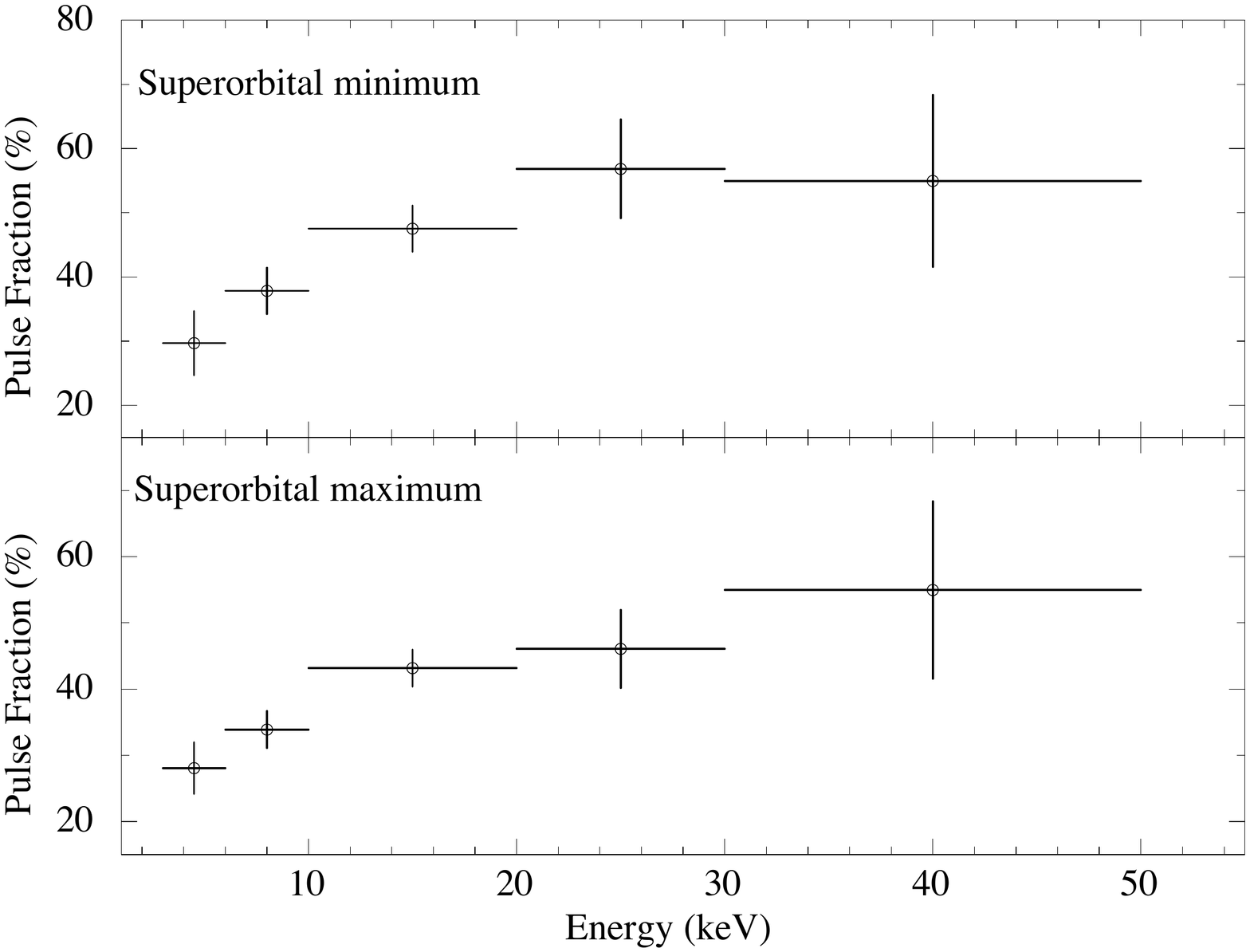}
    \includegraphics[scale=0.3,angle=-90]{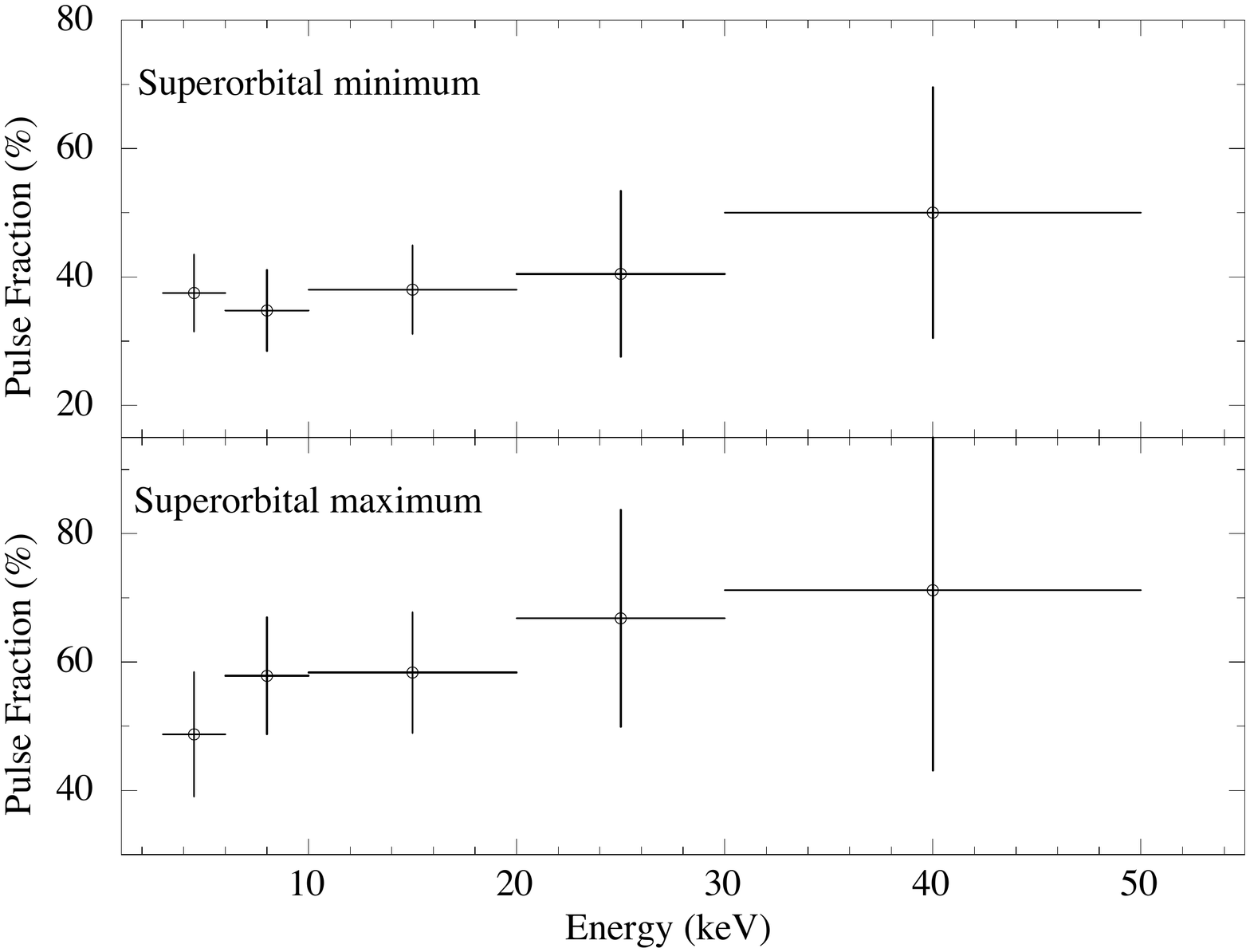}
    \caption{Energy dependence of the pulse fraction for 4U 1909+07 (left panel) and IGR J16418--4532 (right panel) for the NuSTAR observations at the predicted superorbital minimum and maximum phases. }
    \label{pf}
\end{figure}

To search for the pulsations in the NuSTAR observations of the three sgHMXBs, we used the epoch folding and $\chi^{2}$ maximization techniques using the FTOOL {\tt efsearch} \citep{leahy1987}, on the lightcurves binned with 1 s. The errors on the pulse periods are estimated for a 1$\sigma$ confidence interval. The pulsation search was carried out separately for the two NuSTAR observations of 4U 1909+07. The pulse period of the neutron star for the predicted superorbital minimum observation is 602.9$\pm$0.1 s and for the predicted superorbital maximum observation is 603.2$\pm$0.1 s. For IGR J16418--4532, we combined the lightcurves from both observations and calculated the pulse period to be 1208.6$\pm$0.1 s. To evaluate the long-term pulse period changes in IGR J16418--4532, we estimate the pulse periods from a previous NuSTAR observation carried out in 2015 and an XMM observation in 2019 from their lightcurves binned with 1 s. The pulse period estimated from the 2015 NuSTAR observation is 1209.2$\pm$0.1 s and from the XMM observation is 1207.9$\pm$0.2 s.  
\par
Pulsations have not been detected previously in IGR J16479--4514. To search for pulsations in the NuSTAR lightcurves, we used $\chi^{2}$ maximization technique using FTOOLS {\tt efsearch}, power spectrum {\tt powspec} and a Lomb-Scargle periodogram \citep{scargle1982}. We separately checked for pulsations in the large X-ray flares and the non-flaring part of the lightcurves but did not find any. The tentative spin period range estimated from its position on the $P_{\rm{spin}}$ vs $P_{\rm{orbital}}$ ``Corbet diagram" is 500--1000 s \citep{corbet1986}. Since the large X-ray flares last around 3 ks which would be a few pulse period cycles, we only use the non-flaring part to estimate the upper limit on the pulse fraction. We folded the 3--50 keV NuSTAR lightcurves of IGR J16479--4514 for the non-flaring part with 500--1000 s as the tentative spin period with 1 s steps, and estimated the upper limit on the pulse fraction for the putative NS as 16\%.
\par
To investigate the energy dependence of the pulse profiles for 4U 1909+07 and IGR J16418--4532, we extracted the lightcurves from the NuSTAR observations in energy bands of 3--6 keV, 6--10 keV, 10--20 keV, 20--30 keV and 30--50 keV and folded them with their estimated pulse periods to create the energy-resolved pulse profiles. The energy-resolved pulse profiles created from the predicted superorbital minimum and maximum observations are overlaid on the same plot to compare any changes in the shape of the pulse profiles. The peak of the pulse profiles was aligned to phase 0 for the pulse profiles in the 3--50 keV energy band.
\par
The energy-resolved pulse profiles for 4U 1909+07 for both the NuSTAR observations (excluding the short X-ray flare) are shown in the left panel of Figure \ref{pulse}. The energy-resolved pulse profiles below 20 keV are quasi-sinusoidal with a primary peak at phase 0, a dip around phase 0.8 and a broader secondary peak around 0.5. At energies above 20 keV, the profile changes to a single peak sinusoidal profile. The shape of these energy-resolved pulse profiles is similar for both observations. The pulse profiles are similar to those seen in previous studies with RXTE \citep{furst2011}, Suzaku \citep{furst2012, jaisawal2013}, NuSTAR and AstroSat \citep{jaisawal2020}.
\par
The energy-resolved pulse profiles of IGR J16418--4532 in the right panel of Figure \ref{pulse} exhibit a considerable difference between the pulse profiles observed at the predicted superorbital minimum and maximum observations, especially at lower energies. The energy-resolved pulse profiles are complex at energies below 20 keV; the superorbital minimum pulse profile shows a presence of a narrow dip at phase 0.5 and two peaks at phase 0 and 0.8 whereas the narrow dip and the peak at phase 0.8 is absent in predicted superorbital maximum observation. At energies above 20 keV, the pulse profile changes to a single peak sinusoidal profile.  
\par
The pulse fraction is defined as:\\
\begin{equation}
    PF = \frac{F_{max}-F_{min}}{F_{max}+F_{min}} \nonumber
\end{equation}
where $F_{\rm{max}}$ and $F_{\rm{min}}$ are the maximum and minimum count rate in the pulse profiles respectively. Figure \ref{pf} shows the energy dependence of the pulse fraction for the energy-resolved pulse profiles of 4U 1909+07 and IGR J16418--4532 for the NuSTAR observations which increase as a function of energy for both observations.

\subsection{Broadband spectral analysis}
\begin{figure}
    \centering
    \includegraphics[scale=0.32,angle=0]{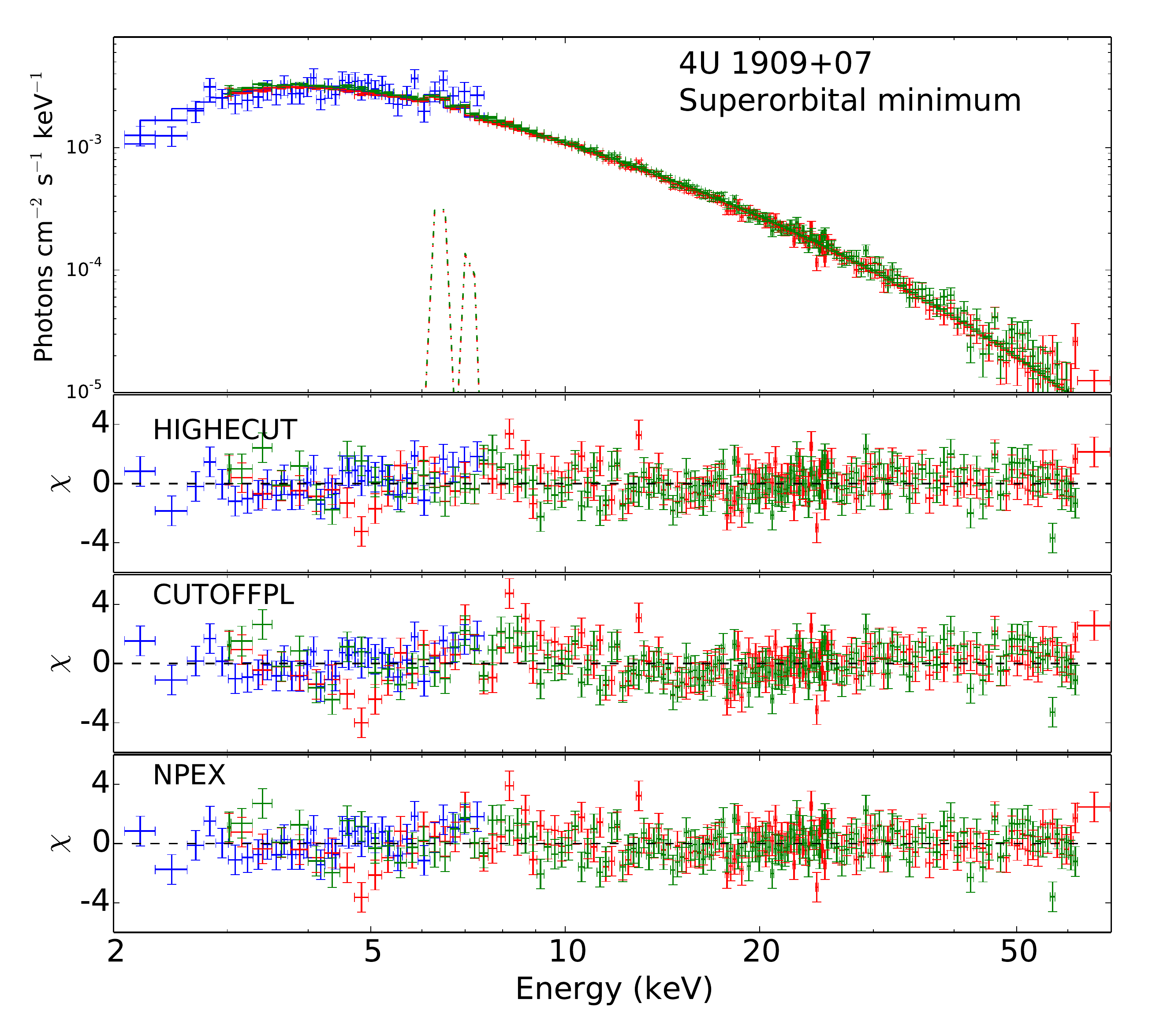}
    \includegraphics[scale=0.32,angle=0]{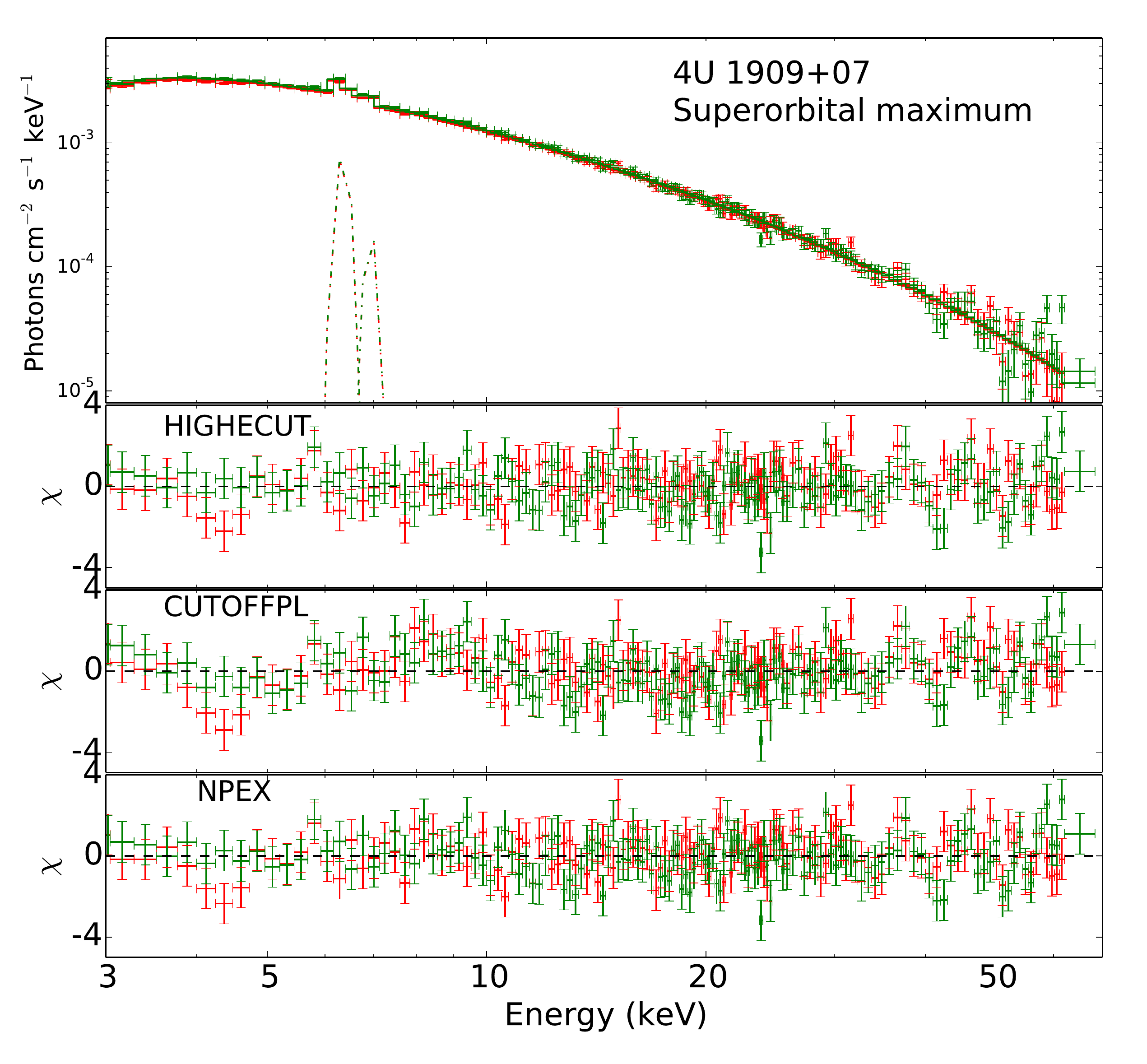}
    \includegraphics[scale=0.32,angle=0]{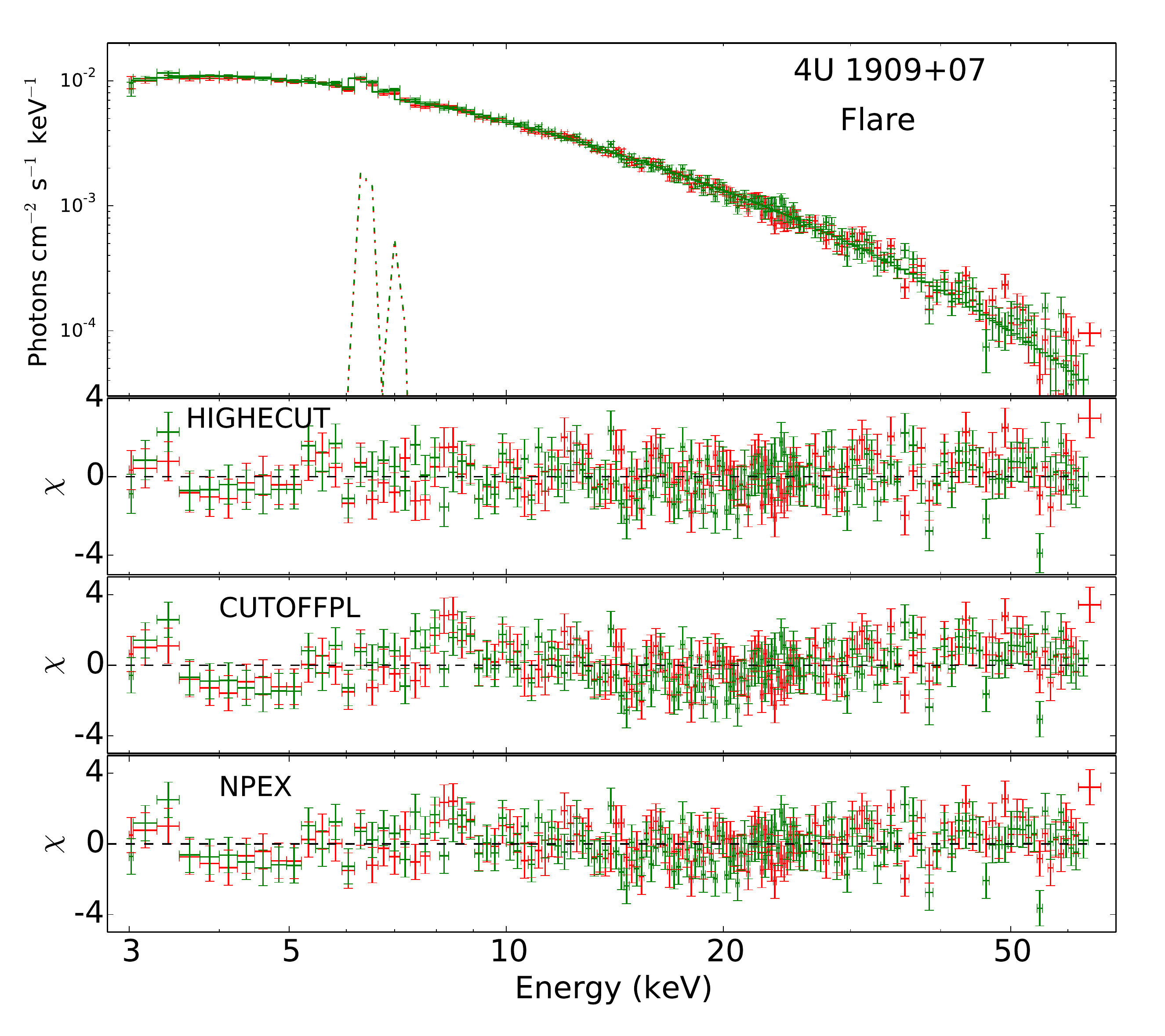}
   \caption{Pulse-phase averaged spectra of 4U 1909+07 at the predicted superorbital minimum (left panel), maximum (right panel) and the X-ray flare (bottom panel). The top panel shows the spectra with Swift XRT (blue), NuSTAR FPMA (red) and FPMB (green). The bottom panels are the residuals to the fit using {\tt HIGHECUT}, {\tt CUTOFFPL} and {\tt NPEX} models. The spectra are rebinned for clarity}
   \label{4U1909_phase_avg}
\end{figure}

\begin{figure}
    \centering
    \includegraphics[scale=0.35,angle=0]{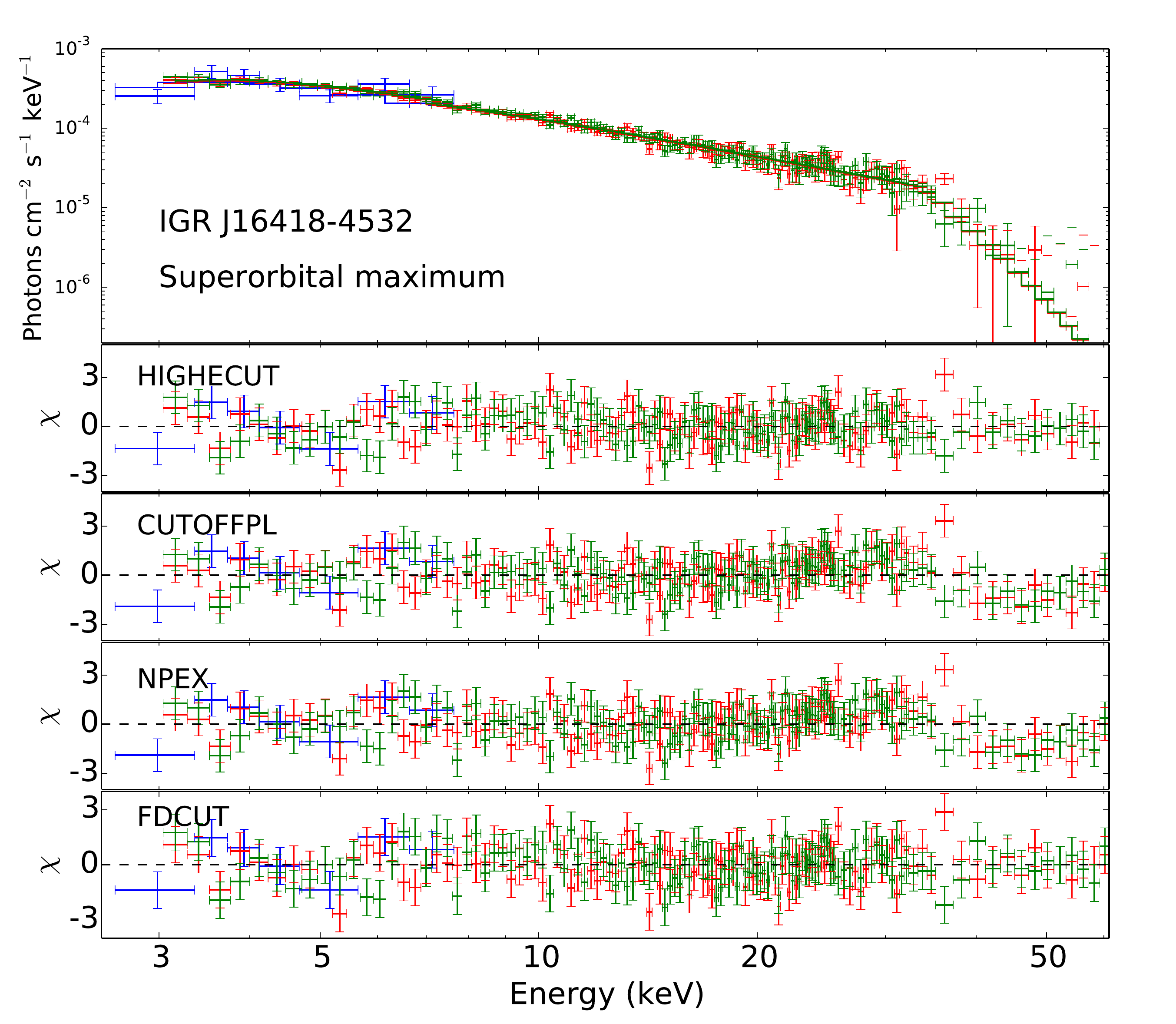}\\
    \includegraphics[scale=0.35,angle=0]{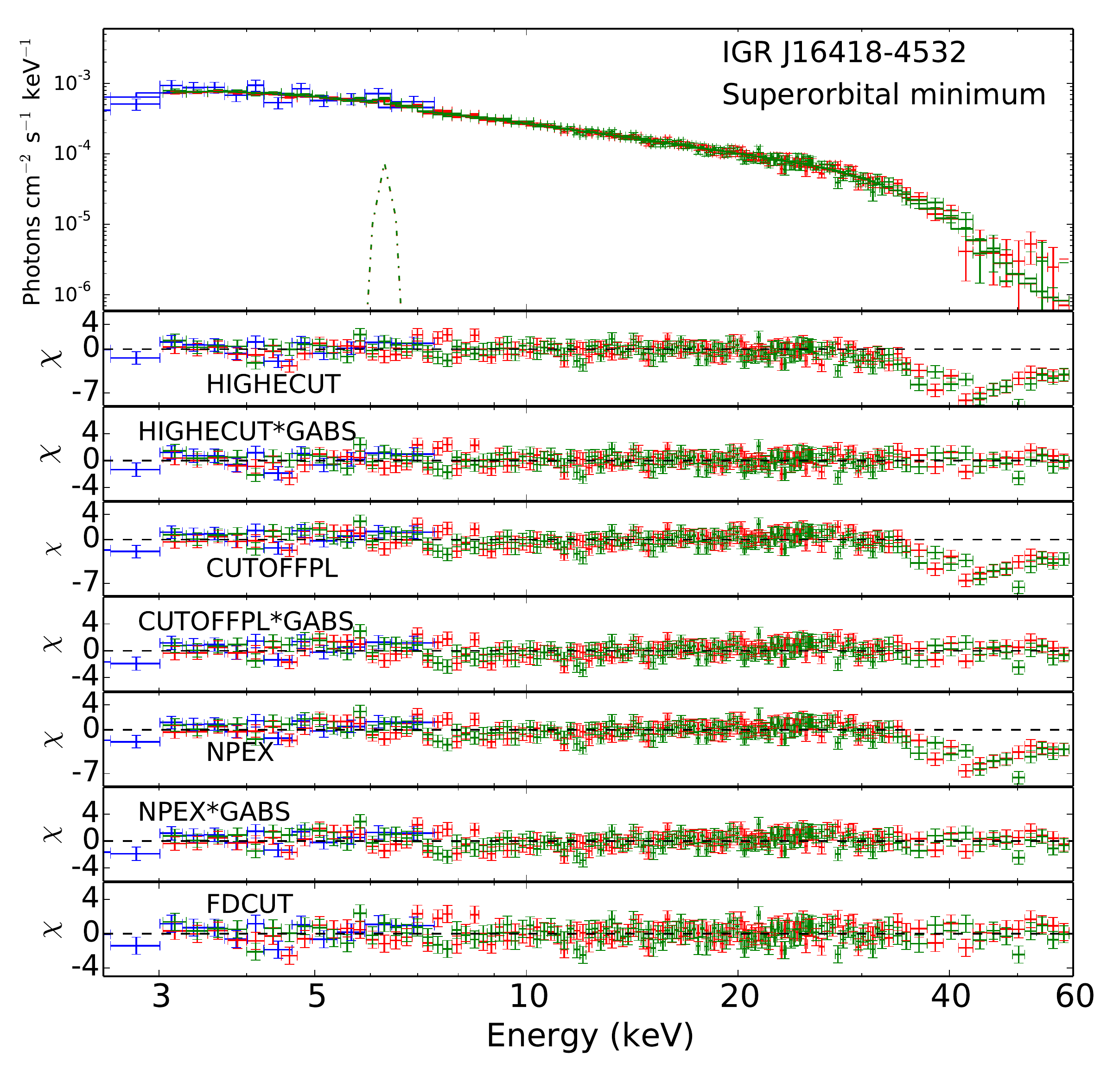}
    \includegraphics[scale=0.35]{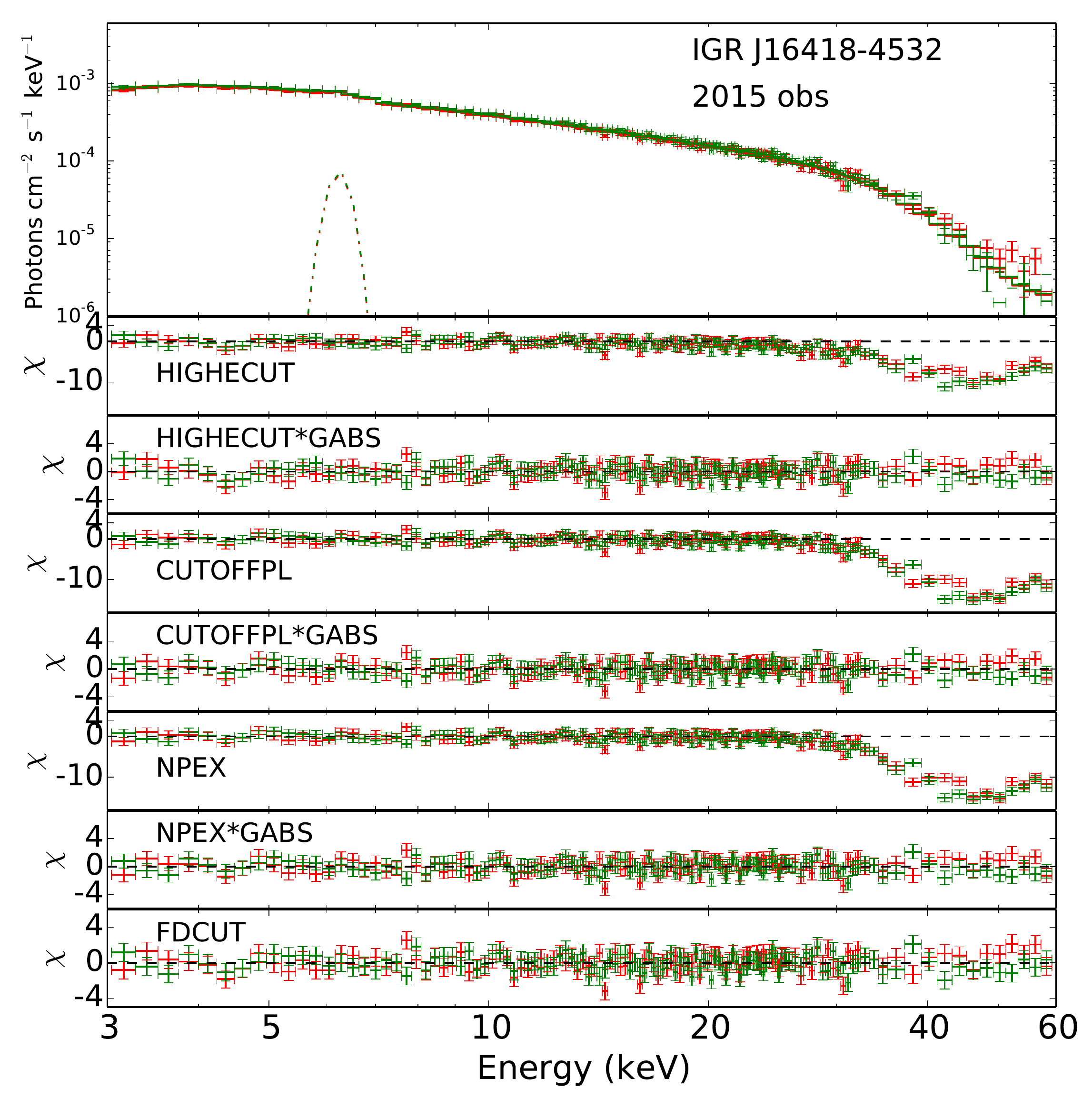}
   \caption{Pulse-phase averaged spectra of IGR J16418--4532 at the predicted superorbital maximum (top panel) minimum (bottom left panel) and during the 2015 NuSTAR observation (bottom right panel); Swift XRT (blue), NuSTAR FPMA (red) and FPMB (green). The bottom panels are the residuals to the fit using {\tt HIGHECUT}, {\tt CUTOFFPL}, {\tt NPEX} and {\tt FDCUT} models. We find absorption-like negative features $\sim$ 50 keV in the residuals to the fit to {\tt HIGHECUT}, {\tt CUTOFFPL} and {\tt NPEX} for the superorbital minimum observation and the 2015 observation (bottom panel), which is fitted with a absorption line with a gaussian optical depth profile {\tt gabs}. The spectra are rebinned for clarity}
   \label{igrj16418_phase_avg}
\end{figure}

\begin{figure}
    \centering
\includegraphics[scale=0.32,angle=0]{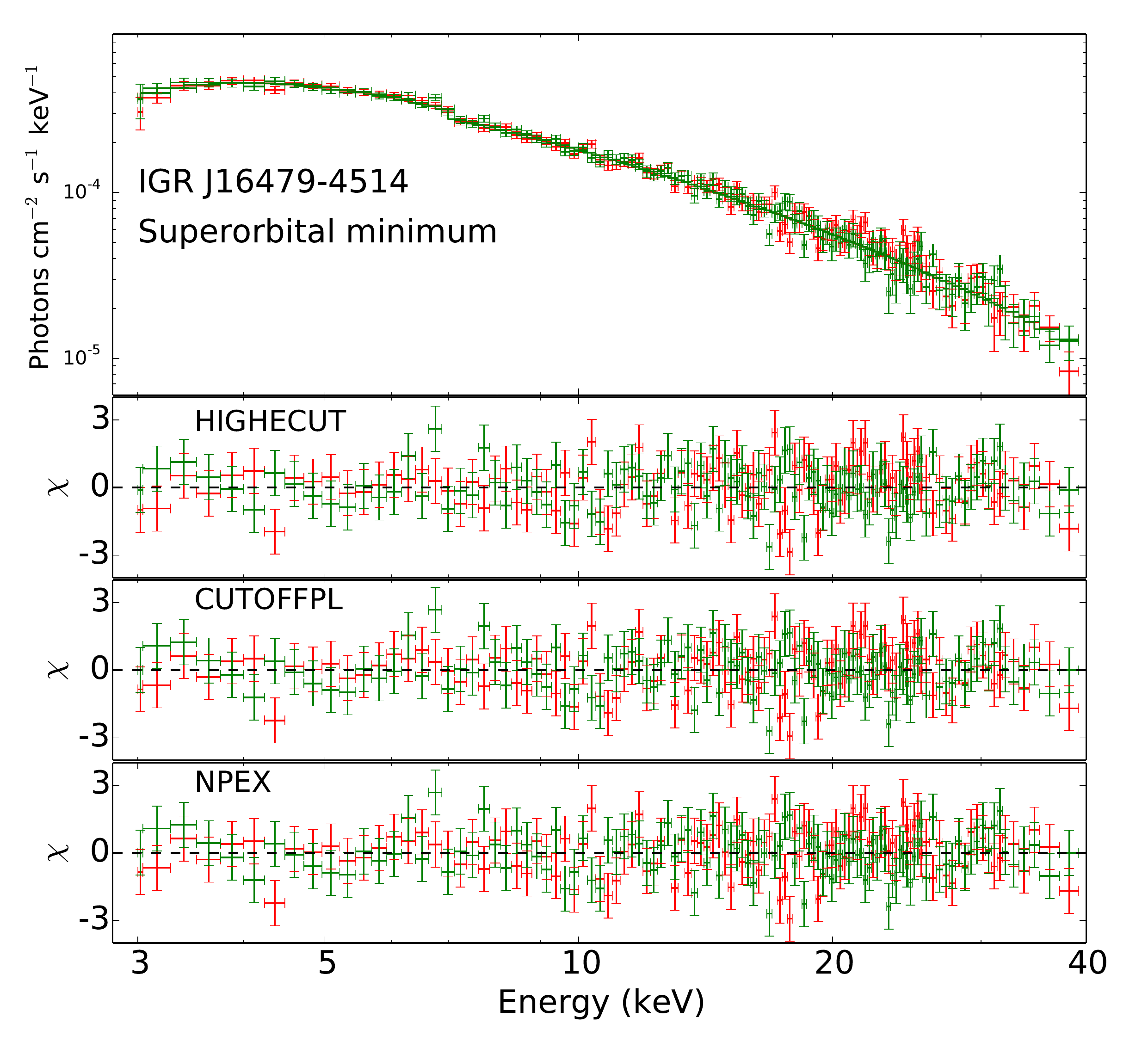}
\includegraphics[scale=0.32,angle=0]{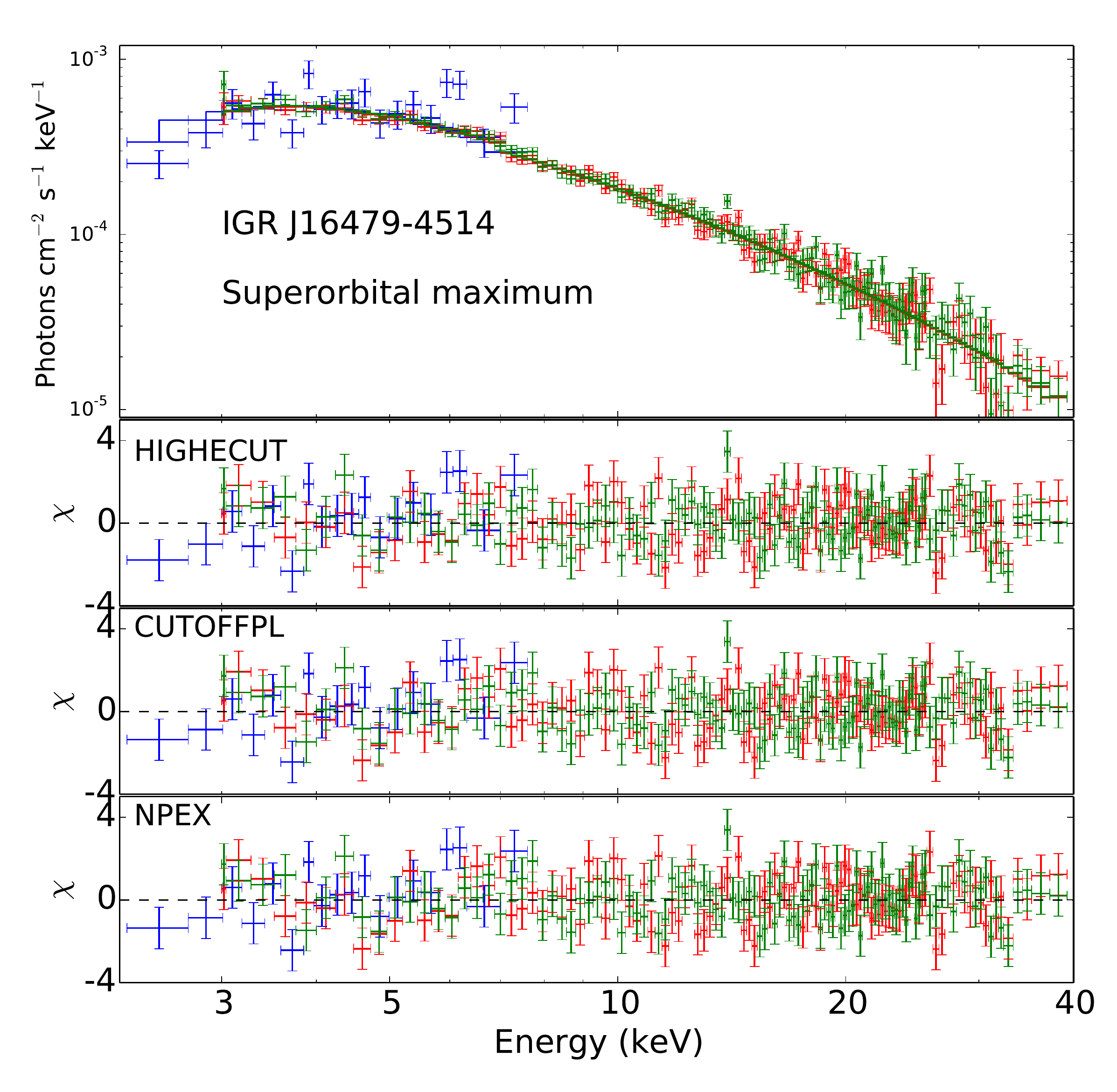}
%\vspace{-20mm}
   \caption{Pulse-phase averaged spectra of IGR J16479--4532 at the predicted superorbital minimum (left panel) and maximum (right panel), excluding the large X-ray flares. The top panel shows the spectra with Swift XRT (blue), NuSTAR FPMA (red) and FPMB (green). The bottom panels are the residuals to the fit using {\tt HIGHECUT}, {\tt CUTOFFPL} and {\tt NPEX} models. The spectra are rebinned for clarity.}
   \label{igrj16479_phase_avg}
\end{figure}

The simultaneous broadband fitting of the Swift XRT and the NuSTAR spectra of the three sources was carried out using empirical spectral models often used for fitting HMXB pulsars: a power-law with high energy cut-off ({\tt HIGHECUT}, \citealt{white1983}), a power-law with high energy exponential cut-off ({\tt CUTOFFPL}), a power-law with a Fermi-Dirac cut-off ({\tt FDCUT}, \citealt{tanaka1986}), and negative and positive power-law exponential ({\tt NPEX}, \citealt{mihara1995}). Different spectral models were used to fit the same spectra to check for the consistency of the results. All spectral models were modified by a photo-absorption component that fully covers the source {\tt tbabs in XSPEC} using {\tt verner} cross-sections and {\tt aspl} abundances \citep{verner1996,wilms2000,asplund2009}. The cross-calibration constants normalized to NuSTAR/FPMA were convolved with the above spectral models to account for the instrumental calibration uncertainties. The X-ray fluxes and their errors were estimated using the {\tt cflux} model in {\tt XSPEC}. All spectral fits were carried out using {\tt XSPEC} v 12.12.0.

\subsubsection{Pulse-phase averaged spectral analysis of 4U 1909+07}

\begin{table}
\footnotesize
\caption{Best fitting parameter values for the Swift XRT + NuSTAR spectra of 4U 1909+07 using the models defined in Section 2.3 for the observations at the predicted superorbital minimum, maximum phases and the X-ray flare. The errors on the parameters are estimated using 90\% confidence limits.}
\begin{tabular}{c c c c c c c c c c}
\hline
 & & HIGHECUT & & & CUTOFFPL & & & NPEX & \\ 
Parameters & Min & Max & Flare & Min & Max & Flare & Min & Max & Flare \\
\hline
$C_{\rm{XRT}}$ & 1.11$\pm$0.06 & - & - & 1.12$\pm$0.06 & - & - & 1.11$\pm$0.06 & - & - \\
$C_{\rm{FPMB}}$ & 1.04$\pm$0.01 & 1.03$\pm$0.01 & 1.02$\pm$0.01 & 1.04$\pm$0.01 & 1.03$\pm$0.01 & 1.02$\pm$0.01 & 1.04$\pm$0.01 & 1.03$\pm$0.01 & 1.03$\pm$0.01 \\
$N_{\rm{H}}^{(a)}$  & 7.2$\pm$0.4 & 6.9$^{+0.5}_{-0.7}$ & 5.9$\pm$0.6 & 7.9$\pm$0.4 & 7.3$\pm$0.5 & 6.3$\pm$0.6 &  6.7$\pm$0.3 & 6.0$\pm$0.4 & 5.5$\pm$0.6 \\
$\Gamma$ & 1.51$\pm$0.04 & 1.4$^{+0.05}_{-0.1}$ & 1.20$\pm$0.06 & 1.34$\pm$0.04 & 1.20$\pm$0.05 & 0.92$\pm$0.06 & 0.90$\pm$0.05 & 0.81$^{+0.08}_{-0.06}$ & 0.63$\pm$0.04 \\
Norm$^{(b)}$  & 4.2$\pm$0.4 & 3.7$\pm$0.5 & 9$\pm$1 & 4.3$\pm$0.3 & 3.5$\pm$0.3 & 7.8$\pm$0.8 & 2.7$\pm$0.2 & 2.2$\pm$0.2 & 5.7$\pm$0.5 \\
Cut-off energy $E_{\rm{C}}^{(c)}$ & 7.8$\pm$0.4 & 8.1$^{+0.5}_{-1.0}$ & 8.4$\pm$0.4 & - & - & - & - & - & - \\
Fold energy $E_{\rm{F}}^{(c)}$ & 24$\pm$1 & 26$\pm$2 & 20$\pm$1 & 19$\pm$1 & 20$\pm$1 & 15.3$\pm$0.9 & 8.9$^{+0.9}_{-0.6}$ & 9.9$^{+1.5}_{-0.9}$ & 9.7$\pm$0.2 \\
Fe K${\alpha}^{(c)}$ & 6.35$\pm$0.06 & 6.35$\pm$0.04 & 6.38$\pm$0.05 & 6.39$\pm$0.05 & 6.37$\pm$0.04 & 6.40$\pm$0.05 & 6.38$\pm$0.05 & 6.35$\pm$0.04 & 6.39$\pm$0.05 \\
Fe K${\beta}^{(c)}$ & 7.1$\pm$0.2 & 6.9$\pm$0.2 & 7.0$\pm$0.3 & 7.5$\pm$0.2 & 7.1$\pm$0.2 & 7.2$\pm$0.2 & 7.5$\pm$0.3 & 6.9$\pm$0.2 & 7.1$\pm$0.3 \\
Fe K$\alpha$ Norm$^{(d)}$ & 1.8$\pm$0.3 & 2.9$\pm$0.4 & 9$\pm$2 & 1.8$\pm$0.3 & 2.9$\pm$0.3 & 8$\pm$2 & 1.7$\pm$0.3 & 2.9$\pm$0.4 & 8$\pm$1 \\
Fe K$\beta$ Norm$^{(d)}$ & 0.7$\pm$0.3 & 0.6$\pm$0.4 & 2$\pm$1 & 1.0$\pm$0.3 & 0.8$\pm$0.3 & 3$\pm$1 & 0.8$\pm$0.3 & 0.7$\pm$0.3 & 2$\pm$1 \\
Fe K$\alpha$ EW$^{(e)}$ & 69$\pm$13 & 109$\pm$15 & 94$\pm$15 & 70$\pm$12 & 108$\pm$13 & 88$\pm$17 & 66$\pm$12 & 108$\pm$13 & 87$\pm$17 \\
Fe K$\beta$ EW$^{(e)}$ & 32$\pm$14 & 27$\pm$17 & 22$\pm$17 & 58$\pm$13 & 34$\pm$13 & 29$\pm$16 & 38$\pm$13 & 29$\pm$15 & 23$\pm$16 \\
$\chi^{2}_{\nu}$/dof & 1.11/318 & 0.91/246 & 1.01/246 & 1.2/308 & 1.13/247 & 1.23/247 & 1.16/307 & 0.93/246 & 1.09/247 \\
Flux$_{\rm{s}}^{(f)}$ & 1.26$\pm$0.03 & - & - & 1.27$\pm$0.01 & - & - & 1.27$\pm$0.01 & - & - \\
Flux$_{\rm{h}}^{(g)}$ & 4.06$\pm$0.02 & 4.99$\pm$0.03 & 18.1$\pm$0.1 & 4.03$\pm$0.02 & 4.90$\pm$0.03 & 17.8$\pm$0.1 & 4.07$\pm$0.02 & 4.96$\pm$0.03 & 18.0$\pm$0.1 \\
\hline
\end{tabular} \\
Notes: (a) in units of $10^{22}$ cm$^{-2}$; (b) in the units of $10^{-2}$ photons  keV$^{-1}$ cm$^{-2}$ s$^{-1}$ at 1 keV; (c) in the units of keV; (d) in the units of $10^{-4}$ photons cm$^{-2}$ s$^{-1}$; (e) EW is the equivalent width in the units of eV; (f) Unabsorbed X-ray flux in 0.5-8.0 keV in $10^{-10}$ ergs s$^{-1}$ cm$^{-2}$; (g) Unabsorbed X-ray flux in 3-70 keV in $10^{-10}$ ergs s$^{-1}$ cm$^{-2}$
\end{table}

For 4U 1909+07, a Swift XRT observation was performed simultaneously with the NuSTAR observation for only the predicted superorbital minimum phase. Therefore, we fitted the Swift XRT + NuSTAR spectra for the predicted superorbital minimum observation, only NuSTAR spectra for the predicted superorbital maximum observation (excluding the X-ray flare), and the short X-ray flare separately with the spectral models described in Section 2.3. The energy band chosen for spectral fitting for Swift XRT is 0.5--8.0 keV, and NuSTAR FPMA and FPMB are 3.0--70.0 keV. Fe fluorescence lines are seen in the spectra at 6.4 keV and 7.1 keV, which were modeled by gaussian lines. The widths of the gaussian lines for the 6.4 keV Fe K$\alpha$ and 7.1 keV Fe K$\beta$ lines were fixed at 0.1 keV. The errors on the equivalent widths of the gaussian lines were propagated from their line fluxes. The spectral parameters especially the cut-off energy $E_{\rm{C}}$ cannot be constrained to a feasible value using the spectral model {\tt FDCUT} and hence we report results of the spectral fits using the spectral models {\tt CUTOFFPL}, {\tt HIGHECUT} and {\tt NPEX} in Table 3. The error bars are reported at 90\% confidence limit. Figure \ref{4U1909_phase_avg} shows the spectral fit to the pulse-phase averaged spectra for the predicted superorbital minimum, maximum observations and the X-ray flare along with the residuals to the best fit to the {\tt HIGHECUT}, {\tt CUTOFFPL} and {\tt NPEX} model for 4U 1909+07. 

\subsubsection{Pulse-phase averaged spectral analysis of IGR J16418--4532}

\begin{sidewaystable}
\footnotesize
\centering
\caption{Best fitting parameter values for the Swift XRT + NuSTAR spectra of IGR J16418--4532 using the models defined in Section 2.3 for the observations at the predicted superorbital minimum, maximum phases and a 2015 observation. The errors on the parameters are estimated using 90\% confidence limits.}
\begin{tabular}{c c c c c c c c c c c c c}
\hline
 & & HIGHECUT & & & CUTOFFPL & & & NPEX & & & FDCUT \\ 
Parameters & Min & Max & 2015 & Min & Max & 2015 & Min & Max & 2015 & Min & Max & 2015 \\
\hline
$C_{\rm {XRT}}$ & 0.93$\pm$0.08 & 0.9$\pm$0.1 & - & 0.87$\pm$0.08 & 0.9$\pm$0.1 & - & 0.89$\pm$0.08 & 0.9$\pm$0.1 & - &  0.93$\pm$0.08 & 0.9$\pm$0.1 & - \\
$C_{\rm {FPMB}}$ & 1.01$\pm$0.01 & 1.03$\pm$0.02 & 1.03$\pm$0.01 & 1.01$\pm$0.1 & 1.03$\pm$0.02 & 1.03$\pm$0.01 & 1.01$\pm$0.01 & 1.03$\pm$0.02 & 1.03$\pm$0.01 & 1.01$\pm$0.01 & 1.03$\pm$0.02 & 1.02$\pm$0.01 \\
$N_{\rm{H}}^{(a)}$ & 5.8$\pm$0.5 & 7.1$\pm$0.8 & 6.7$\pm$0.4 & 3.3$\pm$0.4 & 5$\pm$1 & 5.5$\pm$0.5 & 3.7$\pm$0.4 & 5$\pm$1 & 5.5$\pm$0.5 & 5.6$\pm$0.5 & 6.9$\pm$0.8 & 6.3$\pm$0.4 \\
$\Gamma$ & 1.48$\pm$0.02 & 1.66$\pm$0.04 & 1.39$\pm$0.02 & 1.0$\pm$0.02 & 1.2$\pm$0.1 & 1.18$\pm$0.05 & 1.07$\pm$0.02 & 1.2$\pm$0.1 & 1.18$\pm$0.05 & 1.46$\pm$0.03 & 1.65$\pm$0.05 & 1.35$\pm$0.02 \\
Norm$^{(b)}$ & 8.8$\pm$0.6 & 6.5$\pm$0.6 & 10.3$\pm$0.5 & 4.4$\pm$0.2 & 3.4$\pm$0.6 & 7.5$\pm$0.6 & 4.8$\pm$0.3 & 3.4$\pm$0.6 & 7.5$\pm$0.7 & 8.3$\pm$0.6 & 6.3$\pm$0.7 & 9.5$\pm$0.5 \\
Cut-off energy $E_{\rm{C}}^{(c)}$ & 26$^\pm$2 & 32$\pm$3 & 24$\pm$1 & - & - & - & - & - & 36.6$\pm$0.8 & 40$\pm$2 & 35.7$\pm$0.6 \\
Fold energy $E_{\rm{F}}^{(c)}$ & 40$^{+17}_{-12}$ & 10$\pm$4 & 30$^{+6}_{-4}$ & 30(fixed) & 27$^{+6}_{-5}$ & 57$^{+13}_{-9}$ & 30(fixed) & 27$\pm$5 & 59$^{+14}_{-9}$ & 5.0$\pm$0.7 & 4$\pm$2 & 5.2$\pm$0.5 \\
gabs (LineE)$^{c}$ & 53$^{+24}_{-4}$ & - & 59$^{+46}_{-5}$ & 53$^{+6}_{-3}$ & - & 56$^{+5}_{-3}$ & 52$^{+6}_{-2}$ & - & 56$^{+5}_{-3}$ & - & - & - \\
gabs ($\sigma$)$^{c}$ & 8$^{+7}_{-2}$ & - & 10$^{+10}_{-2}$ & 8$^{+2}_{-1}$ & - & 10$^{+2}_{-1}$ & 8$\pm$2 & - & 10$^{+2}_{-1}$ & - & - & - \\
gabs (Strength) & 33$^{+79}_{-13}$ & - & 77$^{+30}_{-29}$ & 44$^{+51}_{-12}$ & - & 81$^{+53}_{-20}$ & 41$^{+41}_{-11}$ & - & 80$^{+52}_{-21}$ & - & - & - \\ 
Fe K${\alpha}^{(c)}$ & 6.3$\pm$0.1 & 6.4(fixed) & 6.3$\pm$0.1 & 6.2$\pm$0.1 & 6.4(fixed) & 6.3$\pm$0.1 & 6.3$\pm$0.1 & 6.4(fixed) & 6.2$\pm$0.1 & 6.3$\pm$0.1 & 6.4(fixed) & 6.3$\pm$0.1 \\
Fe K$\alpha$ Norm$^{(d)}$ & 2$\pm$1 & $<$0.3 & 4$\pm$1 & 3$\pm$1 & $<$0.6 & 5$\pm$1 & 3$\pm$1 & $<$0.6 & 5$\pm$1 & 3$\pm$1 & $<$0.3 & 4$\pm$1 \\
Fe K$\alpha$ EW$^{(e)}$ & 41$\pm$21 & $<$14 & 49$\pm$15 & 69$\pm$21 & $<$28 & 57$\pm$14 & 61$\pm$20 & $<$28 & 57$\pm$15 & 45$\pm$21 & $<$14 & 51$\pm$14 \\
$\chi^{2}_{\nu}$/dof & 1.00/252 & 0.98/249 & 0.89/240 & 1.15/254 & 1.12/250 & 0.94/240 & 1.15/254 & 1.12/256 & 0.96/239 & 1.03/252 & 0.98/249 & 0.84/240 \\
Flux$_{\rm{s}}^{(f)}$ & 2.64$\pm$0.03 & 1.35$\pm$0.03 & - & 2.71$\pm$0.04 & 1.37$\pm$0.02 & - & 2.71$\pm$0.02 & 1.37$\pm$0.02 & - & 2.67$\pm$0.03 & 1.35$\pm$0.02 & - \\
Flux$_{\rm{h}}^{(g)}$ & 11.9$\pm$0.1 & 5.6$\pm$0.2 & 17.3$\pm$0.1 & 11.8$\pm$0.1 & 5.8$\pm$0.1 & 17.3$\pm$0.1 & 11.7$\pm$0.1 & 5.8$\pm$0.1 & 17.2$\pm$0.1 & 11.8$\pm$0.1 & 5.6$\pm$0.1 & 17.3$\pm$0.1 \\

\hline
\end{tabular} \\
Notes: (a) in units of $10^{22}$ cm$^{-2}$; (b) in the units of $10^{-3}$ photons keV$^{-1}$ cm$^{-2}$ s$^{-1}$ at 1 keV; (c) in the units of keV; (d) in the units of $10^{-5}$ photons cm$^{-2}$ s$^{-1}$; (e) EW is the equivalent width in the units of eV; (f) Unabsorbed X-ray flux in 0.5-8.0 keV in $10^{-11}$ ergs s$^{-1}$ cm$^{-2}$; (g) Unabsorbed X-ray flux in 3-60 keV in $10^{-11}$ ergs s$^{-1}$ cm$^{-2}$
\end{sidewaystable}

For IGR J16418--4532, the background dominates above 60 keV and therefore chose to fit the Swift XRT and NuSTAR spectra in the 0.5--8.0 keV and 3.0--60.0 keV energy bands respectively. The broadband Swift XRT and NuSTAR spectra of the predicted superorbital minimum and maximum observations were fitted with four spectral models described in Section 2.3: {\tt CUTOFFPL}, {\tt HIGHECUT}, {\tt NPEX} and {\tt FDCUT}. A gaussian line was used to model the 6.4 keV Fe K$\alpha$ line present in the spectra for the predicted superorbital minimum observation and its width was fixed at 0.1 keV. There were no strong residuals around the 6.4 keV Fe K$\alpha$ line while fitting the predicted superorbital maximum spectra with the four spectral models. We model a gaussian line with a fixed line center at 6.4 keV and fit the spectra to estimate the upper limits on the equivalent width and the line flux of the Fe K$\alpha$ line, which is reported in Table 4. In the bottom right panel of Figure \ref{igrj16418_phase_avg}, the NuSTAR observation of IGR J16418--4532 at the predicted superorbital minimum phase shows absorption-like features around $\sim$50 keV, for residuals to the spectral fits with {\tt CUTOFFPL}, {\tt HIGHECUT} and {\tt NPEX} models. These absorption-like negative features are not seen in the residuals to the spectral fits to the {\tt FDCUT} model. These negative absorption-like residuals were fitted with an absorption line with a gaussian optical depth profile {\tt gabs} model and the spectral parameters are reported in Table 4. To determine the significance of the inclusion of the possible absorption feature, we carried out {\tt simftest} in {\tt XSPEC} which simulates 10$^{4}$ iterations of the spectra based on the actual data and tests the resulting $\Delta \chi^{2}$ using an additional model component (absorption line {\tt gabs} here; \citealt{protassov2002}). The statistical significance of this absorption feature is low $\leq$25\% which suggests the improvement of the spectral fit from adding an absorption feature to the spectral model is negligible.
\par
The NuSTAR observation at the predicted superorbital maximum does not exhibit any absorption-like negative residuals in the spectral fits (top panel of Figure \ref{igrj16418_phase_avg}) most likely due to the shorter observation time of 20 ks as compared to 31 ks of the predicted superorbital minimum observation. To further investigate these absorption-like features in the spectra of IGR J16418--4532, we analyzed a 40 ks NuSTAR observation carried out on MJD 57,291 (2015-09-26). The NuSTAR data were extracted with the same procedure as described in Section 2.2.1. The energy band chosen for fitting the spectra was 3.0--60 keV since the contribution from the background dominates above 60 keV. We fit the 2015 NuSTAR FPMA and FPMB spectra simultaneously with the four spectral models {\tt CUTOFFPL}, {\tt HIGHECUT}, {\tt NPEX} and {\tt FDCUT}. A 6.4 keV Fe K$\alpha$ line was present in the spectrum which was modeled by a gaussian line. The bottom right panel of Figure \ref{igrj16418_phase_avg} shows the best fit to the spectra for the 2015 NuSTAR observation along with the residuals to the best fit for the {\tt HIGHECUT}, {\tt CUTOFFPL}, {\tt NPEX} and {\tt FDCUT} model. We see a negative absorption-like residual in the best fit for models {\tt HIGHECUT}, {\tt CUTOFFPL}, {\tt NPEX} in spectra at $\sim$ 50 keV, similar to that seen for the predicted superorbital minimum observation, whereas no such residuals were seen with the {\tt FDCUT} model. These negative absorption-like features were fitted with an absorption line with a gaussian optical depth {\tt gabs} model and the spectral parameters are reported in Table 4. From a similar analysis using {\tt simftest}, we found the statistical significance of this absorption feature is $\leq$47\% for the spectral fits to the 2015 observation, which is an improvement over the statistical significance of the absorption feature to the spectral fits for the predicted superorbital minimum observation in 2018. However, we cannot confirm or rule out conclusively the presence of a Cyclotron Scattering Resonance Feature (CRSF) present in the spectrum which could be a possible explanation for such negative absorption-like features in the residuals to the spectral fits. 

\subsubsection{Pulse-phase averaged spectral analysis of IGR J16479--4514}
\begin{table}
\footnotesize
\centering
\caption{Best fitting parameter values for the Swift XRT + NuSTAR spectra of IGR J16479--4514, excluding the large X-ray flares and using the models defined in Section 2.3 for the observations at the predicted superorbital minimum and maximum phases. The errors on the parameters are estimated using 90\% confidence limits.}
\begin{tabular}{c c c c c c c}
\hline
 &  HIGHECUT & & CUTOFFPL & & NPEX & \\ 
Parameters & Min & Max & Min & Max & Min & Max \\
\hline
$C_{\rm{XRT}}$ & - & 1.8$\pm$0.1 & - & 1.8$\pm$0.1 & - & 1.8$\pm$0.1 \\
$C_{\rm{FPMB}}$ & 0.99$\pm$0.02 & 1.01$\pm$0.02 & 0.99$\pm$0.02 & 1.01$\pm$0.02 & 0.99$\pm$0.02 & 1.01$\pm$0.02 \\
$N_{\rm{H}}^{(a)}$ & 7$\pm$1 & 6$\pm$1 & 8.1$\pm$0.9 & 6.8$\pm$0.9 & 8.1$\pm$0.9 & 6.8$\pm$0.9  \\
$\Gamma$ & 1.3$\pm$0.1 & 1.4$\pm$0.1 & 1.28$\pm$0.09 & 1.4$\pm$0.1 & 0.77$\pm$0.06 & 1.4$\pm$0.1  \\
Norm$^{(b)}$ & 5$\pm$1 & 6$\pm$1 & 6$\pm$1 & 7$\pm$1 & 6$\pm$1 & 7$\pm$0.1  \\
Cut-off energy $E_{\rm{C}}^{(c)}$ & 6.5$^{+1.3}_{-0.8}$ & 6$\pm$1 & - & - & - & - \\
Fold energy $E_{\rm{F}}^{(c)}$ & 29$^{+9}_{-5}$ & 32$^{+10}_{-6}$ & 27$^{+6}_{-4}$ & 26$^{+7}_{-5}$ & 27$^{+6}_{-4}$ & 27$^{+7}_{-5}$ \\
$\chi^{2}_{\nu}$/dof & 0.95/222 & 1.14/234 & 0.96/223 & 1.15/235 & 0.97/222 & 1.16/234 \\
Flux$_{\rm{s}}^{(d)}$ & - & 4.39$\pm$0.03 & - & 4.40$\pm$0.02 & - & 4.39$\pm$0.02  \\
Flux$_{\rm{h}}^{(e)}$ & 6.9$\pm$0.1 & 6.8$\pm$0.1 & 6.9$\pm$0.2 & 6.7$\pm$0.1 & 6.9$\pm$0.1 & 6.7$\pm$0.1 \\
\hline
\end{tabular} \\
Notes: (a) in units of $10^{22}$ cm$^{-2}$; (b) in the units of $10^{-3}$ photons keV$^{-1}$ cm$^{-2}$ s$^{-1}$ at 1 keV; (c) in the units of keV; (d) Unabsorbed X-ray flux in 0.5-8.0 keV in $10^{-11}$ ergs s$^{-1}$ cm$^{-2}$; (e) Unabsorbed X-ray flux in 3-40 keV in $10^{-11}$ ergs s$^{-1}$ cm$^{-2}$
\end{table}

The lightcurves of IGR J16479--4514 show the presence of large X-ray flares, one at the superorbital phase $\sim$0.08 and the other at $\sim$0.7. We exclude the X-ray flares and extract the spectrum from the non-flaring part of the lightcurve to investigate the difference in the spectral parameters between the non-flaring part of the superorbital phases, which could be related to the mechanisms driving the superorbital modulation. We note that the Swift XRT observation contemporaneous with the NuSTAR observations at the predicted superorbital minimum phase had a low X-ray count rate (0.03 counts \,s$^{-1}$, given in Table 2) and hence this Swift XRT observation was not used for the broadband spectral fitting. We use the spectral models {\tt HIGHECUT}, {\tt CUTOFFPL} and {\tt NPEX} to fit the spectra from the non-flaring part of the NuSTAR lightcurves. The contribution from the background dominates above 40 keV for the NuSTAR spectra and therefore we chose to fit the Swift XRT and NuSTAR spectra in the 0.5--8.0 keV and 3.0--40.0 keV energy bands respectively. The spectral model {\tt FDCUT} was unable to constrain the cutoff energy ($E_{\rm{cut}}$) of the spectra and hence we do not use this model for the spectral fitting. We report results from the spectral fits to the non-flaring part of the predicted superorbital minimum and maximum observations in Table 5. The cross-normalization constant between Swift XRT and NuSTAR is large which could be due to the intrinsic source variability between the Swift XRT and NuSTAR observations. Figure \ref{igrj16479_phase_avg} shows the spectral fits to the phase-average spectra for the non-flaring part of lightcurves of the predicted superorbital minimum and maximum observation, along with the residuals to the best fit to the {\tt HIGHECUT}, {\tt CUTOFFPL} and {\tt NPEX} model for IGR J16479--4514. A detailed analysis of the large X-ray flares will be presented in a separate paper.

\subsubsection{Pulse-phase resolved spectral analysis of 4U 1909+07 and IGR J16418--4532}

\begin{figure}
    \centering
    \includegraphics[scale=0.35, angle=0]{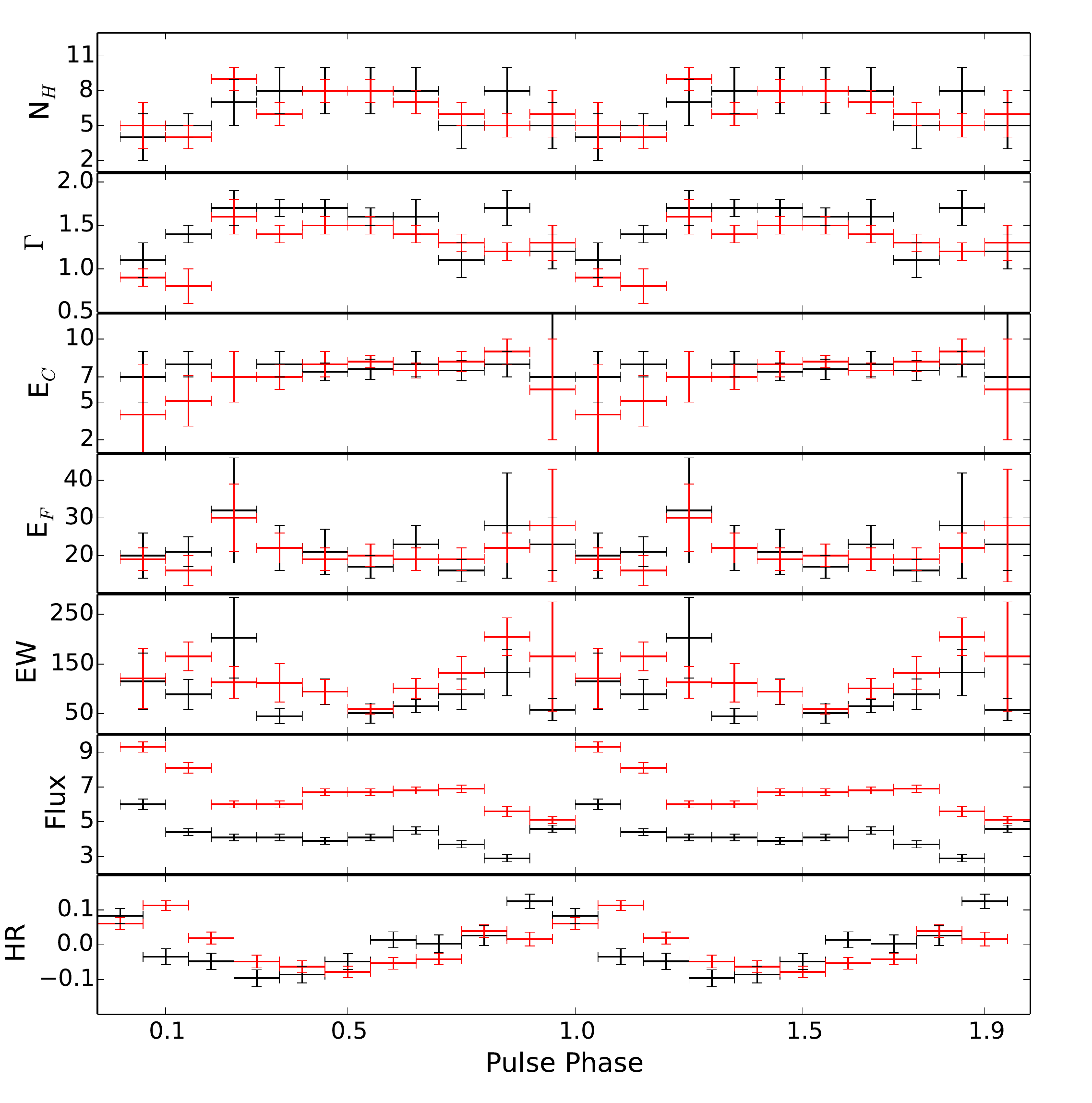}
    \includegraphics[scale=0.35, angle=0]{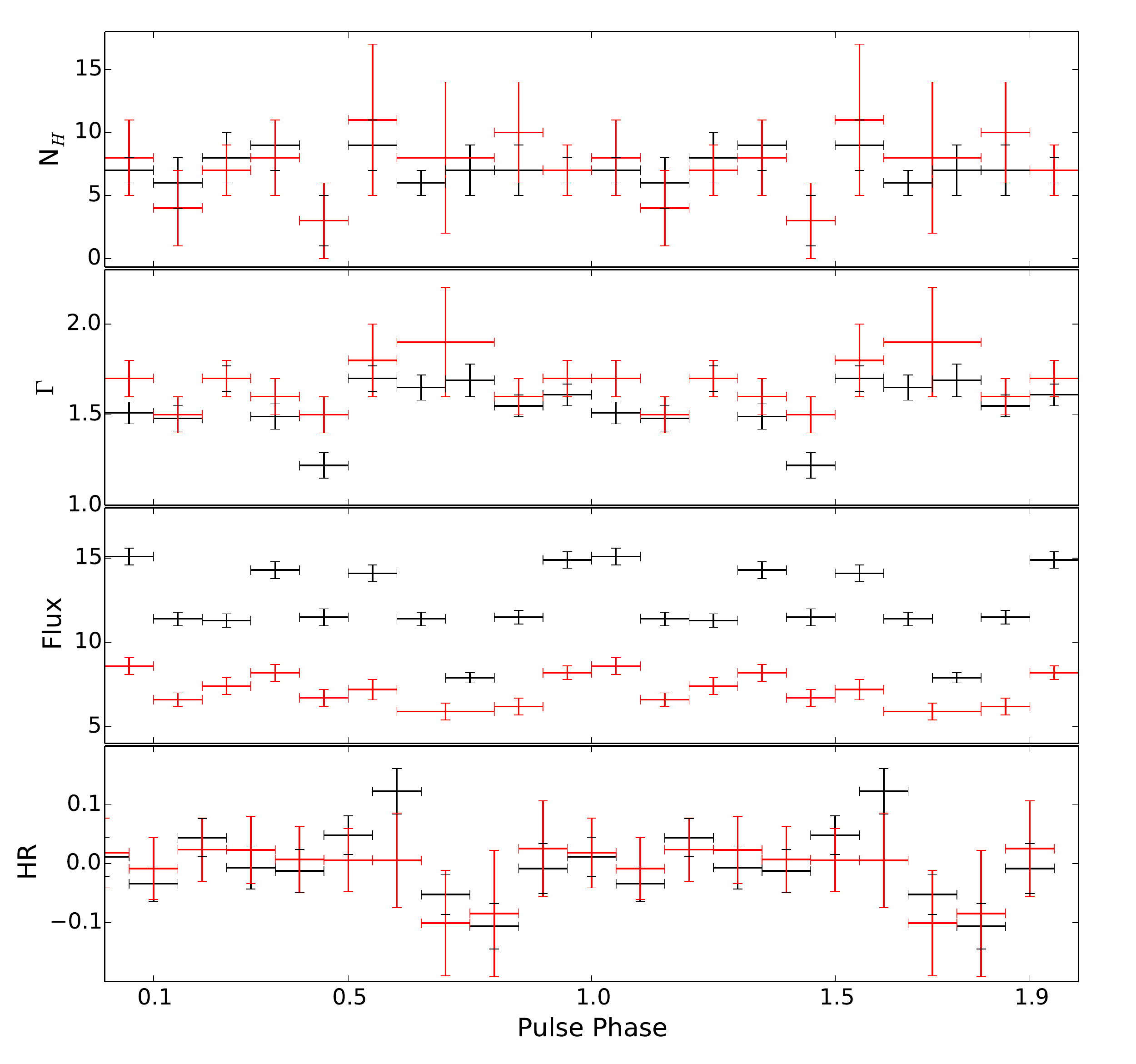}
     \caption{{\it Left panel}: Pulse-phase resolved spectra of 4U 1909+07 at the predicted superorbital minimum (black data-points) and maximum (red data-points), with a {\tt HIGHECUT} model, modified by an absorption component. {\it Right panel}: Pulse-phase resolved spectra of IGR J16418--4532 at the predicted superorbital minimum (black data-points) and maximum (red data-points) with a {\tt HIGHECUT} model, modified by an absorption component. The cut-off energy $E_{\rm{C}}$ and the folding energy $E_{\rm{F}}$ were fixed to its phase-average values. Units of $N_{\rm{H}}$ are in 10$^{22}$ cm$^{-2}$, $E_{\rm{C}}$ in keV, $E_{\rm{F}}$ in keV, equivalent width of an Fe K$\alpha$ line (EW) in eV and the unabsorbed X-ray flux (3--50 keV) in the units of 10$^{-10}$ ergs\, cm$^{-2}$\, s$^{-1}$. The errors on the parameters are estimated using 90\% confidence limits. The hardness ratios were estimated for lightcurves folded with the pulse period in 3--10 keV and 10--50 keV energy bands using Equation (1).}
    \label{phase_res}
\end{figure}

We carried out a pulse-phase resolved spectroscopic analysis of NuSTAR observations of 4U 1909+07 and IGR J16418--4532, to investigate the dependence of the spectral parameters as a function of the pulse phase. The phase bins were chosen to be 0.1 of the pulse period and the NuSTAR spectra were extracted for 10 independent pulse phases. Since the exposures of the Swift XRT observations were short, we did not use them for pulse-phase resolved analysis. For 4U 1909+07, we model the pulse-phase resolved spectra with a {\tt HIGHECUT} model modified by an absorption component, along with a gaussian line to model the Fe K$\alpha$ fluorescence line. There were no strong residuals around the Fe K$\beta$ line at 7.1 keV while fitting the pulse-phase resolved spectra with {\tt HIGHECUT} model and hence we fix this line to its pulse-phase averaged spectral values mentioned in Table 3. The results of the pulse-phase resolved spectral study for the predicted superorbital minimum and maximum observations of 4U 1909+07 are shown in the left panel of Figure \ref{phase_res}. For pulse-phase resolved spectral study of IGR J16418--4532, we use a {\tt HIGHECUT} modified by an absorption component, to model the pulse-phase resolved spectra. The cut-off energy E$_{\rm C}$ and the fold energy E$_{\rm F}$ were fixed to its phase-average values mentioned in Table 4 since these spectral parameters cannot be independently constrained during the spectral fits due to low statistics of the pulse-phase resolved spectra. The results of the pulse-phase resolved spectral study for the predicted superorbital minimum and maximum observations of IGR J16418--4532 are shown in the right panel of Figure \ref{phase_res}. The results of pulse-phase resolved spectral analysis for 4U 1909+07 and IGR J16418--4532 (keeping the HighECut parameter fixed to its phase-average value for IGR J16418--4532) using a {\tt CUTOFFPL} model modified by an absorption component, exhibits a similar variation in its spectral parameters. 

\subsection{Pulse period and Fractional RMS amplitude evolution}
\begin{figure}
    \centering
    \includegraphics[scale=0.25, angle=0]{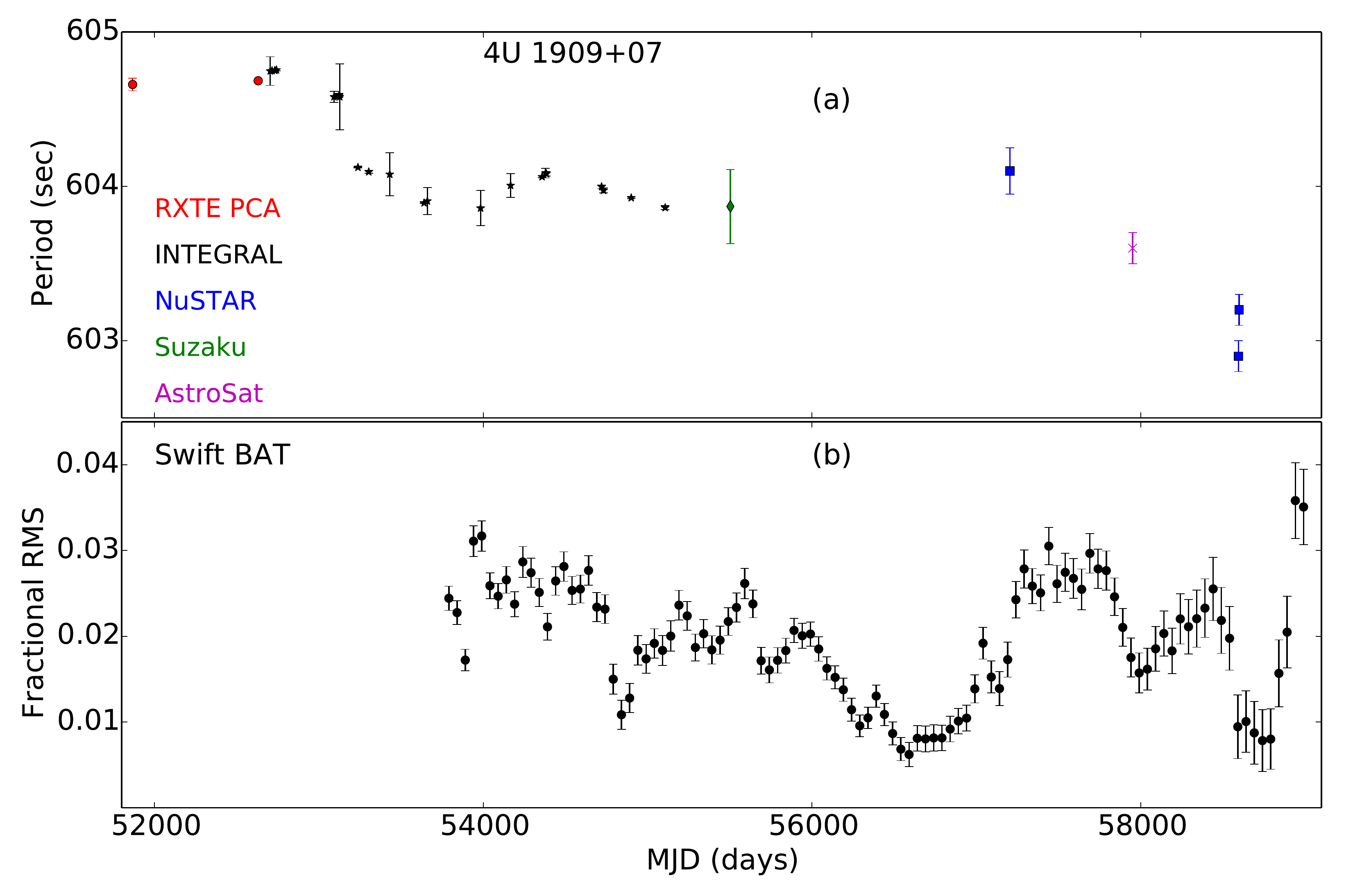}
    \includegraphics[scale=0.25, angle=0]{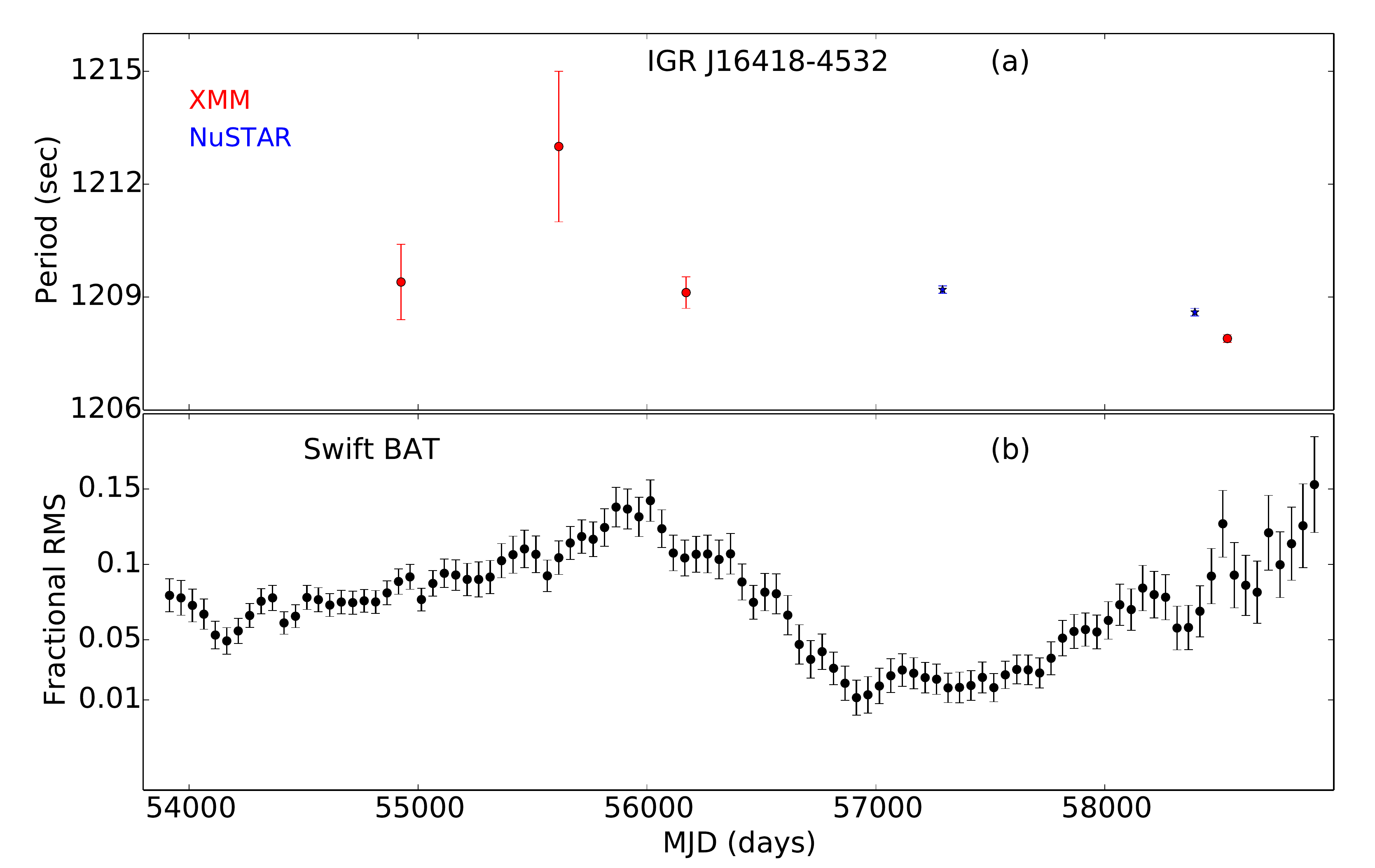}
\caption{{\it Left panel}: Long-term spin period evolution of 4U 1909+07 (a) and the fractional root mean square (RMS) amplitude (b) calculated using Swift BAT lightcurve with 750 day intervals, with 50 day increments in start and end times. The measurement of spin periods of 4U 1909+07 are from previously reported values using RXTE PCA \citep{levine2004}, INTEGRAL \citep{furst2011}, Suzaku \citep{furst2012, jaisawal2013}, NuSTAR and AstroSat observations \citep{jaisawal2020} as well from the current work. {\it Right panel}: Long-term spin period evolution of IGR J16418--4532 (a) and the fractional root mean square (RMS) amplitude (b) calculated using Swift BAT lightcurve with 1000 day intervals, with 50 day increments in start and end times. The measurement of spin periods of IGR J16418--4532 are from XMM observations previously reported in literature \citep{sidoli2012, drave2013} as well as the estimates from the current and a previous 2015 NuSTAR observations and a 2019 XMM observation.}
    \label{p_evol}
\end{figure}

Figure \ref{p_evol} shows the long-term pulse period evolution of 4U 1909+07 and IGR J16418--4532, along with fractional root mean square (RMS) amplitude. The fractional RMS amplitude on the superorbital period and its uncertainty were calculated using 16.7 years Swift BAT lightcurves with 750 day intervals and 1000 day intervals for 4U 1909+07 and IGR J16418--4532 respectively, with 50 days increments in start and end times and using the fractional RMS formulation in \citet{vaughan2003}. The long-term pulse period evolution of 4U 1909+07 is shown in the left panel of Figure \ref{p_evol}a, using measurements from RXTE \citep{levine2004}, INTEGRAL \citep{furst2011}, Suzaku \citep{furst2012, jaisawal2013}, NuSTAR and AstroSat \citep{jaisawal2020}, along with the current NuSTAR measurements. The right panel of Figure \ref{p_evol}a shows the pulse period evolution of IGR J16418--4532 from all previously reported XMM observations \citep{sidoli2012,drave2013}. We do not use the spin period value from \cite{walter2006} due to large error bars of 100 seconds on the measured value of the spin period. As mentioned in Section 2.2.2, we also estimate the pulse period from a previous 2015 NuSTAR observation and a 2019 XMM PN observation of the source to investigate the long-term spin period evolution.

\section{Discussion}
\subsection{Results from the Swift BAT lightcurves}
From the dynamic power spectra constructed using the BAT lightcurves in the left panels of Figures \ref{4U1909_power}a, \ref{igrj16418_power}a and \ref{igrj16479_power}a for 4U 1909+07, IGR J16418--4532 and IGR J16479--4514 respectively, we see two prominent peaks at the fundamental superorbital period and the second harmonic, which strengthen and weaken on the timescales of $\sim$years. The presence of two peaks, one at the fundamental frequency of the superorbital period and the other one at the second harmonic, was seen in the dynamic power spectrum of 4U 1538--52 \citep{corbet2021}, although only a prominent peak at the superorbital period was seen in the dynamic power spectra of 2S 0114+650 \citep{Hu2017} and IGR J16493--4348 \citep{coley2019}. 
\par
We folded the different segments of BAT lightcurve with the superorbital periods of the three sources to illustrate the changes in the shape of the superorbital intensity profiles on the different timescales of the modulations present in the dynamic power spectra. Sections 2.1.1, 2.1.2 and 2.1.3 describe the different segments of the BAT lightcurves of 4U 1909+07, IGR J16418--4532 and IGR J16479--4514 respectively, where either the peak at the fundamental frequency and the second harmonic became stronger or weaker and consistent with the mean power. The right panels of Figures \ref{4U1909_power}, \ref{igrj16418_power} \ref{igrj16479_power} show the superorbital intensity profiles of 4U 1909+07, IGR J16418--4532 and IGR J16479--4514 respectively, at these different time segments. We calculated the fractional RMS $f_{\rm{rms}}$ for the superorbital intensity profiles and found that the $f_{\rm{rms}}$ is lower for the time periods where both the fundamental and harmonic modulations were not strongly present in the dynamic power spectrum.
\par
Figure \ref{p_evol} shows the long-term pulse period evolution of 4U 1909+07 and IGR J16418--4532 using all available spin period measurements from pointed observations (including the current work) and the RMS amplitude on the superorbital period constructed from the long term Swift BAT lightcurves. The pulse period of the NS in 4U 1909+07 spun up from 604.8 s to 602.9 s over $\sim$ 18 years, with an average spin-up rate of 4.7$\times 10^{-9}$ s s$^{-1}$. However, we see different trends of pulse period evolution and the fractional RMS amplitude for 4U 1909+07 on different timescales which could be related to the changes in strengths of the superorbital modulation at the fundamental frequency and the second harmonic. For 4U 1538--52 and 2S 0114+650, two sgHMXBs showing superorbital modulation, the strength of the superorbital modulation in the dynamic power spectra constructed with Swift BAT and RXTE ASM lightcurves, are apparently correlated with the spin period changes ${\dot P_{\rm{spin}}}$ in the source \citep{Hu2017, corbet2021}. For IGR J16493--4348, the decrease in the fractional RMS amplitude was found to be consistent with the weakening of the fundamental frequency of the superorbital modulation seen in the dynamic power spectrum \citep{coley2019}.
\par
For 4U 1909+07, a spin-up trend was observed until MJD 54,000, with an average spin-up rate ${\dot P} = (1.2\pm0.2)\times 10^{-8}$ s s$^{-1}$. The errors on the spin period changes were estimated at 1$\sigma$ confidence. From MJD 54,000 to 56,000, the spin period was steady around $\sim$ 604 s with no significant spin-up or spin-down trend.  The fractional RMS on the superorbital period of 4U 1909+07 (left panel of \ref{p_evol}b), have similar amplitudes from MJD 54,000 to MJD 56,000. After MJD 57,000, the spin-up trend continued, with an average spin-up rate ${\dot P} = (9\pm1) \times 10^{-9}$ s s$^{-1}$. The fractional RMS on the superorbital period of 4U 1909+07 (left panel of \ref{p_evol}b), show a decrease in the amplitude by 60\% from the peak value around MJD 56,000 to MJD 57,000. From the dynamic power spectrum constructed for 4U 1909+07 in the left panel of Figure \ref{4U1909_power}a, the superorbital modulation around the fundamental period became weaker around MJD 55,000 and re-appeared around MJD 57,000. This suggests a correlation between the spin-down trend, fractional RMS amplitude and strength of the superorbital modulations for 4U 1909+07, similar to that seen for 4U 1538--52 and 2S 0114+650. However, the data on the spin period evolution is sparse from MJD 56,000 to MJD 57,000, hence it is difficult to pinpoint the epoch at which the current spin-up trend started.
\par
The average spin-up rate of the pulsar in IGR J16418--4532 is ${\dot P} = (8\pm1) \times 10^{-9}$ s s$^{-1}$. The fractional RMS amplitude on the superorbital period of IGR J16418--4532 (right panel of \ref{p_evol}b), shows a decrease in the amplitude by 80\% from its peak value around MJD 56,500 to MJD 58,000. From the dynamic power spectrum constructed for IGR J16418--4532 in the left panel of Figure \ref{igrj16418_power}a, the superorbital modulation around the fundamental period became weaker around MJD 56,500 and re-appears around MJD 58,000. This suggests a possible correlation between the strength of the fundamental period of the superorbital modulations and the fractional RMS amplitude. The pulse period histories of 4U 1909+07 and IGR J16418--4532 are constructed using sparse pointed observations, unlike 4U 1538--52 and 2S 0114+650 which have been continuously monitored by Fermi GBM, RXTE ASM, Swift BAT. Hence, it is difficult to correlate the changes in the pulse periods with their changes in the strengths of the superorbital modulation due to the lack of regular monitoring.

\subsection{Results from the Swift XRT and NuSTAR observations}
\subsubsection{Lightcurves and energy-resolved pulse profiles}
The simultaneous Swift XRT and NuSTAR observations of the three sources were carried out at their predicted superorbital maximum and minimum phases for a single superorbital cycle using the ephemerides given in \cite{corbet2013}. Unfortunately, these Swift XRT and NuSTAR observations were carried out when the peak of the fundamental frequencies of the superorbital modulations had weakened, as seen in the dynamic spectra of three sources in the left panels of Figure \ref{4U1909_power}a, \ref{igrj16418_power}a and \ref{igrj16479_power}a.
\par
The lightcurves and the hardness ratios of the NuSTAR observations of 4U 1909+07 are shown in Figure \ref{4U1909_lc}. There is a short X-ray flare, about $\sim$ 3.3 ks in the NuSTAR observation at the predicted superorbital maximum phase of 4U 1909+07. The hardness ratios remain constant throughout the observations. The energy-resolved pulse profiles in the left panel of Figure \ref{pulse} exhibit a complex profile with a double-peaked structure and a dip profile at lower energies, evolving to a single-peak profile at higher energies. However, the shape of the energy-resolved pulse profiles in both observations is similar.
\par
The NuSTAR lightcurve of IGR J16418--4532 in Figure \ref{igrj16418_lc} exhibits short X-ray flares lasting a few ks. The hardness ratios indicates a softening of the spectra during these short X-ray flares, which suggests an accretion from a clumpy stellar wind around the SFXT. The energy-resolved pulse profiles shown in the right panel of Figure \ref{pulse} are complex, with the predicted superorbital maximum and minimum observations having significant changes in the pulse profiles at lower energies. The complex behavior of pulse profiles at lower energies is usually indicative of changing circumstellar environment around the NS, especially in the form of pulse phase-locked matter \citep{kretschmar2019}. Variations in the energy-resolved pulse phase profiles at different superorbital phases have been seen in Her X--1 which is attributed to the precessing accretion disks \citep{brumback2021}. However, the superorbital modulation for IGR J16418--4532 was weaker at the time when the NuSTAR observations were carried out. So it is unlikely the changes in the pulse profiles are related directly to the mechanism causing the superorbital modulation. 
\par
The NuSTAR lightcurves of IGR J16479--451 in Figure \ref{igrj16478_lc} show two large X-ray flares lasting a few ks, in addition to short X-ray flares lasting a few 100 s. Such large X-ray flares in IGR J16479--4514 have been seen previously in Suzaku and INTEGRAL observations and are thought to be associated with the orbital phase-locked matter in the stellar wind of the supergiant star, in addition to the clumpy stellar wind of the SFXT system \citep{sidoli2012, sguera2020}. There is a softening of the hardness ratios during the large X-ray flares which suggest an increased mass accretion onto the compact object during these X-ray flares. We do not detect any pulsations from the NuSTAR lightcurves and provide an upper limit on the pulse fraction of 16\% for a pulse period of 500--1000 s, the tentative spin period range estimated from its expected position on the $P_{\rm {spin}}$ vs $P_{\rm {orbital}}$ diagram \citep{corbet1986}. 

\subsubsection{Pulse-phase averaged and pulse-phase resolved spectral analysis}
We jointly fit the NuSTAR spectra with the Swift XRT spectra where applicable, with four empirical spectral models: {\tt HIGHECUT}, {\tt CUTOFFPL}, {\tt NPEX} and {\tt FDCUT}, modified with an absorption component to account for the circumstellar matter surrounding the NS and gaussian lines for the Fe fluorescence emission lines where present. The results from the four empirical spectral models are used to check for the consistency of the results.
\par
The spectra of 4U 1909+07 with two NuSTAR observations along with one simultaneous Swift XRT observation were modeled using three empirical models: {\tt HIGHECUT}, {\tt CUTOFFPL}, {\tt NPEX}. We separately carried out the spectral fitting of the short X-ray flare in the lightcurve of the NuSTAR observation of the superorbital maximum. The spectrum of 4U 1909+07 shows the presence of strong Fe fluorescence lines: at 6.4 keV Fe K$\alpha$ with an equivalent width of $\sim$ 70 eV and 7.1 keV Fe K$\beta$ line with an equivalent width of $\sim$ 30 eV. These strong Fe fluorescence lines have also been previously seen in a Suzaku \citep{furst2012}, NuSTAR and AstroSat spectrum \citep{jaisawal2020} and are attributed to the dense circumstellar medium around the source $N_{\rm{H}} \sim 7 \times 10^{22}$ \,cm$^{-2}$. Analysis of a previous Suzaku observation by \cite{jaisawal2013} detected the presence of a CRSF at $\sim$ 44 keV. However, we do not find the presence of any negative residuals in the spectra from either observation, which is consistent with the null detection of any such negative residuals in the previous NuSTAR spectra by \cite{jaisawal2020}. The values of the spectral parameters absorption column density $N_{\rm{H}}$, photon index $\Gamma$, cut-off energy $E_{\rm{C}}$, folding energy $E_{\rm{F}}$ and the X-ray fluxes are similar for both superorbital minimum and maximum observations. For the spectra of the short X-ray flare, the absorption column density $N_{\rm{H}}$ remains similar; however, the photon-index $\Gamma$ becomes harder with an increase in the X-ray flux. For a distance of 4.85 kpc, the X-ray luminosity for the observations without the X-ray flare in the energy-band of 3--70 keV was $\sim 1.3\times 10^{36}$ ergs\, s$^{-1}$ and that for the X-ray flare was $\sim 5\times 10^{36}$ ergs\, s$^{-1}$. This suggests an increase in mass accretion leading to the X-ray flare as seen for low to intermediate accretion rates in X-ray pulsars \citep{becker2012, ballhausen2017}. 
\par
The broadband Swift XRT and NuSTAR spectra of IGR J16418--4532 were fitted with all four empirical spectral models: {\tt HIGHECUT}, {\tt CUTOFFPL}, {\tt NPEX} and {\tt FDCUT}. The spectra from the predicted superorbital minimum observation and a previous 2015 NuSTAR show a 6.4 keV Fe K$\alpha$ line, which was modeled by a gaussian line. This line had previously been seen in the spectrum from an XMM observation \citep{drave2013}. However no strong residuals around 6.4 keV Fe K$\alpha$ line were seen in the predicted superorbital maximum spectra, mostly likely due to the short observation time compared to the other two observations. We estimated the upper limits of the equivalent width and line flux of an Fe K$\alpha$ line for the spectra of the observation by modelling a gaussian line with a line center fixed at 6.4 keV and fitting the spectra. The absorption column density $N_{\rm{H}}$ and photon index $\Gamma$ are similar for both observations. The X-ray flux was higher for the observation at the predicted superorbital minimum phase than the predicted maximum phase which is most likely due to the longer duration of the observation and hence the presence of more X-ray flares. For a distance of 13 kpc, the X-ray luminosity of the observation at the predicted superorbital maximum phase in the energy band of 3--60 keV was $\sim$ 1.2$\times 10^{36}$ ergs\, s$^{-1}$ and that at the predicted superorbital minimum phase and the 2015 observation was $\sim$ 3$\times 10^{36}$ ergs\, s$^{-1}$. The negative absorption-like features in the residuals to the spectral fits of the predicted superorbital minimum observation and a previous 2015 observation in Figure \ref{igrj16418_phase_avg} seen for the residuals in the spectral fits to {\tt HIGHECUT}, {\tt CUTOFFPL} and {\tt NPEX} models, were modeled by an absorption line with a gaussian optical depth profile {\tt gabs}. The statistical significance of this absorption-like feature is low as mentioned in Section 2.3.2 and hence we do not confirm or firmly rule out the presence of a CRSF in the spectra from these observations.
\par
The non-flaring part of the NuSTAR observation of IGR J16479--4514 along with one Swift XRT observation were fitted with three empirical spectral models: {\tt HIGHECUT}, {\tt CUTOFFPL}, {\tt NPEX}. The large X-ray flares seen in the NuSTAR lightcurves will be studied in detail in a separate paper. The spectral parameters $N_{\rm{H}}$, photon index $\Gamma$, cut-off energy $E_{\rm{C}}$, folding energy $E_{\rm{F}}$ and the X-ray fluxes are similar for both the predicted superorbital minimum and maximum phases, indicating no change in the spectral characteristics at the superorbital phases. For a distance of 4.5 kpc, the X-ray luminosity of the non-flaring part of the observation in the energy band of 3--40 keV was $\sim$ 2$\times 10^{35}$ ergs\, s$^{-1}$ whereas in the large X-ray flares, the X-ray luminosities increased to $\sim 10^{37}$ ergs\, s$^{-1}$.
\par
The pulse-phase resolved spectra of the NuSTAR observations of 4U 1909+07 in the energy range 3--50 keV were modeled with the spectral model {\tt HIGHECUT}, modified with an absorption component and gaussian lines to model the Fe K$\alpha$ and Fe K$\beta$ line. The spectral parameters of the Fe K$\beta$ line were kept fixed to the phase-averaged values. The left panel of Figure \ref{phase_res} shows the pulse phase evolution of the spectral parameters of 4U 1909+07: absorption column density $N_{\rm{H}}$, photon-index $\Gamma$, cut-off energy $E_{\rm{C}}$, folding energy $E_{\rm{F}}$ and the unabsorbed X-ray fluxes (in 3-50 keV). The hardness ratios were estimated for lightcurves folded with the pulse period in 3-10 keV and 10-50 keV using Equation (1). The unabsorbed X-ray flux is higher for the predicted superorbital maximum observation due to the inclusion of the X-ray flare when carrying out pulse-phase resolved spectral analysis. The hardness ratios indicate a harder spectrum around the primary peak at pulse phase 0, and a softer spectrum around the secondary broader peak at pulse phase 0.5. We see a variation in the absorption column density $N_{\rm{H}}$, photon-index $\Gamma$ and the equivalent width of an Fe K$\alpha$ line as a function of the pulse phase, similar to the results of pulse-phase resolved spectroscopic studies using a previous Suzaku observation by \cite{furst2012} and NuSTAR observation by \cite{jaisawal2020}. A harder photon-index $\Gamma \sim$ 1 is seen at the pulse phase 0--0.2 for a lower value of $N_{\rm{H}} \sim 5 \times 10^{22}$ cm$^{-2}$ at higher unabsorbed X-ray fluxes, which suggests higher accretion rates at those pulse phases. However, the spectral evolution as a function of the pulse phase remains similar for both observations, indicating no change in the emission properties at the predicted superorbital maximum and minimum phase. 
\par
For IGR J16418--4532, the energy-resolved pulse profiles exhibit a complex morphology at lower energies (left panel of Figure \ref{pulse}) which are uncorrelated to its superorbital phase. These structures in the pulse profiles at lower energies are most likely related to a variation of the absorption column densities and/or photon-index $\Gamma$ as a function of the pulse phase. The pulse-phase resolved NuSTAR spectra of IGR J16418--4532 in the energy range 3-50 keV were modeled with a {\tt HIGHECUT} model with the cut-off energy E$_{\rm{C}}$ and the fold energy E$_{\rm{F}}$ fixed to its phase-average values. The right panel of Figure \ref{phase_res} shows the pulse phase evolution of the spectral parameters absorption column density $N_{\rm{H}}$, photon index $\Gamma$ and unabsorbed X-ray flux (in 3-50 keV). The hardness ratios were estimated for lightcurves folded with the pulse period in the energy bands 3-10 and 10-50 keV using Equation (1). As seen in the phase-averaged spectral analysis, the X-ray flux during the predicted superorbital minimum observation is higher than that during the predicted superorbital maximum observation. The hardness ratios indicate a spectral softening around pulse phase 0.7-0.9, which corresponds to the trough in the pulse profiles. However due to limited statistics of the pulse-phase resolved spectra, we cannot infer any strong variations in the absorption column density $N_{\rm{H}}$ and photon-index $\Gamma$ as a function of a pulse phase. The spectral evolution as a function of pulse phase remains similar for both observations, indicating no change in the emission properties at the predicted superorbital maximum and minimum phases. 

\subsection{Comparison with different models of superorbital modulations}

\subsubsection{Precession of accretion disks}
The superorbital modulations seen in Her X--1, SMC X--1 and LMC X--4 have been attributed to the precession of the warped and tilted accretion disk which also leads to changes in the superorbital periods; for example the superorbital modulation in SMC X--1 varies from 39--72 days \citep{ogilvie2001,clarkson2003a}. The quasi-stable nature of the superorbital modulation seen in the dynamic power spectra with long-term Swift BAT lightcurves for 4U 1909+07, IGR J16418--4532 and IGR J16479--4514 (seen in left panels of Figure \ref{4U1909_power}a, \ref{igrj16418_power}a and \ref{igrj16479_power}a) suggests the absence of stable accretion disks in the system. Transient accretion disks have been observed in a few sgHMXBs such as OAO 1657--415 \citep{jenke2012} and 2S 0114+650 \citep{Hu2017}. In such systems, the neutron star is expected to spin up when a transient accretion disk is present. The spin period evolution seen in 2S 0114+650 is found to be correlated with the strength of its superorbital modulations and anti-correlated with the X-ray fluxes, indicating most likely a transition from a direct stellar wind to accretion disk-stellar wind accretion \citep{Hu2017}. The transient accretion disk lasting for 1000-1500 days could also explain the weakening or strengthening of the superorbital modulations in 2S 0114+650 although the mechanism triggering the formation of such a transient accretion disk is unclear.
\par
However, there have been no indications of the presence of a transient accretion disk in any of the five other sgHMXB systems where superorbital modulation has been conclusively detected: 4U 1909+07, IGR J16418--4532, IGR J16479--4514, IGR J16493--4348 and 4U 1538--522. For 4U 1538--522, the torque reversals are found to be uncorrelated with the X-ray fluxes \citep{corbet2021}. For many of these sgHMXBs showing superorbital modulation, there were times seen in their dynamic power spectra where the second harmonic was stronger than the fundamental frequency of the superorbital period. The current model of the precessing accretion disks is unable to account for such changes. 

\subsubsection{Tidally induced oscillations}

Tidal oscillations from a non-synchronously rotating donor star can exhibit different periodicities of the superorbital modulations \citep{koenigsberger2006,toledano2007}. Such oscillations could produce a localized structured stellar wind from the donor star which could lead to superorbital modulation seen when this stellar wind interacts with the NS and gets accreted. Tidal oscillations are predicted for systems with circular orbits, which is the case for many of the sgHMXBs except for 4U 1538--52 which might have an eccentric orbit \citep{clark2000,hemphill2019}. However the tidal circularization timescales for asynchronously rotating stars \citep{zahn1977, hut1981, toledano2007} fail to account for the quasi-stable nature of the superorbital modulations on the timescale of years seen in the Swift BAT dynamic power spectra of 4U 1909+07, IGR J16418--4532, IGR J16479--4514 (left panels of Figures \ref{4U1909_power}a, \ref{igrj16418_power}a, \ref{igrj16479_power}a), IGR J16493--4348 \citep{coley2019} and 4U 1538--52 \citep{corbet2021}. Tidal oscillations can be present in sources with eccentric orbits; however, the model predicts that the modulation will be observed near the binary orbital period \citep{moreno2011}.

\subsubsection{Corotating Interaction Regions in the supergiant wind}
A large-scale CIRs formed in the stellar wind of the supergiant stars can account for the observed superorbital variability in sgHMXBs. The quasi-stable nature of the modulations and the changing shapes of the superorbital intensity profiles provide a stronger justification for the CIR model as the mechanism driving these superorbital modulations in sgHMXBs. The clumpy stellar winds in sgHMXBs are the main reason behind the X-ray variabilities seen in sgHMXBs and its subclass SFXTs \citep{martinez2017}. The presence of large structures in the stellar winds of OB supergiant stars in the form of Discrete Absorption Components (DACs) has been confirmed using optical/UV spectroscopy \citep{underhill1975, howarth1984}. These larger structures are thought to be formed due to irregularities on the stellar surface related either to dark/bright spots, magnetic loops, or non-radial pulsations causing spiral-shaped density and velocity perturbations in the stellar wind up to several tens of stellar radii. The interactions of these CIRs with the neutron star have been proposed to be the driving mechanisms behind the superorbital modulations \citep{bozzo2017}. The quasi-stable nature of the superorbital modulations seen in the sgHMXBs provides constraints on the formation and dissipation of the CIRs in the stellar winds of the OB supergiant stars.
\par
\cite{corbet2021} showed a tight correlation between the orbital period of a CIR with that of the orbital period of the sgHMXBs, indicating a pseudo-synchronization of the CIRs with the orbit of the NS. This pseudo-synchronization could be driven by the tidal synchronization or differential rotation of the primary star. A change in the superorbital modulation profile with the harmonic becoming stronger or weaker as seen in the dynamic power spectra of 4U 1909+07, IGR J16418--4532 and IGR J16479--4514 suggests a change in the CIR structures in the stellar winds of these supergiant stars. One favorable scenario could be the appearance of multiple CIR structures in the supergiant stellar winds \citep{bozzo2017}. As shown in Figure 3 of \cite{bozzo2017}, the NS in the sgHMXB system crossing two CIRs would produce a double-peaked superorbital modulation profile. A single CIR would explain the presence of only the fundamental frequency of superorbital modulations in 2S 0114+650 \citep{Hu2017} and IGR J16493--4348 \citep{coley2019}. 

\section{Summary}
In this paper, we have investigated the superorbital modulations seen in the three sgHMXBs: 4U 1909+07, IGR J16418--4532 and IGR J16479--4514 using long-term Swift BAT lightcurves and pointed Swift XRT and NuSTAR observations. The Swift BAT lightcurves do not exhibit any significant long-term intensity variations between the different superorbital cycles for any of the three sources. The power spectra constructed from the Swift BAT lightcurves show the peak at the fundamental frequency and the second harmonic of the superorbital modulations for all the three sources. The strengths of the peaks at fundamental frequencies of the superorbital modulation and the second harmonics vary on the typical timescales of years as seen in their dynamic power spectra. The results suggest a presence of multiple CIRs in the stellar winds of the supergiant stars. However the presence of structured wind due to tidal oscillations cannot be ruled out conclusively.
\par
The pointed Swift XRT and NuSTAR observations for the three sources were scheduled to be carried out at the predicted superorbital maximum and minimum of the same superorbital cycle using the ephemerides given in \cite{corbet2013}. However, the observations occurred at the times when the superorbital modulation at the fundamental frequency was no longer strongly present for all the three sources. Hence, we do not find any significant changes between the spectral parameters of the predicted superorbital maximum and minimum phase. The NuSTAR spectra of 4U 1909+07 show the presence of a strong Fe K$\alpha$ line at $\sim$6.4 keV and Fe K$\beta$ line at $\sim$ 7.1 keV. A 3 ks X-ray flare is seen in the lightcurve of 4U 1909+07, with an unabsorbed X-ray flux of $\sim$ 2$\times 10^{-9}$ ergs \, cm$^{-2}$ \, s$^{-1}$. The spectral hardening during the X-ray flare without a change in the absorption column density $N_{\rm{H}}$ suggests an increase in mass accretion at low or intermediate accretion rates. The lightcurves of IGR J16418--4532 exhibits short X-ray flares and the hardness ratios indicate spectral softening during the flares. The negative absorption-like residuals are seen in the spectral fits to the spectral models (except {\tt FDCUT} model) for the predicted superorbital minimum observation and a previous 2015 observation, could be an artifact of spectral modelling or a CRSF which could not be confirmed or ruled out in our analysis. The lightcurves of IGR J16479--4514 shows large X-ray flares which could be related to the orbital phase-locked X-ray flares present in this system. We do not find pulsations in the lightcurves and provide an upper limit of 16\% on the pulse fraction for the putative NS in the pulse period range of 500--1000 s. 
\par
The energy-resolved pulse profiles of 4U 1909+07 and IGR J16418--4532 in Figure \ref{pulse}, exhibit a complex profile at energies$\leq$ 20 keV, which evolves to a single peak profile at higher energies. However, these changes in the energy-resolved pulse profiles are unrelated to their superorbital modulations. The evolution of spectral parameters from pulse-phase resolved spectroscopic studies of 4U 1909+07 and IGR J16418--4532 were similar for both the predicted superorbital minimum and maximum observations. The pulse period evolution of 4U 1909+07 in the left panel of Figure \ref{p_evol}a shows a possible correlation between the spin period changes, fractional RMS amplitude and changes in the strength of the superorbital modulations, similar to that seen for 2S 0114+650 and 4U 1538--52 \citep{Hu2017,corbet2021}. Regular monitoring programs of 4U 1909+07 and IGR J16418--4532 with more sensitive all sky monitors such as the enhanced X-ray Timing and Polarimetry mission (eXTP; \citealt{zhang2016}) or with pointed observatories such as NICER, NuSTAR and AstroSat will be crucial in understanding the correlations between the spin period changes, changes in the spectral parameters and the strength of the superorbital modulations.

\section*{Acknowledgement}
We thank the referee for insightful comments which has helped us improve the manuscript. The scientific results reported here are based on observations made by the NuSTAR X-ray observatory and we thank the NuSTAR Operations, Software and Calibration teams for scheduling and the execution of these observations. We have used the public lightcurves from the Swift BAT transient monitor provided by the Swift BAT team. Support for this work was provided by the NASA through Grant Number 80NSSC20K0036 and NASA award number 80GSFC21M0002.  This research has made use of software provided by the XRT Data Analysis Software (XRTDAS) developed under the responsibility of the ASI Science Data Center (ASDC), Italy and the NuSTAR Data Analysis Software (NuSTARDAS) jointly developed by the ASI Science Science Data Center (ASDC, Italy) and the California Institute of Technology. This research has made use of data and/or software provided by the High Energy Astrophysics Science Archive Research Center (HEASARC), which is a service of the Astrophysics Science Division at NASA/GSFC.

\software{XSPEC (v12.12.0; \citealt{arnaud1996}); ROBOT \citep{corbet1992}}

\bibliography{bibtex}

\begin{thebibliography}{}
\expandafter\ifx\csname natexlab\endcsname\relax\def\natexlab#1{#1}\fi
\providecommand{\url}[1]{\href{#1}{#1}}
\providecommand{\dodoi}[1]{doi:~\href{http://doi.org/#1}{\nolinkurl{#1}}}
\providecommand{\doeprint}[1]{\href{http://ascl.net/#1}{\nolinkurl{http://ascl.net/#1}}}
\providecommand{\doarXiv}[1]{\href{https://arxiv.org/abs/#1}{\nolinkurl{https://arxiv.org/abs/#1}}}

\bibitem[{{Arnaud}(1996)}]{arnaud1996}
{Arnaud}, K.~A. 1996, in Astronomical Society of the Pacific Conference Series,
  Vol. 101, Astronomical Data Analysis Software and Systems V, ed. G.~H.
  {Jacoby} \& J.~{Barnes}, 17

\bibitem[{{Asplund} {et~al.}(2009){Asplund}, {Grevesse}, {Sauval}, \&
  {Scott}}]{asplund2009}
{Asplund}, M., {Grevesse}, N., {Sauval}, A.~J., \& {Scott}, P. 2009, \araa, 47,
  481, \dodoi{10.1146/annurev.astro.46.060407.145222}

\bibitem[{{Ballhausen} {et~al.}(2017){Ballhausen}, {Pottschmidt}, {F{\"u}rst},
  {Wilms}, {Tomsick}, {Schwarm}, {Stern}, {Kretschmar}, {Caballero},
  {Harrison}, {Boggs}, {Christensen}, {Craig}, {Hailey}, \&
  {Zhang}}]{ballhausen2017}
{Ballhausen}, R., {Pottschmidt}, K., {F{\"u}rst}, F., {et~al.} 2017, \aap, 608,
  A105, \dodoi{10.1051/0004-6361/201730845}

\bibitem[{{Barthelmy} {et~al.}(2005){Barthelmy}, {Barbier}, {Cummings},
  {Fenimore}, {Gehrels}, {Hullinger}, {Krimm}, {Markwardt}, {Palmer},
  {Parsons}, {Sato}, {Suzuki}, {Takahashi}, {Tashiro}, \&
  {Tueller}}]{barthelmy2005}
{Barthelmy}, S.~D., {Barbier}, L.~M., {Cummings}, J.~R., {et~al.} 2005, \ssr,
  120, 143, \dodoi{10.1007/s11214-005-5096-3}

\bibitem[{{Becker} {et~al.}(2012){Becker}, {Klochkov}, {Sch{\"o}nherr},
  {Nishimura}, {Ferrigno}, {Caballero}, {Kretschmar}, {Wolff}, {Wilms}, \&
  {Staubert}}]{becker2012}
{Becker}, P.~A., {Klochkov}, D., {Sch{\"o}nherr}, G., {et~al.} 2012, \aap, 544,
  A123, \dodoi{10.1051/0004-6361/201219065}

\bibitem[{{Bozzo} {et~al.}(2009){Bozzo}, {Giunta}, {Stella}, {Falanga},
  {Israel}, \& {Campana}}]{bozzo2009}
{Bozzo}, E., {Giunta}, A., {Stella}, L., {et~al.} 2009, \aap, 502, 21,
  \dodoi{10.1051/0004-6361/200912131}

\bibitem[{{Bozzo} {et~al.}(2017){Bozzo}, {Oskinova}, {Lobel}, \&
  {Hamann}}]{bozzo2017}
{Bozzo}, E., {Oskinova}, L., {Lobel}, A., \& {Hamann}, W.~R. 2017, \aap, 606,
  L10, \dodoi{10.1051/0004-6361/201731930}

\bibitem[{{Brightman} {et~al.}(2019){Brightman}, {Harrison}, {Bachetti}, {Xu},
  {F{\"u}rst}, {Walton}, {Ptak}, {Yukita}, \& {Zezas}}]{brightman2019}
{Brightman}, M., {Harrison}, F.~A., {Bachetti}, M., {et~al.} 2019, \apj, 873,
  115, \dodoi{10.3847/1538-4357/ab0215}

\bibitem[{{Brumback} {et~al.}(2020){Brumback}, {Hickox}, {F{\"u}rst},
  {Pottschmidt}, {Tomsick}, \& {Wilms}}]{brumback2020}
{Brumback}, M.~C., {Hickox}, R.~C., {F{\"u}rst}, F.~S., {et~al.} 2020, \apj,
  888, 125, \dodoi{10.3847/1538-4357/ab5b04}

\bibitem[{{Brumback} {et~al.}(2021){Brumback}, {Hickox}, {F{\"u}rst},
  {Pottschmidt}, {Tomsick}, {Wilms}, {Staubert}, \& {Vrtilek}}]{brumback2021}
---. 2021, \apj, 909, 186, \dodoi{10.3847/1538-4357/abe122}

\bibitem[{{Burrows} {et~al.}(2005){Burrows}, {Hill}, {Nousek}, {Kennea},
  {Wells}, {Osborne}, {Abbey}, {Beardmore}, {Mukerjee}, {Short}, {Chincarini},
  {Campana}, {Citterio}, {Moretti}, {Pagani}, {Tagliaferri}, {Giommi},
  {Capalbi}, {Tamburelli}, {Angelini}, {Cusumano}, {Br{\"a}uninger}, {Burkert},
  \& {Hartner}}]{burrows2005}
{Burrows}, D.~N., {Hill}, J.~E., {Nousek}, J.~A., {et~al.} 2005, \ssr, 120,
  165, \dodoi{10.1007/s11214-005-5097-2}

\bibitem[{{Chaty} {et~al.}(2008){Chaty}, {Rahoui}, {Foellmi}, {Tomsick},
  {Rodriguez}, \& {Walter}}]{chaty2008}
{Chaty}, S., {Rahoui}, F., {Foellmi}, C., {et~al.} 2008, \aap, 484, 783,
  \dodoi{10.1051/0004-6361:20078768}

\bibitem[{{Chou} \& {Grindlay}(2001)}]{chou2001}
{Chou}, Y., \& {Grindlay}, J.~E. 2001, \apj, 563, 934, \dodoi{10.1086/324038}

\bibitem[{{Clark}(2000)}]{clark2000}
{Clark}, G.~W. 2000, \apjl, 542, L131, \dodoi{10.1086/312926}

\bibitem[{{Clarkson} {et~al.}(2003){Clarkson}, {Charles}, {Coe}, {Laycock},
  {Tout}, \& {Wilson}}]{clarkson2003a}
{Clarkson}, W.~I., {Charles}, P.~A., {Coe}, M.~J., {et~al.} 2003, \mnras, 339,
  447, \dodoi{10.1046/j.1365-8711.2003.06176.x}

\bibitem[{{Coleiro} {et~al.}(2013){Coleiro}, {Chaty}, {Zurita Heras}, {Rahoui},
  \& {Tomsick}}]{coleiro2013}
{Coleiro}, A., {Chaty}, S., {Zurita Heras}, J.~A., {Rahoui}, F., \& {Tomsick},
  J.~A. 2013, \aap, 560, A108, \dodoi{10.1051/0004-6361/201322382}

\bibitem[{{Coley} {et~al.}(2019){Coley}, {Corbet}, {F{\"u}rst}, {Huxtable},
  {Krimm}, {Pearlman}, \& {Pottschmidt}}]{coley2019}
{Coley}, J.~B., {Corbet}, R. H.~D., {F{\"u}rst}, F., {et~al.} 2019, \apj, 879,
  34, \dodoi{10.3847/1538-4357/ab223c}

\bibitem[{{Coley} {et~al.}(2015){Coley}, {Corbet}, \& {Krimm}}]{coley2015}
{Coley}, J.~B., {Corbet}, R. H.~D., \& {Krimm}, H.~A. 2015, \apj, 808, 140,
  \dodoi{10.1088/0004-637X/808/2/140}

\bibitem[{{Corbet} {et~al.}(2006){Corbet}, {Barbier}, {Barthelmy}, {Cummings},
  {Fenimore}, {Gehrels}, {Hullinger}, {Krimm}, {Markwardt}, {Palmer},
  {Parsons}, {Sakamoto}, {Sato}, {Tueller}, \& {Remillard}}]{corbet2006}
{Corbet}, R., {Barbier}, L., {Barthelmy}, S., {et~al.} 2006, The Astronomer's
  Telegram, 779, 1

\bibitem[{{Corbet} {et~al.}(2007){Corbet}, {Markwardt}, {Barbier}, {Barthelmy},
  {Cummings}, {Gehrels}, {Krimm}, {Palmer}, {Sakamoto}, {Sato}, {Tueller}, \&
  {Swift/Bat Survey Team}}]{corbet2007}
{Corbet}, R., {Markwardt}, C., {Barbier}, L., {et~al.} 2007, Progress of
  Theoretical Physics Supplement, 169, 200, \dodoi{10.1143/PTPS.169.200}

\bibitem[{{Corbet}(1986)}]{corbet1986}
{Corbet}, R.~H.~D. 1986, \mnras, 220, 1047, \dodoi{10.1093/mnras/220.4.1047}

\bibitem[{{Corbet} {et~al.}(2021){Corbet}, {Coley}, {Krimm}, {Pottschmidt}, \&
  {Roche}}]{corbet2021}
{Corbet}, R. H.~D., {Coley}, J.~B., {Krimm}, H.~A., {Pottschmidt}, K., \&
  {Roche}, P. 2021, \apj, 906, 13, \dodoi{10.3847/1538-4357/abc477}

\bibitem[{{Corbet} \& {Krimm}(2013)}]{corbet2013}
{Corbet}, R. H.~D., \& {Krimm}, H.~A. 2013, \apj, 778, 45,
  \dodoi{10.1088/0004-637X/778/1/45}

\bibitem[{{Corbet} {et~al.}(1992){Corbet}, {Larkin}, \& {Nousek}}]{corbet1992}
{Corbet}, R.~H.~D., {Larkin}, C., \& {Nousek}, J.~A. 1992, in Astronomical
  Society of the Pacific Conference Series, Vol.~25, Astronomical Data Analysis
  Software and Systems I, ed. D.~M. {Worrall}, C.~{Biemesderfer}, \&
  J.~{Barnes}, 106

\bibitem[{{Corbet} {et~al.}(2008){Corbet}, {Sokoloski}, {Mukai}, {Markwardt},
  \& {Tueller}}]{corbet2008}
{Corbet}, R.~H.~D., {Sokoloski}, J.~L., {Mukai}, K., {Markwardt}, C.~B., \&
  {Tueller}, J. 2008, \apj, 675, 1424, \dodoi{10.1086/526337}

\bibitem[{{Drave} {et~al.}(2013{\natexlab{a}}){Drave}, {Bird}, {Goossens},
  {Sidoli}, {Sguera}, {Fiocchi}, \& {Bazzano}}]{drave2013a}
{Drave}, S.~P., {Bird}, A.~J., {Goossens}, M.~E., {et~al.} 2013{\natexlab{a}},
  The Astronomer's Telegram, 5131, 1

\bibitem[{{Drave} {et~al.}(2013{\natexlab{b}}){Drave}, {Bird}, {Sidoli},
  {Sguera}, {McBride}, {Hill}, {Bazzano}, \& {Goossens}}]{drave2013}
{Drave}, S.~P., {Bird}, A.~J., {Sidoli}, L., {et~al.} 2013{\natexlab{b}},
  \mnras, 433, 528, \dodoi{10.1093/mnras/stt754}

\bibitem[{{Ducci} {et~al.}(2010){Ducci}, {Sidoli}, \& {Paizis}}]{ducci2010}
{Ducci}, L., {Sidoli}, L., \& {Paizis}, A. 2010, \mnras, 408, 1540,
  \dodoi{10.1111/j.1365-2966.2010.17216.x}

\bibitem[{{Farrell} {et~al.}(2006){Farrell}, {Sood}, \&
  {O'Neill}}]{farrell2006}
{Farrell}, S.~A., {Sood}, R.~K., \& {O'Neill}, P.~M. 2006, \mnras, 367, 1457,
  \dodoi{10.1111/j.1365-2966.2006.10150.x}

\bibitem[{{Farrell} {et~al.}(2008){Farrell}, {Sood}, {O'Neill}, \&
  {Dieters}}]{farrell2008}
{Farrell}, S.~A., {Sood}, R.~K., {O'Neill}, P.~M., \& {Dieters}, S. 2008,
  \mnras, 389, 608, \dodoi{10.1111/j.1365-2966.2008.13588.x}

\bibitem[{{F{\"u}rst} {et~al.}(2011){F{\"u}rst}, {Kreykenbohm}, {Suchy},
  {Barrag{\'a}n}, {Wilms}, {Rothschild}, \& {Pottschmidt}}]{furst2011}
{F{\"u}rst}, F., {Kreykenbohm}, I., {Suchy}, S., {et~al.} 2011, \aap, 525, A73,
  \dodoi{10.1051/0004-6361/201015636}

\bibitem[{{F{\"u}rst} {et~al.}(2012){F{\"u}rst}, {Pottschmidt}, {Kreykenbohm},
  {M{\"u}ller}, {K{\"u}hnel}, {Wilms}, \& {Rothschild}}]{furst2012}
{F{\"u}rst}, F., {Pottschmidt}, K., {Kreykenbohm}, I., {et~al.} 2012, \aap,
  547, A2, \dodoi{10.1051/0004-6361/201219845}

\bibitem[{{Harrison} {et~al.}(2013){Harrison}, {Craig}, {Christensen},
  {Hailey}, {Zhang}, {Boggs}, {Stern}, {Cook}, {Forster}, {Giommi},
  {Grefenstette}, {Kim}, {Kitaguchi}, {Koglin}, {Madsen}, {Mao}, {Miyasaka},
  {Mori}, {Perri}, {Pivovaroff}, {Puccetti}, {Rana}, {Westergaard}, {Willis},
  {Zoglauer}, {An}, {Bachetti}, {Barri{\`e}re}, {Bellm}, {Bhalerao},
  {Brejnholt}, {Fuerst}, {Liebe}, {Markwardt}, {Nynka}, {Vogel}, {Walton},
  {Wik}, {Alexander}, {Cominsky}, {Hornschemeier}, {Hornstrup}, {Kaspi},
  {Madejski}, {Matt}, {Molendi}, {Smith}, {Tomsick}, {Ajello}, {Ballantyne},
  {Balokovi{\'c}}, {Barret}, {Bauer}, {Blandford}, {Brandt}, {Brenneman},
  {Chiang}, {Chakrabarty}, {Chenevez}, {Comastri}, {Dufour}, {Elvis}, {Fabian},
  {Farrah}, {Fryer}, {Gotthelf}, {Grindlay}, {Helfand}, {Krivonos}, {Meier},
  {Miller}, {Natalucci}, {Ogle}, {Ofek}, {Ptak}, {Reynolds}, {Rigby},
  {Tagliaferri}, {Thorsett}, {Treister}, \& {Urry}}]{harrison2013}
{Harrison}, F.~A., {Craig}, W.~W., {Christensen}, F.~E., {et~al.} 2013, \apj,
  770, 103, \dodoi{10.1088/0004-637X/770/2/103}

\bibitem[{{Hemphill} {et~al.}(2019){Hemphill}, {Rothschild}, {Cheatham},
  {F{\"u}rst}, {Kretschmar}, {K{\"u}hnel}, {Pottschmidt}, {Staubert}, {Wilms},
  \& {Wolff}}]{hemphill2019}
{Hemphill}, P.~B., {Rothschild}, R.~E., {Cheatham}, D.~M., {et~al.} 2019, \apj,
  873, 62, \dodoi{10.3847/1538-4357/ab03d3}

\bibitem[{{Horne} \& {Baliunas}(1986)}]{horne1986}
{Horne}, J.~H., \& {Baliunas}, S.~L. 1986, \apj, 302, 757,
  \dodoi{10.1086/164037}

\bibitem[{{Howarth} {et~al.}(1984){Howarth}, {Prinja}, \&
  {Willis}}]{howarth1984}
{Howarth}, I.~D., {Prinja}, R.~K., \& {Willis}, A.~J. 1984, \mnras, 208, 525,
  \dodoi{10.1093/mnras/208.3.525}

\bibitem[{{Hu} {et~al.}(2017){Hu}, {Chou}, {Ng}, {Lin}, \& {Yen}}]{Hu2017}
{Hu}, C.-P., {Chou}, Y., {Ng}, C.~Y., {Lin}, L. C.-C., \& {Yen}, D. C.-C. 2017,
  \apj, 844, 16, \dodoi{10.3847/1538-4357/aa79a3}

\bibitem[{{Hu} {et~al.}(2019){Hu}, {Mihara}, {Sugizaki}, {Ueda}, \&
  {Enoto}}]{Hu2019}
{Hu}, C.-P., {Mihara}, T., {Sugizaki}, M., {Ueda}, Y., \& {Enoto}, T. 2019,
  \apj, 885, 123, \dodoi{10.3847/1538-4357/ab48e4}

\bibitem[{{Hut}(1981)}]{hut1981}
{Hut}, P. 1981, \aap, 99, 126

\bibitem[{{Jain} {et~al.}(2009){Jain}, {Paul}, \& {Dutta}}]{jain2009}
{Jain}, C., {Paul}, B., \& {Dutta}, A. 2009, \mnras, 397, L11,
  \dodoi{10.1111/j.1745-3933.2009.00668.x}

\bibitem[{{Jaisawal} {et~al.}(2020){Jaisawal}, {Naik}, {Ho}, {Kumari}, {Epili},
  \& {Vasilopoulos}}]{jaisawal2020}
{Jaisawal}, G.~K., {Naik}, S., {Ho}, W. C.~G., {et~al.} 2020, \mnras, 498,
  4830, \dodoi{10.1093/mnras/staa2604}

\bibitem[{{Jaisawal} {et~al.}(2013){Jaisawal}, {Naik}, \&
  {Paul}}]{jaisawal2013}
{Jaisawal}, G.~K., {Naik}, S., \& {Paul}, B. 2013, \apj, 779, 54,
  \dodoi{10.1088/0004-637X/779/1/54}

\bibitem[{{Jenke} {et~al.}(2012){Jenke}, {Finger}, {Wilson-Hodge}, \&
  {Camero-Arranz}}]{jenke2012}
{Jenke}, P.~A., {Finger}, M.~H., {Wilson-Hodge}, C.~A., \& {Camero-Arranz}, A.
  2012, \apj, 759, 124, \dodoi{10.1088/0004-637X/759/2/124}

\bibitem[{{Koenigsberger} {et~al.}(2006){Koenigsberger}, {Georgiev}, {Moreno},
  {Richer}, {Toledano}, {Canalizo}, \& {Arrieta}}]{koenigsberger2006}
{Koenigsberger}, G., {Georgiev}, L., {Moreno}, E., {et~al.} 2006, \aap, 458,
  513, \dodoi{10.1051/0004-6361:20065305}

\bibitem[{{Kretschmar} {et~al.}(2019){Kretschmar}, {F{\"u}rst}, {Sidoli},
  {Bozzo}, {Alfonso-Garz{\'o}n}, {Bodaghee}, {Chaty}, {Chernyakova},
  {Ferrigno}, {Manousakis}, {Negueruela}, {Postnov}, {Paizis}, {Reig},
  {Rodes-Roca}, {Tsygankov}, {Bird}, {Bissinger n{\'e} K{\"u}hnel}, {Blay},
  {Caballero}, {Coe}, {Domingo}, {Doroshenko}, {Ducci}, {Falanga}, {Grebenev},
  {Grinberg}, {Hemphill}, {Kreykenbohm}, {Kreykenbohm n{\'e} Fritz}, {Li},
  {Lutovinov}, {Mart{\'\i}nez-N{\'u}{\~n}ez}, {Mas-Hesse}, {Masetti},
  {McBride}, {Neronov}, {Pottschmidt}, {Rodriguez}, {Romano}, {Rothschild},
  {Santangelo}, {Sguera}, {Staubert}, {Tomsick}, {Torrej{\'o}n}, {Torres},
  {Walter}, {Wilms}, {Wilson-Hodge}, \& {Zhang}}]{kretschmar2019}
{Kretschmar}, P., {F{\"u}rst}, F., {Sidoli}, L., {et~al.} 2019, \nar, 86,
  101546, \dodoi{10.1016/j.newar.2020.101546}

\bibitem[{{Krimm} {et~al.}(2013{\natexlab{a}}){Krimm}, {Mangano}, {Romano},
  {Sbarufatti}, {Evans}, {Kennea}, {Barthelmy}, {Burrows}, {Esposito},
  {Gehrels}, \& {Vercellone}}]{krimm2013a}
{Krimm}, H.~A., {Mangano}, V., {Romano}, P., {et~al.} 2013{\natexlab{a}}, The
  Astronomer's Telegram, 5398, 1

\bibitem[{{Krimm} {et~al.}(2013{\natexlab{b}}){Krimm}, {Holland}, {Corbet},
  {Pearlman}, {Romano}, {Kennea}, {Bloom}, {Barthelmy}, {Baumgartner},
  {Cummings}, {Gehrels}, {Lien}, {Markwardt}, {Palmer}, {Sakamoto},
  {Stamatikos}, \& {Ukwatta}}]{krimm2013}
{Krimm}, H.~A., {Holland}, S.~T., {Corbet}, R.~H.~D., {et~al.}
  2013{\natexlab{b}}, \apjs, 209, 14, \dodoi{10.1088/0067-0049/209/1/14}

\bibitem[{{Lang} {et~al.}(1981){Lang}, {Levine}, {Bautz}, {Hauskins}, {Howe},
  {Primini}, {Lewin}, {Baity}, {Knight}, {Rotschild}, \&
  {Petterson}}]{lang1981}
{Lang}, F.~L., {Levine}, A.~M., {Bautz}, M., {et~al.} 1981, \apjl, 246, L21,
  \dodoi{10.1086/183545}

\bibitem[{{Leahy}(1987)}]{leahy1987}
{Leahy}, D.~A. 1987, \aap, 180, 275

\bibitem[{{Leahy}(2002)}]{leahy2002}
---. 2002, \mnras, 334, 847, \dodoi{10.1046/j.1365-8711.2002.05547.x}

\bibitem[{{Levine} {et~al.}(2004){Levine}, {Rappaport}, {Remillard}, \&
  {Savcheva}}]{levine2004}
{Levine}, A.~M., {Rappaport}, S., {Remillard}, R., \& {Savcheva}, A. 2004,
  \apj, 617, 1284, \dodoi{10.1086/425567}

\bibitem[{{Mart{\'\i}nez-N{\'u}{\~n}ez}
  {et~al.}(2015){Mart{\'\i}nez-N{\'u}{\~n}ez}, {Sander},
  {G{\'\i}menez-Garc{\'\i}a}, {G{\'o}nzalez-Gal{\'a}n}, {Torrej{\'o}n},
  {G{\'o}nzalez-Fern{\'a}ndez}, \& {Hamann}}]{martinez2015}
{Mart{\'\i}nez-N{\'u}{\~n}ez}, S., {Sander}, A., {G{\'\i}menez-Garc{\'\i}a},
  A., {et~al.} 2015, \aap, 578, A107, \dodoi{10.1051/0004-6361/201424823}

\bibitem[{{Mart{\'\i}nez-N{\'u}{\~n}ez}
  {et~al.}(2017){Mart{\'\i}nez-N{\'u}{\~n}ez}, {Kretschmar}, {Bozzo},
  {Oskinova}, {Puls}, {Sidoli}, {Sundqvist}, {Blay}, {Falanga}, {F{\"u}rst},
  {G{\'\i}menez-Garc{\'\i}a}, {Kreykenbohm}, {K{\"u}hnel}, {Sander},
  {Torrej{\'o}n}, \& {Wilms}}]{martinez2017}
{Mart{\'\i}nez-N{\'u}{\~n}ez}, S., {Kretschmar}, P., {Bozzo}, E., {et~al.}
  2017, \ssr, 212, 59, \dodoi{10.1007/s11214-017-0340-1}

\bibitem[{{Mihara}(1995)}]{mihara1995}
{Mihara}, T. 1995, PhD thesis, -

\bibitem[{{Moreno} {et~al.}(2011){Moreno}, {Koenigsberger}, \&
  {Harrington}}]{moreno2011}
{Moreno}, E., {Koenigsberger}, G., \& {Harrington}, D.~M. 2011, \aap, 528, A48,
  \dodoi{10.1051/0004-6361/201015874}

\bibitem[{{Ogilvie} \& {Dubus}(2001)}]{ogilvie2001}
{Ogilvie}, G.~I., \& {Dubus}, G. 2001, \mnras, 320, 485,
  \dodoi{10.1046/j.1365-8711.2001.04011.x}

\bibitem[{{Postnov} {et~al.}(2013){Postnov}, {Shakura}, {Staubert},
  {Kochetkova}, {Klochkov}, \& {Wilms}}]{postnov2013}
{Postnov}, K., {Shakura}, N., {Staubert}, R., {et~al.} 2013, \mnras, 435, 1147,
  \dodoi{10.1093/mnras/stt1363}

\bibitem[{{Protassov} {et~al.}(2002){Protassov}, {van Dyk}, {Connors},
  {Kashyap}, \& {Siemiginowska}}]{protassov2002}
{Protassov}, R., {van Dyk}, D.~A., {Connors}, A., {Kashyap}, V.~L., \&
  {Siemiginowska}, A. 2002, \apj, 571, 545, \dodoi{10.1086/339856}

\bibitem[{{Rajoelimanana} {et~al.}(2011){Rajoelimanana}, {Charles}, \&
  {Udalski}}]{rajoelimanana2011}
{Rajoelimanana}, A.~F., {Charles}, P.~A., \& {Udalski}, A. 2011, \mnras, 413,
  1600, \dodoi{10.1111/j.1365-2966.2011.18243.x}

\bibitem[{{Romano} {et~al.}(2008){Romano}, {Sidoli}, {Mangano}, {Vercellone},
  {Kennea}, {Cusumano}, {Krimm}, {Burrows}, \& {Gehrels}}]{romano2008}
{Romano}, P., {Sidoli}, L., {Mangano}, V., {et~al.} 2008, \apjl, 680, L137,
  \dodoi{10.1086/590082}

\bibitem[{{Romano} {et~al.}(2012{\natexlab{a}}){Romano}, {Mangano}, {Ducci},
  {Esposito}, {Evans}, {Vercellone}, {Kennea}, {Burrows}, \&
  {Gehrels}}]{romano2012}
{Romano}, P., {Mangano}, V., {Ducci}, L., {et~al.} 2012{\natexlab{a}}, \mnras,
  419, 2695, \dodoi{10.1111/j.1365-2966.2011.19916.x}

\bibitem[{{Romano} {et~al.}(2012{\natexlab{b}}){Romano}, {Barthelmy}, {Kennea},
  {Esposito}, {Evans}, {Mangano}, {Palmer}, {Sakamoto}, {Burrows}, {Chester},
  {Krimm}, {Vercellone}, \& {Gehrels}}]{romano2012a}
{Romano}, P., {Barthelmy}, S.~D., {Kennea}, J.~A., {et~al.} 2012{\natexlab{b}},
  The Astronomer's Telegram, 4148, 1

\bibitem[{{Romano} {et~al.}(2015){Romano}, {Barthelmy}, {Krimm}, {Evans},
  {Gehrels}, {Kennea}, {Gronwall}, {Malesani}, {Page}, {Sbarufatti}, \&
  {Siegel}}]{romano2015}
{Romano}, P., {Barthelmy}, S.~D., {Krimm}, H.~A., {et~al.} 2015, The
  Astronomer's Telegram, 7454, 1

\bibitem[{{Scargle}(1982)}]{scargle1982}
{Scargle}, J.~D. 1982, \apj, 263, 835, \dodoi{10.1086/160554}

\bibitem[{{Sguera} {et~al.}(2020){Sguera}, {Tiengo}, {Sidoli}, \&
  {Bird}}]{sguera2020}
{Sguera}, V., {Tiengo}, A., {Sidoli}, L., \& {Bird}, A.~J. 2020, arXiv
  e-prints, arXiv:2007.15329.
\newblock \doarXiv{2007.15329}

\bibitem[{{Sguera} {et~al.}(2006){Sguera}, {Bazzano}, {Bird}, {Dean},
  {Ubertini}, {Barlow}, {Bassani}, {Clark}, {Hill}, {Malizia}, {Molina}, \&
  {Stephen}}]{sguera2006}
{Sguera}, V., {Bazzano}, A., {Bird}, A.~J., {et~al.} 2006, \apj, 646, 452,
  \dodoi{10.1086/504827}

\bibitem[{{Sguera} {et~al.}(2008){Sguera}, {Bassani}, {Landi}, {Bazzano},
  {Bird}, {Dean}, {Malizia}, {Masetti}, \& {Ubertini}}]{sguera2008}
{Sguera}, V., {Bassani}, L., {Landi}, R., {et~al.} 2008, \aap, 487, 619,
  \dodoi{10.1051/0004-6361:20079195}

\bibitem[{{Sidoli} {et~al.}(2012){Sidoli}, {Mereghetti}, {Sguera}, \&
  {Pizzolato}}]{sidoli2012}
{Sidoli}, L., {Mereghetti}, S., {Sguera}, V., \& {Pizzolato}, F. 2012, \mnras,
  420, 554, \dodoi{10.1111/j.1365-2966.2011.20063.x}

\bibitem[{{Sidoli} {et~al.}(2013){Sidoli}, {Esposito}, {Sguera}, {Bodaghee},
  {Tomsick}, {Pottschmidt}, {Rodriguez}, {Romano}, \& {Wilms}}]{sidoli2013}
{Sidoli}, L., {Esposito}, P., {Sguera}, V., {et~al.} 2013, \mnras, 429, 2763,
  \dodoi{10.1093/mnras/sts559}

\bibitem[{{Staubert} {et~al.}(2009){Staubert}, {Klochkov}, {Postnov},
  {Shakura}, {Wilms}, \& {Rothschild}}]{staubert2009}
{Staubert}, R., {Klochkov}, D., {Postnov}, K., {et~al.} 2009, \aap, 494, 1025,
  \dodoi{10.1051/0004-6361:200810743}

\bibitem[{{Tanaka}(1986)}]{tanaka1986}
{Tanaka}, Y. 1986, {Observations of Compact X-Ray Sources}, ed. D.~{Mihalas} \&
  K.-H.~A. {Winkler}, Vol. 255, 198

\bibitem[{{Toledano} {et~al.}(2007){Toledano}, {Moreno}, {Koenigsberger},
  {Detmers}, \& {Langer}}]{toledano2007}
{Toledano}, O., {Moreno}, E., {Koenigsberger}, G., {Detmers}, R., \& {Langer},
  N. 2007, \aap, 461, 1057, \dodoi{10.1051/0004-6361:20065776}

\bibitem[{{Torrej{\'o}n} {et~al.}(2010){Torrej{\'o}n}, {Schulz}, {Nowak}, \&
  {Kallman}}]{torrejon2010}
{Torrej{\'o}n}, J.~M., {Schulz}, N.~S., {Nowak}, M.~A., \& {Kallman}, T.~R.
  2010, \apj, 715, 947, \dodoi{10.1088/0004-637X/715/2/947}

\bibitem[{{Townsend} \& {Charles}(2020)}]{townsend2020}
{Townsend}, L.~J., \& {Charles}, P.~A. 2020, \mnras, 495, 139,
  \dodoi{10.1093/mnrasl/slaa078}

\bibitem[{{Underhill}(1975)}]{underhill1975}
{Underhill}, A.~B. 1975, \apj, 199, 691, \dodoi{10.1086/153738}

\bibitem[{{Vaughan}(2005)}]{vaughan2005}
{Vaughan}, S. 2005, \aap, 431, 391, \dodoi{10.1051/0004-6361:20041453}

\bibitem[{{Vaughan} {et~al.}(2003){Vaughan}, {Edelson}, {Warwick}, \&
  {Uttley}}]{vaughan2003}
{Vaughan}, S., {Edelson}, R., {Warwick}, R.~S., \& {Uttley}, P. 2003, \mnras,
  345, 1271, \dodoi{10.1046/j.1365-2966.2003.07042.x}

\bibitem[{{Verner} {et~al.}(1996){Verner}, {Ferland}, {Korista}, \&
  {Yakovlev}}]{verner1996}
{Verner}, D.~A., {Ferland}, G.~J., {Korista}, K.~T., \& {Yakovlev}, D.~G. 1996,
  \apj, 465, 487, \dodoi{10.1086/177435}

\bibitem[{{Walter} {et~al.}(2006){Walter}, {Zurita Heras}, {Bassani},
  {Bazzano}, {Bodaghee}, {Dean}, {Dubath}, {Parmar}, {Renaud}, \&
  {Ubertini}}]{walter2006}
{Walter}, R., {Zurita Heras}, J., {Bassani}, L., {et~al.} 2006, \aap, 453, 133,
  \dodoi{10.1051/0004-6361:20053719}

\bibitem[{{Walton} {et~al.}(2016){Walton}, {F{\"u}rst}, {Bachetti}, {Barret},
  {Brightman}, {Fabian}, {Gehrels}, {Harrison}, {Heida}, {Middleton}, {Rana},
  {Roberts}, {Stern}, {Tao}, \& {Webb}}]{walton2016}
{Walton}, D.~J., {F{\"u}rst}, F., {Bachetti}, M., {et~al.} 2016, \apjl, 827,
  L13, \dodoi{10.3847/2041-8205/827/1/L13}

\bibitem[{{White} {et~al.}(1983){White}, {Swank}, \& {Holt}}]{white1983}
{White}, N.~E., {Swank}, J.~H., \& {Holt}, S.~S. 1983, \apj, 270, 711,
  \dodoi{10.1086/161162}

\bibitem[{{Wilms} {et~al.}(2000){Wilms}, {Allen}, \& {McCray}}]{wilms2000}
{Wilms}, J., {Allen}, A., \& {McCray}, R. 2000, \apj, 542, 914,
  \dodoi{10.1086/317016}

\bibitem[{{Wojdowski} {et~al.}(1998){Wojdowski}, {Clark}, {Levine}, {Woo}, \&
  {Zhang}}]{wojdowski1998}
{Wojdowski}, P., {Clark}, G.~W., {Levine}, A.~M., {Woo}, J.~W., \& {Zhang},
  S.~N. 1998, \apj, 502, 253, \dodoi{10.1086/305893}

\bibitem[{{Zahn}(1977)}]{zahn1977}
{Zahn}, J.~P. 1977, \aap, 500, 121

\bibitem[{{Zhang} {et~al.}(2016){Zhang}, {Feroci}, {Santangelo}, {Dong},
  {Feng}, {Lu}, {Nandra}, {Wang}, {Zhang}, {Bozzo}, {Brandt}, {De Rosa}, {Gou},
  {Hernanz}, {van der Klis}, {Li}, {Liu}, {Orleanski}, {Pareschi}, {Pohl},
  {Poutanen}, {Qu}, {Schanne}, {Stella}, {Uttley}, {Watts}, {Xu}, {Yu}, {in 't
  Zand}, {Zane}, {Alvarez}, {Amati}, {Baldini}, {Bambi}, {Basso},
  {Bhattacharyya S.}, {}, {Belloni}, {Bellutti}, {Bianchi}, {Brez}, {Bursa},
  {Burwitz}, {Budtz-J{\o}rgensen}, {Caiazzo}, {Campana}, {Cao}, {Casella},
  {Chen}, {Chen}, {Chen}, {Chen}, {Chen}, {Chen}, {Civitani}, {Coti Zelati},
  {Cui}, {Cui}, {Dai}, {Del Monte}, {de Martino}, {Di Cosimo}, {Diebold},
  {Dovciak}, {Donnarumma}, {Doroshenko}, {Esposito}, {Evangelista}, {Favre},
  {Friedrich}, {Fuschino}, {Galvez}, {Gao}, {Ge}, {Gevin}, {Goetz}, {Han},
  {Heyl}, {Horak}, {Hu}, {Huang}, {Huang}, {Hudec}, {Huppenkothen}, {Israel},
  {Ingram}, {Karas}, {Karelin}, {Jenke}, {Ji}, {Korpela}, {Kunneriath},
  {Labanti}, {Li}, {Li}, {Li}, {Liang}, {Limousin}, {Lin}, {Ling}, {Liu},
  {Liu}, {Liu}, {Lu}, {Lund}, {Lai}, {Luo}, {Luo}, {Ma}, {Mahmoodifar},
  {Marisaldi}, {Martindale}, {Meidinger}, {Men}, {Michalska}, {Mignani},
  {Minuti}, {Motta}, {Muleri}, {Neilsen}, {Orlandini}, {Pan}, {Patruno},
  {Perinati}, {Picciotto}, {Piemonte}, {Pinchera}, {Rachevski A.}, {Rapisarda},
  {Rea}, {Rossi}, {Rubini}, {Sala}, {Shu}, {Sgro}, {Shen}, {Soffitta}, {Song},
  {Spandre}, {Stratta}, {Strohmayer}, {Sun}, {Svoboda}, {Tagliaferri},
  {Tenzer}, {Hong}, {Taverna}, {Torok}, {Turolla}, {Vacchi}, {Wang}, {Walton},
  {Wang}, {Wang}, {Wang}, {Wang}, {Weng}, {Wilms}, {Winter}, {Wu}, {Wu},
  {Xiong}, {Xu}, {Xue}, {Yan}, {Yang}, {Yang}, {Yang}, {Yuan}, {Yuan}, {Yuan},
  {Zampa}, {Zampa}, {Zdziarski}, {Zhang}, {Zhang}, {Zhang}, {Zhang}, {Zhang},
  {Zhang}, {Zheng}, {Zhou}, \& {Zhou X.~L.}}]{zhang2016}
{Zhang}, S.~N., {Feroci}, M., {Santangelo}, A., {et~al.} 2016, in Society of
  Photo-Optical Instrumentation Engineers (SPIE) Conference Series, Vol. 9905,
  Space Telescopes and Instrumentation 2016: Ultraviolet to Gamma Ray, ed.
  J.-W.~A. {den Herder}, T.~{Takahashi}, \& M.~{Bautz}, 99051Q

\end{thebibliography}
\end{document}